\documentclass[aps,prd,amsmath,amsfonts,amssymb,eqsecnum,nofootinbib,
  floatfix,secnumarabic,preprintnumbers,nobalancelastpage,
 onecolumn,notitlepage]{revtex4-1}

\pdfoutput=1
\usepackage[utf8x]{inputenc}
\usepackage[section]{placeins}
\usepackage{subfigure,color,graphicx,hyperref,dsfont,slashed,multirow}
\usepackage{fancyvrb}
\usepackage{color}
\usepackage{hyperref}
\usepackage{fontenc}
\usepackage{pbox}
\usepackage{textcomp}

\usepackage{feynmp}

\usepackage{mathtools,nccmath}%
\usepackage{ etoolbox, xparse} 

\usepackage{array}
\usepackage[section]{placeins}
\newcolumntype{P}[1]{>{\centering\arraybackslash}p{#1}}

\graphicspath{{figs/},{figs/2l/},{figs/4l/},{figs/6l/}}

\DeclarePairedDelimiterX{\abs}[1]\lvert\rvert{\ifblank{#1}{\,\cdot\,}{#1}}

\let\oldabs\abs
\def\abs{\futurelet\testchar\MaybeOptArgAbs}
\def\MaybeOptArgAbs{\ifx[\testchar\let\next\OptArgAbs
	\else \let\next\NoOptArgAbs\fi \next}
\def\OptArgAbs[#1]#2{\oldabs[#1]{#2}}
\def\NoOptArgAbs#1{\ifblank{#1}{\oldabs{}}{\oldabs[\big]{#1}}}

\DeclarePairedDelimiterX{\set}[1]\{\}{\setargs{#1}}
\NewDocumentCommand{\setargs}{>{\SplitArgument{1}{;}}m}
{\setargsaux#1}
\NewDocumentCommand{\setargsaux}{mm}
{\IfNoValueTF{#2}{#1}{\nonscript\,#1\nonscript\;\delimsize\vert\nonscript\:\allowbreak #2\nonscript\,}}
\let\oldset\set
\def\set{\futurelet\testchar\MaybeOptArgSet}
\def\MaybeOptArgSet{\ifx[\testchar \let\next\OptArgSet
	\else \let\next\NoOptArgSet \fi \next}
\def\OptArgSet[#1]#2{\oldset[#1]{#2}}
\def\NoOptArgSet#1{\OptArgSet[\big]{#1}}

\newcommand{\tens}[1]{%
  \mathbin{\mathop{\otimes}\displaylimits_{#1}}%
}
\newcommand{\be}{\begin{equation}}
\newcommand{\ee}{\end{equation}}
\def\bsp#1\esp{\begin{split}#1\end{split}}
\renewcommand{\figureautorefname}{Fig.}

\def\sectionautorefname~#1\null{Sec.~(#1)\null}
\def\subsectionautorefname~#1\null{sub--Sec.~(#1)\null}
\def\figureautorefname~#1\null{Fig.~#1\null}
\def\tableautorefname~#1\null{Table~#1\null}
\def\equationautorefname~#1\null{Eq.~#1\null}

\def\to{\rightarrow}

\def\bsg{\ifmmode B\to X_s\gamma\else $B\to X_s\gamma$\fi}
\def\bsll{\ifmmode B\to X_s\ell^+\ell^-\else $B\to X_s\ell^+\ell^-$\fi}
\def\bstt{\ifmmode B\to X_s\tau^+\tau^-\else $B\to X_s\tau^+\tau^-$\fi}
\def\shat{\ifmmode \hat{s}\else $\hat{s}$\fi}

\newcommand{\nc}{\newcommand}
\nc{\postscript}[2] 
{\setlength{\epsfxsize}{#2\hsize}\centerline{\epsfbox{#1}}}
\nc{\non}{\nonumber}
\nc{\hc}{\hbox {h.c.}} \nc{\re}{\hbox {Re}} 
\nc{\mev}{\hbox {MeV}} \nc{\gev}{\;\hbox {GeV}} \nc{\tev}{\;\hbox {TeV}}
\def\lsim{\mathrel{\raise.3ex\hbox{$<$\kern-.75em\lower1ex\hbox{$\sim$}}}}
\def\gsim{\mathrel{\raise.3ex\hbox{$>$\kern-.75em\lower1ex\hbox{$\sim$}}}}

\nc{\etal}{{\it et al.}}
\nc{\Lsp}{\;\;\;\;\;\;\;\;\;\;}  \nc{\LLLsp}{\lspace \lspace}
\nc{\lsp}{\;\;\;\;\;\;}
\nc{\spac}{\;\;\;}
\nc{\noi}{\noindent}
\nc{\beq}{\begin{equation}}   \nc{\eeq}{\end{equation}}
\nc{\bea}{\begin{eqnarray}}   \nc{\eea}{\end{eqnarray}}
\nc{\baa}{\begin{array}}      \nc{\eaa}{\end{array}}
\nc{\bit}{\begin{itemize}}    \nc{\eit}{\end{itemize}}
\nc{\ben}{\begin{enumerate}}  \nc{\een}{\end{enumerate}}
\nc{\bce}{\begin{center}}     \nc{\ece}{\end{center}}

\def\sq2{\sqrt{2}}

\def\ph{\varphi}

\def\m4{m^4(\ph)}
\def\mn2{m_n^2}

\def\v5{V^{(5)}}

\def\lsim{\mathrel{\raise.3ex\hbox{$<$\kern-.75em\lower1ex\hbox{$\sim$}}}}
\def\gsim{\mathrel{\raise.3ex\hbox{$>$\kern-.75em\lower1ex\hbox{$\sim$}}}}

\def\slashchar#1{\setbox0=\hbox{$#1$}           % set a box for #1
   \dimen0=\wd0                                 % and get its size
   \setbox1=\hbox{/} \dimen1=\wd1               % get size of /
   \ifdim\dimen0>\dimen1                        % #1 is bigger
      \rlap{\hbox to \dimen0{\hfil/\hfil}}      % so center / in box
      #1                                        % and print #1
   \else                                        % / is bigger
      \rlap{\hbox to \dimen1{\hfil$#1$\hfil}}   % so center #1
      /                                         % and print /
   \fi}                                         %
\catcode`@=11
\long\def\@caption#1[#2]#3{\par\addcontentsline{\csname
  ext@#1\endcsname}{#1}{\protect\numberline{\csname
  the#1\endcsname}{\ignorespaces #2}}\begingroup
    \small
    \@parboxrestore
    \@makecaption{\csname fnum@#1\endcsname}{\ignorespaces #3}\par
  \endgroup}
\catcode`@=12
\begin{document}

%\preprint{CUMQ/HEP 207}

\title{Restricting the parameter space of Type-II Two Higgs Doublet Models with CP violation}

\author{Mariana Frank$^1$\footnote{Email: mariana.frank@concordia.ca}}
\author{Eric Gyabeng Fuakye$^2$\footnote{Email: ericgyabeng2012@gmail.com}}
\author{Manuel Toharia$^3$\footnote{Email: mtoharia@dawsoncollege.qc.ca}}
\affiliation{ $^1$Department of Physics,  
Concordia University, 7141 Sherbrooke St. West ,
Montreal, Qc, Canada H4B 1R6,}
\affiliation{$^2$Department of Physics, University of Regina, 3737 Wascana Pkwy, Regina, SK, Canada S4S 0A2,}
\affiliation{$^3$Department of Physics, Dawson College, 3040 Sherbrooke St W, Montreal, Qc, Canada H3Z 1A4
}

\date{\today}
\begin{abstract}
We explore  the parameter space of the Type-II Two Higgs Doublet Model with
softly broken $Z_2$ symmetry, allowing for CP violation in the scalar
potential. Imposing theory-motivated and experimentally-driven
constraints, we show that as the CP-violating phases are increased,
 only small regions of parameter space survive, including
regions slightly away from the alignment 
limit. Electroweak oblique parameters and electric dipole moments
emerge as most restrictive constraints.  We show that
imposing these constraints (as well as theoretical bounds such as perturbativity)
 the masses of the charged and heavy neutral Higgs,    unlike in the CP conserving case, are  bound from above and below. In particular, the heavy neutral Higgs masses are 
almost degenerate. In this parameter
space region we highlight the relevant decay signals of the heavy
neutral Higgs, involving both $Zh$ and $WW/ZZ$, 
indicative of CP-violation in the model.   
\end{abstract}

%%%%%%%%%%%%%%
%\pacs{12.60.Cn,12.60.Jv,14.80.Ly.}
\keywords{}%Use showkeys class option if keyword

\maketitle
%\flushbottom

%%%%%%%%%%%%%%%%%%%%%%%%%%%%%%%%%%%%%%%%%%%%%%%%%%%%%%%%%%%%%%%%%%%%%%%%%%%%%%
%%%%%%%%%%%%%%%%%%%%%%%%%%%%%%%%%%%%%%%%%%%%%%%%%%%%%%%
\section{Introduction and Motivation}
\label{sec:intro}
Our present day knowledge of particle physics, which seeks to address
the fundamental  building blocks of matter and how they interact, is
encapsulated in the Standard Model (SM). 
The Standard Model is a relativistic quantum field theory based on
$SU(3)_C\tens{} SU(2)_L\tens{} U(1)_Y$ gauge symmetry, where the group
$SU(2)_L\tens{} U(1)_Y$ is responsible for electroweak unification,
and $SU(3)_C$ provides the description of strong interactions
\cite*{Pich:2012sx, Furey:2018drh, Diaz:2003dk, Novaes:1999yn,
  Logan:2014jla}. In 2012, the LHC discovered
a spin 0 scalar boson with mass about 125 -
126 GeV \cite{Aad:2012tfa,Chatrchyan:2012xdj}, and with properties in
remarkable agreement with the Higgs boson predicted by the SM,  was
especially important,  as it confirmed experimentally  one of the key
mechanisms of the SM,  responsible for giving masses to the  SM
particles. The dynamics of this mechanism, known as electroweak
symmetry breaking (EWSB), can have 
a significant effect on the properties of SM particles. Unfortunately,
despite its remarkable success, the SM cannot be the ultimate
theory. Experimentally, it lacks a dark matter candidate, a mechanism
for neutrino masses and mixing (neutrino oscillations) or for
sufficient generation of matter-antimatter asymmetry, to name a
few. From a theoretical point of view, despite 
being the most successful model to date, the SM is inherently an
incomplete theory \cite{Lykken:2010mc}. Thus a lot of effort from
phenomenologists and theorists has been invested in models
beyond the SM (now interpreted as a low-energy effective theory), all
of which call for extending the particle content of the SM and
sometimes, of the gauge symmetry of the model.  

Extension of the particle content includes, in its simplest
realization, extending the Higgs boson content.  One of the simplest, and
most widely studied  extension of the SM Higgs sector is the Two Higgs
doublet model (2HDM), where a second scalar Higgs doublet is added,
with the same quantum numbers as the existing  one \cite{Gunion_2018, *HABER198575, *Gunion:1984yn, Branco:2011iw}. There are several
versions of the model, depending on the choice of Yukawa couplings of
the two doublets. One of these 2HDM is adopted by supersymmetry, where
up-and down-type quarks cannot acquire masses by coupling to the same
Higgs doublet, see for example \cite{Nilles:1983ge}. 

With the addition of the new doublet, the model could provide a new
source of charge-parity (CP) violation (\cite{,Accomando:2006ga,Pilaftsis:1999qt, Haber:2006ue, Branco:2011iw})
which is  fundamental  to explaining  matter-antimatter asymmetry of
the Universe \cite{Sakharov:1967dj, Kuzmin:1985mm}.

%% \cite*{Dorsch:2016nrg, Hou:2011df, Gavela:1994dt, Pilaftsis:1999qt,
%%   Carena:2015uoe, Bian:2016awe, Basler:2019iuu}.  Progress on
%% understanding this asymmetry was made %in 1967,  
%% when Sakharov proposed that CP violation was part of the three 
%% necessary conditions needed for the generation of the
%% matter-antimatter asymmetry \cite{Sakharov:1967dj}, while later %in 1985
%% , Kuzmin discovered that the CP violation generated in the SM is
%% insufficient for this purpose \cite{Kuzmin:1985mm}.

%Two versions of the 2HDMs have been studied,  one CP conserving (CPC) and one CP
%violating (CPV).
A significant difference between the CP conserving (CPC) and the CP
violating (CPV) 2HDMs is that, in a CPC 2HDM, one of the three scalar physical states,
identified as the CP-odd Higgs boson remains decoupled to the other
neutral scalars, whereas in the latter all the three neutral Higgs
states mix. One of these states is identified with the 125 GeV Higgs
boson, and all three scalar states interact with the gauge bosons.
Many detailed studies of CPC 2HDMs have been presented, for example in
\cite{Gunion:2002zf,Branco:2011iw,Haber:2015pua},  and other works such as
\cite{Grzadkowski:2015zma,Keus:2015hva,Darvishi:2016gvm,Emmanuel-Costa:2016opd}
have looked at CPV 2HDMs using CP-odd weak-basis invariants. Also,
recent studies of charged Higgs bosons phenomenology in the context of the
CPV 2HDMs have been presented in \cite{Pich:2009sp, Arhrib:2010ju, Jung:2010ik, Bao:2010sz,
  Basso:2012st} and a CPV 2HDM analysis of viable parameter space
regions which pass experimental constraints has been given in
\cite{Jung:2013hka, Brod:2013cka, Inoue:2014nva, Gaitan:2015hga,
  Chen:2015gaa}. 
 
 %In contrast to the SM Yukawa Lagrangian, 
The 2HDM Yukawa Lagrangian entails additional features with respect to
the SM Lagrangian, which are strongly constrained by experimental
data. The main difference is that, in the most general 2HDM Yukawa
Lagrangian, flavour changing neutral currents (FCNC) arise at the
tree-level \cite{Glashow:1976nt, Georgi:1978ri}%,
%Keus:2015hva,Gori:2017qwg}
, and these are strongly constrained by experimental data.
One can eliminate these dangerous FCNCs by imposing a $Z_2$ symmetry
on the scalar potential and assigning $Z_2$ charges to the
fermions. Various versions of 2HDMs were studied, which 
can be classified into four different classes based on which type of
fermions couple to a specific doublet. These classes are usually
grouped as Type I, Type II, X and Y \cite{Barger:1989fj,
  Grossman:1994jb,Akeroyd:1996he, Branco:2011iw} {\footnote{Type-X and Type-Y 2HDMs
  are often known as  lepton-specific and flipped 2HDMs
  respectively.}}.

In this paper we perform a detailed study of the CPV 2HDMs with
Type-II Yukawa couplings. In this model, the couplings of the down-type
quarks and charged leptons are proportional to $\tan\beta$, the ratio
of the two neutral Higgs bosons vacuum expectation values (VEVs), while the top quark couplings are proportional to $\cot \beta$, and
thus the model is significantly constrained by flavor physics and
direct searches of extra Higgs bosons.

Due to its connection with tree-level MSSM, the Type-II 2HDM is 
a popular extension of the Higgs sector \cite{Gunion_2018,
  Djouadi:2005gj}. While at tree-level, the MSSM Higgs sector is
strictly CP-conserving, it has been shown that under certain
circumstances, CP-violating effective Higgs sectors could arise via
loop corrections \cite{Pilaftsis:1999qt,Demir:1999hj,Choi:2000wz,Carena:2000yi,Carena:2001fw, King:2015oxa}. CPV 2HDMs effects in the 
Higgs sector have been studied extensively in the MSSM limit of the
2HDM in, for example, \cite{Choi:1999aj,Choi:2001pg,Choi:2002zp,Lee:2008eqa, Bechtle:2013wla}.

Here we analyse the consistency of the Type-II CPV 2HDM
with Higgs data, electroweak precision and perturbativity and
unitarity constraints, as well as CPV in electric dipole moments. 
Our aim is to investigate in detail the allowed parameter space in the
light of the most recent LHC results, in particular the
constraints imposed by LHC-Higgs data. In order to study 
 %have an intuition about  
the impact of CP violation on the Higgs bosons and their respective
decay channels, we consider the CPV 2HDM potential parameters
\cite{Chen:2015gaa} in the approximate $Z_2$ case
\cite{Kraml:2019sis}.
We study the various real and reduced couplings in a basis
containing the physical states with CP violation turned on by
fixing the real and imaginary couplings in the scalar potential.
The size of these couplings is constrained by perturbativity,
unitarity, as well as by electroweak precision parameters, and the
imaginary part of these couplings will be 
further constrained by Electric Dipole Moment (EDM) measurements.

As we increase the amount of CPV by turning on the phases in the Higgs
potential, we show that constraints from LHC Higgs data still allow
some regions of parameter space slightly away from the alignment
limit (the region where the SM Higgs boson decouples from the rest of 
the scalar states). In those viable regions, the masses of 
the heavy neutral Higgs bosons are tightly constrained by the theoretical
and experimental bounds.
These constraints prove to be very stringent also in regards to the 
rest of parameters such as $\tan \beta$, or the mixing angle of the
Higgs sector $\alpha$.
Production and decay rates of the heavy neutral Higgs bosons, constrained from these considerations, at the
LHC are presented and CPV specific signals are highlighted.  

Our paper is organized as follows: in Section~\ref{sec:model} we
present the Type-II Two Higgs Doublet Model with CP violation, concentrating
on the parameters in the scalar potential and their effects on the
neutral Higgs states.  In Section~\ref{sec:constraints} we explore the
bounds on the parameters in the context of both theoretical and
experimental constraints, from perturbativity and unitarity of the
potential, the Higgs data, B-decays, as well as the oblique parameters
and the EDMs.  In Section~\ref{sec:allowed}
we present the main results, showing the restrictions on the parameter
space on the charged and neutral Higgs masses, and in Section
\ref{sec:results} we analyze their production cross sections and plot
their decay patterns. We summarize and conclude in
Section~\ref{sec:summary}. 

%%%%%%%%%%%%%%%%%%%%%%%%%%%%%%%%%%%%%%%%%%%%%%%%%%%%%
\section{ Two Higgs Doublet Model with CP violation}
\label{sec:model}
%%%%%%%%%%%%%%%%%%%%%%%%%%%%%%%%
In this section we review the 2HDM considered in this study.
We first introduce the most general Two Higgs doublet scalar potential  which breaks $SU(2)_L\times U(1)$ to $U(1)_{EM}$,
\begin{eqnarray}                           %\label{pot}
V(\Phi_d, \Phi_u)&=&-\frac{1}{2}\left[m_{11}^2(\Phi_d^\dagger\Phi_d)
+\left(m_{12}^2 (\Phi_d^\dagger\Phi_u)+{\rm h.c.}\right)
+m_{22}^2(\Phi_u^\dagger\Phi_u)\right] 
\nonumber \\
&&+ \frac{\lambda_1}{2}(\Phi_d^\dagger\Phi_d)^2
+\frac{\lambda_2}{2}(\Phi_u^\dagger\Phi_u)^2+\lambda_3(\Phi_d^\dagger\Phi_d) (\Phi_u^\dagger\Phi_u) 
+\lambda_4(\Phi_d^\dagger\Phi_u) (\Phi_u^\dagger\Phi_d) 
\nonumber \\
&&+\frac{1}{2}\left[\lambda_5(\Phi_d^\dagger\Phi_u)^2 + \lambda_6 (\Phi_d^\dagger\Phi_u) (\Phi_d^\dagger\Phi_d) 
+ \lambda_7 (\Phi_d^\dagger\Phi_u) (\Phi_u^\dagger\Phi_u) +{\rm h.c.}\right] \ . 
\label{eq:pot_gen}
\end{eqnarray}
where $m_{11}, m_{22}$ and $\lambda_1,  \lambda_2, \lambda_3, \lambda_4 $  are real parameters,  while $m^2_{12},  \lambda_6$ and  $\lambda_7$ are complex, and where the Higgs doublets are parametrized as 
\begin{eqnarray}
\Phi_d=\left (\begin{array}{c} \Phi_d^+\\ \Phi_d^0\end{array} \right), \qquad \Phi_u=\left (\begin{array}{c} \Phi_u^+\\ \Phi_u^0\end{array} \right)\, .
\end{eqnarray}
To avoid tree-level  flavor changing neutral currents, one imposes a $Z_2$ symmetry with Higgs bosons obeying 
the transformation properties,
\begin{equation}
\Phi_d\rightarrow -\Phi_d\qquad \Phi_u\rightarrow \Phi_u\, .
\label{eq:z2sym}
\end{equation}
Eq.~(\ref{eq:z2sym}) implies $\lambda_6=\lambda_7=0${\footnote{This avoids hard-breaking of $Z_2$ symmetry, while soft breaking is still  allowed.}}, while we allow a non-zero $m_{12}$ to break softly the  $Z_2$ symmetry of Eq.~(\ref{eq:pot_gen}).  

After electroweak symmetry breaking, the Higgs doublets can be written in terms of the physical states as,
\begin{eqnarray}\label{VEVs}
\Phi_d=\begin{pmatrix}
-\sin\beta \, H^+ \\
\frac{1}{\sqrt2} (v \cos\beta + h_1^0 - i \sin\beta A^0)
\end{pmatrix}, \ \ 
\Phi_u=e^{i\xi} \begin{pmatrix} 
\cos\beta \, H^+ \\
\frac{1}{\sqrt2} (v \sin\beta + h_2^0 + i \cos\beta A^0) 
\end{pmatrix} \, ,
\end{eqnarray}
where $\tan\beta=v_u/v_d$,  $v=\sqrt{\mid v_u\mid^2+\mid v_d\mid^2}=246$ GeV and $H^+$ is the physical charged Higgs boson with mass 
$M_{H^+}$.  In the neutral scalar sector, there is mixing of the three states, $h_1^0, h_2^0$ and $A_0$ due to  
 two independent angles and  phases from the 
 the imaginary parts of $m_{12}$ and $\lambda_5$. 
The mixing among  the three neutral scalars can be parametrized by an orthogonal matrix ${\cal R}$ \cite{Chen:2015gaa}, 
\begin{eqnarray}\label{eq:R}
{\cal R} =\begin{pmatrix}
-\sin\alpha \cos \alpha_b & \cos \alpha \cos \alpha_b & \sin \alpha_b \\
\sin \alpha \sin \alpha_b \sin \alpha_c - \cos \alpha \cos \alpha_c & -\sin \alpha \cos \alpha_c - \cos \alpha \sin \alpha_b \sin \alpha_c & \cos \alpha_b \sin \alpha_c \\
\sin \alpha \sin \alpha_b \cos \alpha_c + \cos \alpha \sin \alpha_c & \sin \alpha \sin \alpha_c - \cos \alpha \sin  \alpha_b \cos \alpha_c & \cos \alpha_b \cos \alpha_c
\end{pmatrix} \ .
\end{eqnarray}
 with the convention:
\begin{equation}
-{\pi\over 2}< \alpha_b\le {\pi\over 2} \qquad -{\pi/ 2}\le\alpha_c\le {\pi\over 2}\, .
\end{equation}
This form of parametrization is transparent: $\alpha$  mixes the lightest Higgs  with the heavier Higgs 
(as in the CP-conserving case), $\alpha_b$  mixes the lightest Higgs
with the CP-odd component, while  $\alpha_c$  mixes the heavier Higgs
with the CP-odd component. The physical mass eigenstates are then
defined as $(h, H_2, H_3)^T = {\cal R} (h_1^0, h_2^0, A^0)^T$, with
$h$ assumed to the the SM-like boson with $M_h= 125$ GeV. This
diagonalization procedure yields seven linearly independent equations
that link the physical parameters to the parameters in the scalar
potential (see for example \cite{Inoue:2014nva,Chen:2015gaa}). 
\begin{eqnarray}
  \label{eq:lambda1}
\lambda_1 &=& \frac{M_{h}^2 \sin^2\alpha \cos^2\alpha_b + M_{H_2}^2 {\cal R}_{21}^2 
+ M_{H_3}^2 {\cal R}_{31}^2}{v^2 \cos\beta^2} - \nu \tan^2\beta \ , \\
  \label{eq:lambda2}
\lambda_2 &=& \frac{M_{h}^2 \cos^2\alpha \cos^2\alpha_b + M_{H_2}^2 {\cal R}_{22}^2 
  + M_{H_3}^2 {\cal R}_{32}^2}{v^2 \sin\beta^2} - \nu \cot^2\beta \ , \\
\label{eq:lambda_3}
\lambda_3 &=& \nu - \frac{M_{h}^2 \sin\alpha \cos\alpha \cos^2\alpha_b
  - M_{H_2}^2{\cal R}_{21} {\cal R}_{22} - M_{H_3}^2 {\cal R}_{31}
  {\cal R}_{32}}{v^2\sin\beta\cos\beta} - \lambda_4 - {\Re}\lambda_5
\ , \\
\label{eq:lambda_4}
\lambda_4 &=& 2 \nu - {\Re}\lambda_5 - \frac{2 M_{H^+}^2}{v^2} \ , \\
\label{eq:lambda5}
      {\Re}\lambda_5 &=& \nu - \frac{M_{h}^2 \sin^2\alpha_b + \cos^2\alpha_b (M_{H_2}^2 \sin^2\alpha_c + M_{H_3}^2 \cos^2\alpha_c)}{v^2} \ , \\
\label{eq:Imlambda_5}
{\Im}\lambda_5 &=& \frac{2 \cos\alpha_b \left[ (M_{H_2}^2-M_{H_3}^2) \cos\alpha \sin\alpha_c \cos\alpha_c +  (M_{h}^2 - M_{H_2}^2 \sin^2\alpha_c-M_{H_3}^2\cos^2\alpha_c) \sin\alpha \sin\alpha_b \right]}{v^2 \sin\beta}   \ , 
 \\ 
\tan\beta &=& \frac{(M_{H_2}^2 -M_{H_3}^2) \cos\alpha_c \sin\alpha_c + (M_{h}^2 -M_{H_2}^2 \sin^2\alpha_c-M_{H_3}^2 \cos^2\alpha_c) \tan\alpha \sin\alpha_b}
{(M_{H_2}^2 -M_{H_3}^2) \tan\alpha \cos\alpha_c \sin\alpha_c - (M_{h}^2 -M_{H_2}^2 \sin^2\alpha_c-M_{H_3}^2 \cos^2\alpha_c) \sin\alpha_b} \ ,
\label{eq:tanbeta}
\end{eqnarray}
where $\nu = {\Re}(m_{12})^2/({{v^2\sin2\beta}})$. 
  Note that using Eq.~(\ref{eq:tanbeta}) in
  Eq.~(\ref{eq:Imlambda_5}) we obtain a much simpler expression for the
  quartic  ${\Im}\lambda_5 $ {\it i.e.}
  \bea
\label{eq:Imlambda_52}
  {\Im}\lambda_5 &=& \frac{1}{v^2}(M_{H_2}^2  \sin^2\alpha_c + M_{H_3}^2
  \cos^2\alpha_c -M_h^2) \frac{\sin(2
    \alpha_b)}{\cos(\alpha+\beta)}\ \ \ \ \ \ \ \ \  {\rm for}
  \ \ \  (\alpha_b\neq 0).
  \eea
In the CP conserving version of the 2HDM, $\alpha_b=\alpha_c=0$, ${\cal R}$
is block diagonal, and $h$ and $H_2$ have no pseudoscalar component.

%------------------------------------------------------------------------------------
\subsection{CP Violation and Mixing in the Neutral Higgs Sector}
\label{sec:CPV}
%-------------------------------------------------------------------------------------
CP violation can occur in the 2HDM in the scalar sector, and the effects can be considerable. Setting  $\lambda_6= \lambda_7=0$ terms, to avoid hard breaking of $Z_2$ symmetry, we allow it to be only  softly broken  by the $m_{12}^2$ term, and using the tadpole equations (obtained by minimizing the Higgs potential Eq. \ref{eq:pot_gen}), we are left with 8 independent physical parameters
\begin{itemize}
\item Three scalar masses,  $M_h$, $M_{H_2}$,  and $M_{H^\pm}$;
\item Three mixing angle $\alpha$, $\alpha_b$, $\alpha_c$ in the neutral  scalar sector (one from the CP-conserving sector, two from allowing CP violation);
\item $\tan\beta= v_u/v_d$, the ratio of VEVs;
\item  ${\Re}(m_{12}^2)$, or $\nu \equiv {\Re}(m_{12})^2/({{v^2\sin2\beta}})$. 
\end{itemize}
%In this paper, we use the input parametrization in terms of the mixing angles $\alpha\, , \alpha_b\, ,\alpha_c$ as in Eq. \ref\label{eq:R} to express
%\begin{align}
%   & M_{H_2}\eqcomma            &
%   & M_{H_3}\eqcomma            &
%   & m_{H^\pm}\eqcomma          &
%   & \alpha \eqcomma &
%   & \alpha_b\eqcomma &
%   & \alpha_c\eqcomma &
%   & \tan\beta\eqcomma          &
%   & \Re(m_{12}^2)\eqcomma      &
%   & v = v_\text{EW}\eqcomma
%\end{align}
Here $M_{H_{2,3}}$ are two of the neutral Higgs masses, $M_{H^\pm}$ is the
charged Higgs mass,  the neutral-sector input
mixing angles  are $\alpha$, $\alpha_b$, $\alpha_c$,  %$\tan\beta=v_2/v_1$ is the ratio of the VEVs 
and  $\Re(m_{12}^2)$ is the real part of the soft $Z_2$ breaking parameter. 

We fix $M_h=125 GeV$. As well the third neutral scalar
mass $M_{H_3}$ is calculated from the two other ones using matrix
elements of ${\cal R}$~\cite{Fontes:2017zfn} 
\begin{equation}
  M_{H_3}^2 = \frac{M_{h}^2 {\cal R}_{13}({\cal R}_{12}\tan\beta - {\cal R}_{11}) + M_{H_2}^2 {\cal R}_{23}({\cal R}_{22}\tan\beta-{\cal R}_{21})}{{\cal R}_{33}({\cal R}_{31}-{\cal R}_{32}\tan\beta)}\, ,
\label{eq:mh3}
\end{equation}
which reduces the number of independent parameters to 7. 
In this model, $H_3$ is, in the limit of no CPV, the CP odd Higgs,
while  in the same limit $H_2$ is the heavy  CP-even Higgs. We  require
the lightest Higgs properties to be consistent with those of the
SM-like boson observed at the LHC, and in particular, set its mass  to
$ M_h=125.35 \pm 0.15$ GeV \cite{Sirunyan:2020xwk}. We proceed by
listing the restrictions imposed on the parameter space of the
model. The imaginary part of $\lambda_5$, which is a source of CP
violation is given in Eq. \ref{eq:Imlambda_5}. 

%%----------------------------------------------------------
%\subsection{Reduced Couplings with fermions and bosons of CPV admixtures}
%\label{subsec:reducedcoup}
%%-------------------------------------------------------------

We concentrate on  2HDMs with a $Z_2$ symmetry in the Yukawa sector and where  $\Phi_d$ and $\Phi_u$  give mass to only up or down type fermions, respectively. This corresponds to a CPV version of Type-II 2HDMs, leading to
suppressed tree-level flavor changing processes mediated by the neutral Higgs scalars. In general, the imposed $Z_2$ symmetry in the Yukawa sector is not preserved by renormalization, as
 terms from the Higgs potential will induce couplings of $\Phi_d$, $\Phi_u$ to both up and down type quarks. However, this does not reintroduce any tree level flavor changing effects because the induced Yukawa matrices are still proportional to the corresponding fermion mass matrices.
Neglecting mixing from the CKM matrix, the Yukawa Lagrangian is
\begin{eqnarray}                           
    \label{Yuk}
\mathcal{L}_Y= 
-\biggl(  \displaystyle \frac{\cos\alpha}{\sin\beta}\frac{m_u}{v} \biggr)\overline Q_L (i\tau_2) \Phi_u^* u_R 
+\biggl( \frac{\sin\alpha}{\cos\beta}\frac{m_d}{v} \biggr)\overline Q_L \Phi_d d_R +\biggl( \frac{\sin\alpha}{\cos\beta}\frac{m_\ell}{v} \biggr)\overline L_L \Phi_d e_R
+ {\rm h.c.} 
\end{eqnarray}
where $Q_L^T=(u_L,d_L), L_L^T=(\nu_l, \ell_L)$, \ $u$, $d$ and $\ell$ stand for up and down-type quarks and leptons, respectively.
%We also assume that the charged lepton Yukawa couplings have the same form as for $d$-type  quarks.
%Under the $Z_2$ symmetry, $Q_L, u_R, \Phi_u$ are always even, $\Phi_d$ is always odd, and $d_R$ is odd in 
%Type II models.
The couplings between neutral Higgs bosons and the fermions and gauge bosons are
\begin{eqnarray}
\mathcal{L}_{NC} = \sum_{i=1}^3 \left[-m_f\left( c_{H_i ff} \bar f f+ \tilde c_{H_i ff} \bar f i\gamma_5 f  \right) \frac{H_i}{v} + \left( 2 a_{H_i} m_W^2 W_\mu W^\mu + a_{H_i} m_Z^2 Z_\mu Z^\mu \right) \frac{H_i}{v}\right] \ .
\label{coup_f}
\end{eqnarray}
where $H_i=h, H_2, H_3$ and  $c_{H_i ff}\, , {\tilde c}_{H_i ff} \neq 0$ or $a_{H_i}\,  ,{\tilde c}_{H_i ff}\neq 0$ indicates CP-violation, that is, the mass eigenstate $H_i$ couples to both CP even and CP odd operators.
The coefficients $c_{H_i ff}$, $\tilde c_{H_i ff}$ and $a_{H_i}$ are obtained using the matrix ${\cal R}$ defined above in terms of $\alpha$, $\alpha_b$, $\alpha_c$ and $\tan\beta$ and are given  in Table \ref{tab:Hcouplings}.
%\begin{eqnarray}\label{Hcouplings}
\begin{table}[h]
\centering{\begin{tabular}{|c|c|c|c|c|}
\hline
 $c_{H_i tt}$ & $c_{H_i bb}=c_{H_i \ell \ell}$ & $\tilde c_{H_i tt}$ & $\tilde c_{H_i bb}=\tilde c_{H_i \ell \ell}$ & $a_{H_i}$ \\
%\hline
%Type I & $R_{i2}/\sin\beta$ & $R_{i2}/\sin\beta$ & $-R_{i3}\cot\beta$ & $R_{i3}\cot\beta$ & 
%$R_{i2}\sin\beta+R_{i1}\cos\beta$ \\
\hline
%Type II & 
${\cal R}_{i2}/\sin\beta$ & ${\cal R}_{i1}/\cos\beta$ & $-{\cal R}_{i3}\cot\beta$ & $-{\cal R}_{i3}\tan\beta$ & 
${\cal R}_{i2}\sin\beta+{\cal R}_{i1}\cos\beta$ \\
\hline
\end{tabular}} \ \ \ 
\caption{Fermion and gauge boson couplings to Higgs mass eigenstates.}
\label{tab:Hcouplings}
\end{table}

Thus we can express the couplings presented in Table~ \ref{tab:Hcouplings}~ in terms of the  mixing angles in the scalar sector of the CPV 2HDM,  $\alpha$ (CPC), $\alpha_b$ and $\alpha_c$ (CPV). They represent the $12$, $13$ and $23$ angles from the $3\times3$ orthogonal diagonalization matrix ${\cal R}$ given in Eq. \ref{eq:R}. For the SM-like  Higgs boson, the mixing elements are
\begin{equation}
{\cal R}_{11}= - \sin \alpha \cos \alpha_b\, , \qquad
{\cal R}_{12}= \cos \alpha \cos \alpha_b\, , \qquad
{\cal R}_{13}= \sin \alpha_b
\end{equation}
%In the case of the fermionic sector, the general vector and axial--vector structure of the Higgs coupling to fermions has been expressed in \cite{Bernon:2015hsa} as 
%\begin{eqnarray}
%   H_i f\bar f: \quad   -\bar f (\Ree(c_{H_i ff})+i\Imm(c_{H_i ff})\gamma_5) f \, \frac{gm_f}{2M_W} \,,
%\label{eq:hfit-CP-Hff}
%\end{eqnarray}
%where in the SM one has $\Ree(c_{h ff})=1$ and $\Imm(c_{h ff})=0$, while a purely CP-odd Higgs would have $\Ree(c_{H_i ff})=0$ and $\Imm(c_{H_i ff})=1$.
For example, the normalized coupling of top quarks to the SM-like Higgs
has the simple expression \cite{Muhlleitner:2020wwk} (in terms of $\alpha, \beta$ and $\alpha_b$){\footnote {Note that the Yukawa couplings for the neutral Higgs bosons $H_i$ with the up-type quarks in the CPV 2HDM are $\displaystyle \frac{R_{i2}}{\sin \beta}- i \frac{R_{i3}}{\tan \beta}$, with the first term from the CP-even Higgs coupling, and the second from the CP-odd coupling.}}
:
%used as free parameter  is
\begin{eqnarray}
|C_{htt}|^2 &=&
\left(\frac{{\cal R}_{12}}{\sin \beta}\right)^2+\left(\frac{{\cal R}_{13}}{\tan \beta}\right)^2
=\left(\frac{\cos \alpha \cos \alpha_b}{\sin \beta}\right)^2+\left(\frac{\sin \alpha_b}{\tan \beta}\right)^2.%\nonumber \\
\end{eqnarray}
%using $\displaystyle \frac{\cos \alpha }{\sin \beta} =
%\sin(\beta-\alpha) + \frac{\cos(\beta-\alpha)}{\tan \beta}$, 
%\bea
%|c_{tt}|^2 
%\eea
%which explicitly depends on $\tan \beta$, $\cos(\alpha-\beta)$ and $\sin \alpha_b$.
%The same way we obtain the normalized coupling of Higgs to vector bosons as
%\begin{equation}
%|c_{H_i VV}|^2 = \left(R_{i2} s_\beta+R_{i1} c_\beta\right)^2= \begin{qrray (1-c^2_{\beta\alpha})  (1-s^2_{\alpha_b}) 
%\label{eq:cvv}
%\end{equation}
%where $V=W, Z$ and which also explicitly depends on $\tan \beta$, $\cos(\alpha-\beta)$ and $\sin \alpha_b$ .
%\bea
%R_{23} = \sqrt{1-s^2_{\alpha_b}}  s_{\alpha_c}
%\eea
When the Higgs boson mixes with both the CP-even and CP-odd states, the Higgs coupling to vector bosons $V$ is 
$\displaystyle
 a_{h}\, \frac{gm_V^2}{M_W}\, g^{\mu \nu} \,,
$ \cite{Bernon:2015hsa}, 
where as above $a_{h}$  measures the departure from the SM,  $a_{h}=1$ for a pure CP-even state with SM-like couplings and $a_{H_i}=0$ for a pure CP-odd state.   

The  effects of CP mixing will appear also at the loop level,
especially in the $gg\to h$ and $h\to \gamma\gamma$ rates. At leading
order, the Higgs production rates normalized to the SM expectations
can be written as \cite{Bernon:2015hsa, Djouadi:2013qya} 
\begin{eqnarray}
\frac{\Gamma( h \to \gamma\gamma)}{\Gamma^{\rm SM}( h \to \gamma\gamma)} 
&  \simeq & 
\frac{\left| \frac{1}{4} a_{h} A^+_1[m_W] + \left(\frac{2}{3}\right)^2
  {\Re}(C_{h tt}) +  \left(\frac{1}{3}\right)^2  A^S_{1/2}[m_b] {\Re}(C_{hbb})  \right|^2
+ \left| \left(\frac{2}{3}\right)^2 \frac{3}{2} {\Im}(C_{h tt})  +  \left(\frac{1}{3}\right)^2  A^P_{1/2}[m_b] {\Im}(C_{hbb})        \right|^2} 
{\left| \frac{1}{4} A^+_1[m_W] + \left(\frac{2}{3}\right)^2  \right|^2 } \,  
\nonumber \\ 
\frac{\sigma( gg \to h)}{\sigma^{\rm SM}( gg \to h)} & = & 
\frac{\Gamma( h \to gg)}{\Gamma^{\rm SM}( h \to gg)}  \simeq 
\left| {\Re}(C_{htt})+A^S_{1/2}[m_b] {\Re}(C_{hbb})  \right|^2 + \left|\frac{3}{2} {\Im}(C_{h tt})  +A^P_{1/2}[m_b] {\Im}(C_{hbb}) 
\right|^2 \,, 
\label{eq:hfit-widthsCP} 
\end{eqnarray}
with the loop form factors given  by \cite{Gunion:2002zf} $A^+_1[m_W] \simeq -8.32$, $A^S_{1/2}[m_b]\simeq  -0.063+0.090\ i
$ and $A^P_{1/2}[m_b]\simeq -0.072+0.090\ i$  for $M_h = 125$ GeV.
%%%%%%%%%%%%%%%%%%%%%%%%%%%%%%%%%%%%%%%%%%%%%%%%%%%%%
\section{Constraints}
\label{sec:constraints}
%%%%%%%%%%%%%%%%%%%%%%%%%%%%%%%%%%%%%%%%%%%%%%%%%%%%%%%%
The CP-violating 2HDM  is constrained from various theoretical observations and from measurements at collider
and non-collider experiments. We first discuss constraints on model parameters from theoretical considerations, then those from the Higgs data, 
from $B$-physics measurements, from precision data (the $S, T, U$ oblique parameters) and from measurements
of EDMs. We restrict ourselves to a brief discussion, and refer to explicit expressions that have appeared before.
%---------------------------------------------------------------------------------
\subsection{Perturbativity, vacuum stability and unitarity}
\label{subsec:theory}
%---------------------------------------------------------------------------------
Constraining the theory to be perturbative imposes constraints on all the couplings in the scalar potential
\begin{equation}
|\lambda_i| \lsim 8 \pi\, .
\label{eq:th_const1}
\end{equation}
In addition, vacuum stability requires  the scalar potential,
Eq. \ref{eq:pot_gen},  to be be positive for large values of $\Phi_d\,
, \Phi_u$ 
\begin{equation}
\lambda_1>0, \qquad \lambda_2>0, \qquad \lambda_3+\sqrt{\lambda_1 \lambda_2} >0, \qquad \lambda_3+\lambda_4-|\lambda_5| +\sqrt{\lambda_1 \lambda_2}>0\, .
\label{eq:th_const2}
\end{equation}
Both unitarity and perturbativity requirements.
lead to the constraints   \cite{Branco:2011iw,Ginzburg:2004vp}:
\begin{eqnarray}
\label{eq:th_const3}
\left|~\frac{1}{2}\left( \lambda_1+\lambda_2\pm \sqrt{(\lambda_1-\lambda_2)^2+4 |\lambda_5|^2}\right)\right| &<& 8\pi\,,\\ \nonumber
\left|~\frac{1}{2}\left(\lambda_1+\lambda_2\pm \sqrt{(\lambda_1-\lambda_2)^2+4\lambda_4^2} \right) \right| &<& 8 \pi\,, \\ \nonumber
\left|~\frac{3(\lambda_1+\lambda_2)\pm\sqrt{9(\lambda_1-\lambda_2)^2+4(2\lambda_3+\lambda_4)^2)} }{2} \right| &<& 8 \pi\,, \\\nonumber
|~\lambda_3\pm\lambda_4|<8 \pi\,, \quad |~\lambda_3\pm |\lambda_5||<8 \pi\,,\quad |~\lambda_3+2 \lambda_4\pm 3 |\lambda_5|| &<& 8 \pi\,.
%\label{eq:th_const3}
\end{eqnarray}
As the Eqs.~(\ref{eq:th_const1}) - (\ref{eq:th_const3}) only depend on the absolute value of
$\lambda_5$, they do not constrain the CPV phases. 

These conditions are very important for the CP-violating model, as they provide bounds on the masses of the heavy neutral and charged Higgs.  From Eqs. \ref{eq:lambda_4} - \ref{eq:lambda5},
% \begin{eqnarray}
% \lambda_4 &=& 2 \nu - {\Re}\lambda_5 - \frac{2 M_{H^+}^2}{v^2} \ , \nonumber  \\
%{\Re}\lambda_5 &=& \nu - \frac{M_{h}^2 \sin^2\alpha_b + \cos^2\alpha_b (M_{H_2}^2 \sin^2\alpha_c + M_{H_3}^2 \cos^2\alpha_c)}{v^2}. \nonumber
%\end{eqnarray}
requiring $-8 \pi \le \lambda_4,  {\Re}\lambda_5 \le 8\pi$, we obtain the from-above and from-below bounds for $M^2_{H^\pm}$

\begin{eqnarray} 
 && M_{h}^2 + \cos^2\alpha_b\ (\tilde{M}^2)  -8 \pi v^2 \le
  M_{H^\pm}^2 \le   M_{h}^2 + \cos^2\alpha_b\ (\tilde{M}^2)+ 8 \pi v^2  
\end{eqnarray}
where we have introduced the mass combination $\tilde{M}^2 =  M_{H_2}^2
\sin^2\alpha_c + M_{H_3}^2 \cos^2\alpha_c -M_h^2 $, which is 
positive, as long as $M_h^2<M^2_{H_i}$. If we now consider
Eq. \ref{eq:Imlambda_52} and require $-8 \pi \le {\Im}\lambda_5 \le
8\pi$, we obtain an upper bound for $\tilde{M}^2$
\begin{eqnarray} 
\label{eq:mtilde}
  \tilde{M}^2  &\le&  8 \pi v^2 \frac{|\cos(\alpha+\beta)|}{|\sin2\alpha_b|}.  
\end{eqnarray}
And since $\tilde{M}^2$ is present in the inequality involving $M_{H^\pm}^2$,
we can finally obtain the upper bound
\begin{eqnarray} 
  M_{H^\pm}^2  \le   M_{h}^2 + 4 \pi v^2 \left(2+ \frac{|\cos(\alpha+\beta)|}{|\tan{\alpha_b}|}\right)      \ \ \ \ \ \ \ \ {\rm for}
  \ \ \ \ (\alpha_b \neq 0) .
\end{eqnarray}
Note that this bound comes from requiring perturbativity on the
quartic couplings $\lambda_4$,  $\Re\lambda_5$  and $\Im\lambda_5$ and so is not necessarily
the strongest bound.  Similar upper bounds can be obtained for both
$(M_{H_2}^2+M_{H_3}^2)$ and $\nu$.
From Eq. \ref{eq:lambda5} we obtain
\bea
 8\pi \left(\frac{|\cos(\alpha+\beta)|}{|\sin2\alpha_b|}-1\right) + \frac{M_h^2}{v^2}
 \le\  \nu\  \le
 8\pi \left(\frac{|\cos(\alpha+\beta)|}{|\sin2\alpha_b|}+1
\right) + \frac{M_h^2}{v^2}\ \ \ \ \ \ \ \ {\rm for}
  \ \ \ \ (\alpha_b \neq 0) .
\eea
From Eqs. \ref{eq:lambda1} - \ref{eq:lambda2}, imposing $-8 \pi \le \lambda_1, \lambda_2 \le 8\pi$, we obtain
  \begin{equation}
 ( \nu-8\pi ) v^2 \le M_{H_2}^2 +M_{H_3}^2 -\cos^2 \alpha_b \left
    (\tilde{M}^2  \right) \le  ( \nu+8\pi) v^2 \,
  \end{equation}
and finally, using the previous bounds on $\tilde{M}^2$ we obtain
\bea
(M_{H_2}^2 +M_{H_3}^2 ) \le  M_h^2 + 8\pi v^2 \left(2+  \frac{|\cos(\alpha+\beta)|}{|\sin2\alpha_b|}(1+\cos\alpha_b^2)   \right) \ \ \ \ \ \ \ \ {\rm for}
  \ \ \ \ (\alpha_b \neq 0) .
\eea

%---------------------------------------------------
\subsection{SM Higgs data}
\label{subsec:higgs}
%------------------------------------------------------

The global data set on the SM Higgs boson includes results from LEP,
the Tevatron and the more recent LHC experiments.
 In order to properly combine the constraints coming from the LHC-Higgs data, we used {\tt Lilith}
\cite{Bernon:2015hsa, Bertrand:2020lyb}. {\tt Lilith} is an open-source library written in {\tt Python},  which can be  used
in any {\tt Python} script as well as in {\tt C} \cite{kernighan2006c}
and {\tt C++}\cite{stroustrup2000c++}/{\tt ROOT}\cite{brun1997root}
codes, with a command-line interface. At present, {\tt Lilith} includes the complete data on Higgs measurements at the Tevatron and LHC  Run II at 36 fb$^{-1}$, and only the Higgs production  and diboson decay at 139 fb$^{-1}$. {\footnote{Other codes such as  {\tt
  HiggsBounds}\cite{Bechtle:2013wla} or {\tt HiggsSignals}
\cite{Bechtle:2015pma} perform similar data analysis but {\tt Lilith} uses
as  primary inputs results where  production and decay modes are unfolded
from experimental categories, while for example {\tt HiggsSignals} code uses signal strengths for individual
measurements, taking into account the associated efficiencies. The
user inputs in  {\tt Lilith} are given in terms of reduced couplings
or signal strengths for one or multiple Higgs states.} }

The SM-like Higgs signal rates and masses are compared
with the  signal rate measurements, with the additional requirement
that the lightest Higgs has a same mass and production and decay
properties as  the observed Higgs peak in the channels with high mass
resolutions $h \to ZZ^\star \to 4 \ell$ and $h \to \gamma \gamma$. The
signal strength for the 2HDM versus the SM is defined as 
\begin{equation}
\mu_i =\frac{\sigma \left [( gg \to h) \times BR(h \to X_i) \right] _{\rm 2HDM}}{\sigma \left [( gg \to h) \times BR(h \to X_i) \right]_{\rm SM}} \times \omega_i \, ,
\end{equation}
where $\omega_i$  are the experimental weights of the measurement which includes the experimental efficiency.

%------------------------------------------------------
\subsection{$B$ Physics constraints}
\label{subsec:bphysics}
%------------------------------------------------------
The charged Higgs bosons in the 2HDM contribute to the decays of $B$ mesons and thus $B$-physics data set can be used to constrain their masses and couplings. Note that, since the couplings of the
charged Higgs bosons are not sensitive to the parameters in the neutral sector, these constraints
are independent of the amount of CP violation in the model.
The most stringent constraints on model parameters emerge from the $B \to X_s \gamma$ \cite{Amhis:2019ckw}
\begin{equation}
BR(B \to  X_s \gamma) \le (3.32 \pm 0.15) \times 10^{-4}\, ,
\end{equation}
and the most restrictive conditions on  charged Higgs masses are for $\tan \beta=1$, where $M_{H^\pm} \ge 580$ GeV \cite{Antusch:2020ngh}.

%------------------------------------------------------------
\subsection{S, T, U parameters}
\label{subsec:stu}
%------------------------------------------------------------
This theoretical constraint comes from the oblique parameters, constrained by the global fit \cite{Baak:2012kk,Enomoto:2015wbn,ParticleDataGroup:2020ssz} to be
\begin{equation}
S=0.00 \pm0.07 \, , \qquad T=0.05 \pm 0.06\, , \qquad U=0.03 \pm 0.10\, .
\label{eq:stu}
\end{equation}
The oblique parameters $S, T$, and $U$  are observables that combine electroweak precision data to quantify deviation from the SM and thus are used in any  BSM to ensure that the model is consistent with the data. %These parameters were calculated at one loop in the basis-independent CP-violating 2HDM in \cite{Funk:2011ad}. 

An alternative way to evaluate the oblique parameters is to fix $U = 0$,
motivated by the fact that $U$ is suppressed by an additional factor $M_Z^2/M_H^2$ compared
$S$ and $T$, which  improves the precision on $S$ and particularly on the $T$ parameter, yielding
$S=0.00\pm 0.07$, $T=0.05 \pm 0.06$.

The implication of electroweak precision on the mass spectrum of heavier neutral and on the charged Higgs was thoroughly investigated before \cite{Funk:2011ad}, in the context of CPV 2HDM. It was shown that, except for the $T$ parameter, $S$ and $U$  fall well within the experimental bounds. Thus working in the limit $U$=0 is more conservative, as it restricts the $T$ parameter more. Allowing $U \ne 0$ indicates a strong correlation between $T$ and $U$ parameters, with values of $T>1$ corresponding to values of $U>0.01$. In general, the 2HDM  can generically produce values for $T$ that exceed the experimental bounds, limiting much of the parameter space.

However, imposing mass splitings  \cite{Chen:2015gaa}:

$-80$ GeV $\le M_{H^\pm} -M_{H_2} \le$ 100 GeV, and 

$-600$ GeV $\le M_{H^\pm} -M_{H_3} \le$ 100 GeV,

insures that the electroweak precision lie in the allowed ranges.

We shall see that our  parameter space surviving all constraints is highly degenerate and thus the electroweak corrections are well within experimental bounds.
The $S, T$ and $U$ parameters finite and in principle, observable. Their expressions in are found it, e.g., ~\cite{Grimus:2007if}, so we do not repeat them here.

%------------------------------------------------------------
\subsection{Electric Dipole Moments}    
\label{subsec:edm}
%------------------------------------------------------------
The electron EDM measurement places a very strict constraint on the complex Yukawa couplings in most models. In general, the electric dipole moment of a fermion $f$ corresponds to the  imaginary parts of the Wilson coefficient $d_f$ of the effective operator
\begin{eqnarray}
\mathcal{L}_{eff} = d_f \bar f_L \sigma_{\mu\nu} f_R F^{\mu\nu} +{\rm h.c.} \, .
\end{eqnarray}
 Complete expressions for the one-loop evaluation of the EDMs can be found in \cite*{Abe:2013qla, Inoue:2014nva,BowserChao:1997bb}. In addition to one-loop contributions, EDMs can originate from two-loop Barr Zee contributions \cite{Barr:1990vd}, which are proportional to a single power of the electron Yukawa coupling \cite{Chun:2019oix,Altmannshofer:2020shb}. These contributions are more important for Type-X (lepton-specific) 2HDM, and still, even in Type-X, agreement with low-energy
data for the muon anomalous magnetic moment (and significant EDM contributions) require very light pseudoscalar masses and very large values of $\tan \beta$ \cite{Wang:2018hnw,Chun:2015xfx,Wang:2014sda,Broggio:2014mna,Ilisie:2015tra,Abe:2015oca}, inconsistent with the parameter space analyzed in this model.  It has also been shown that, in unlike in the Type-I 2HDM, the electron and mercury EDMs are not able to probe the parameter space  when $\tan \beta$ is close to 1, which will be our case, due to the cancellation in Barr-Zee diagrams \cite{Chen:2017com}. We have not considered these contributions here.

A recent electron EDM measurement was performed using the ThO molecule in \cite{Andreev:2018ayy}. The constrained quantity is
\begin{equation}
|d_e^{\rm eff}| <1.1 \times 10^{-29} e \cdot{\rm cm} \, .
\end{equation}
In 2HDM the CP violating  phase  strongly affects the magnitude of the
EDM through the modified couplings of the Higgs bosons. In Type-II
2HDM the EDM is also expected to be enhanced by $\tan \beta$
\cite{Amhis:2019ckw}. By investigating CP violating effects in
extended Higgs sector, the electric dipole moment  and  its effects on
produced particles via protons, photons, or electron and positron
collisions have all been studied \cite{Mendez:1991gp,
  Bernreuther:1993hq, Ogreid:2018xdx}. We impose these constraints and
explore their effects further in our numerical explorations.

%%%%%%%%%%%%%%%%%%%%%%%%%%%%%%%%%%%%
\section{Allowed Parameter Space}
\label{sec:allowed}
%%%%%%%%%%%%%%%%%%%%%%%%%%%%%%%%%%%%%%

%%%%%%%%%%%%%%%%%%%%%%%%%%%%%%%%%%%%%%%%%%%%%%%%%%%%%%%%%%%%%
\begin{figure}[htbp]
\begin{centering}
\includegraphics[width=8.1cm]{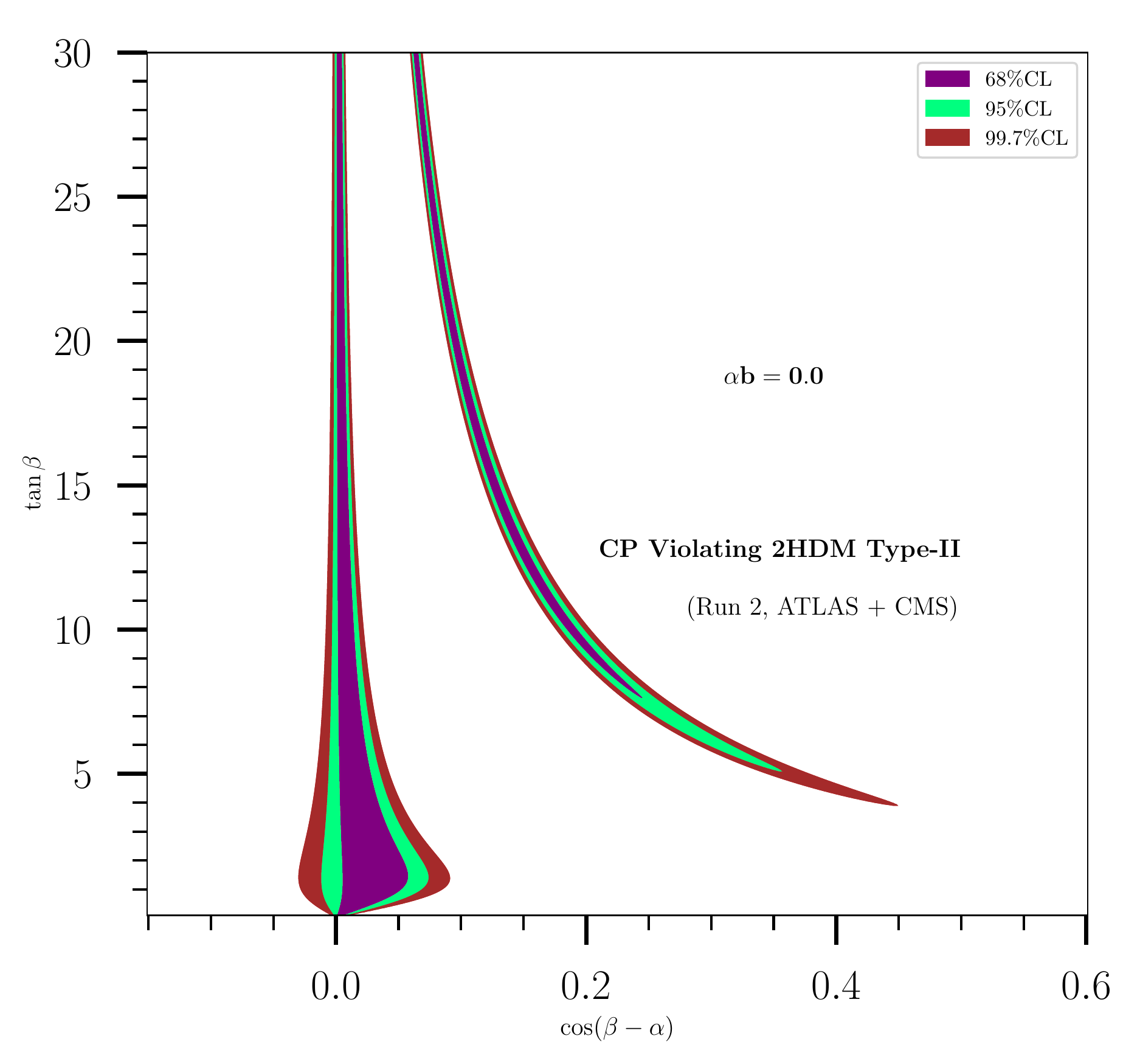}
\includegraphics[width=8.1cm]{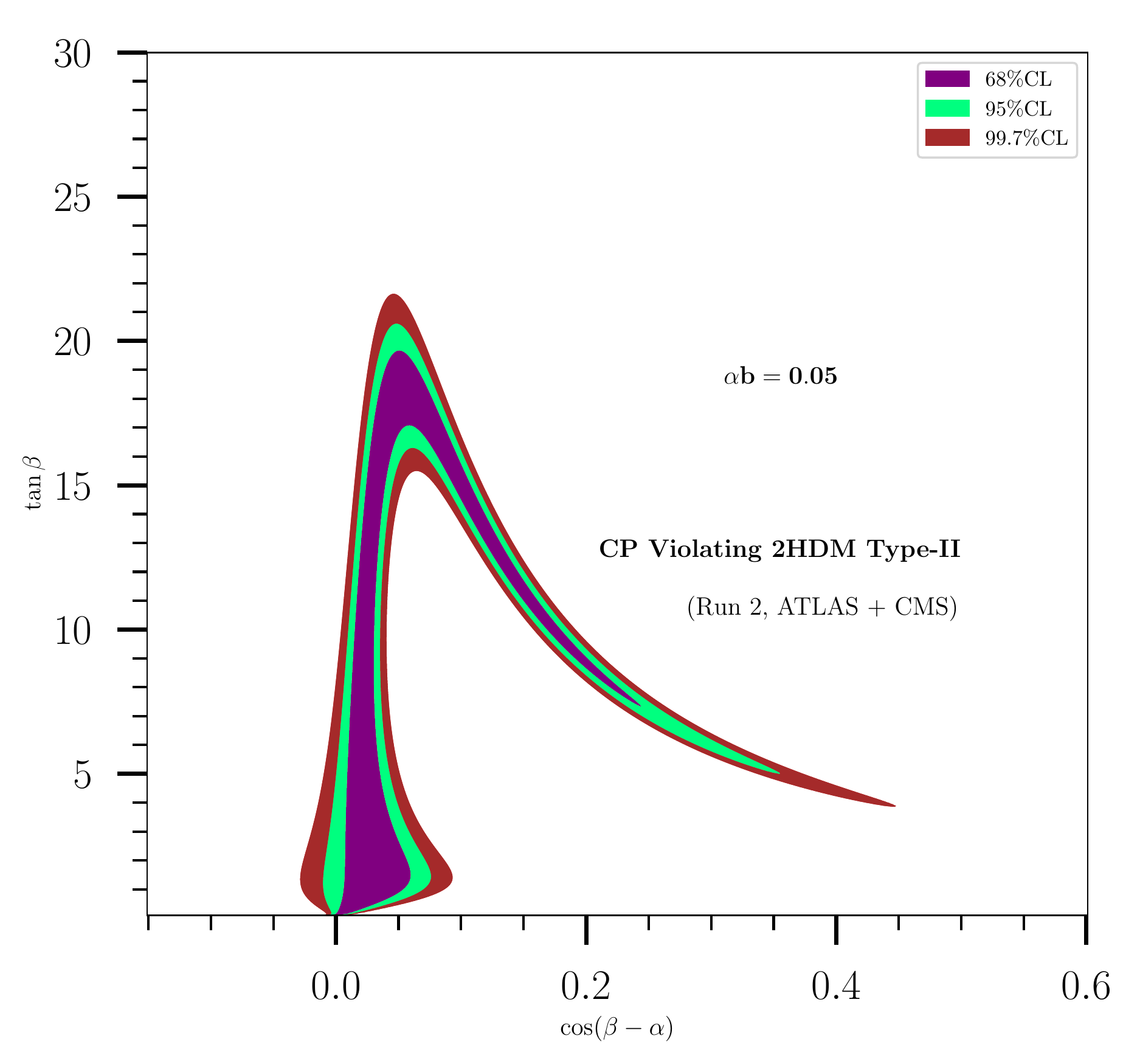}
\includegraphics[width=8.1cm]{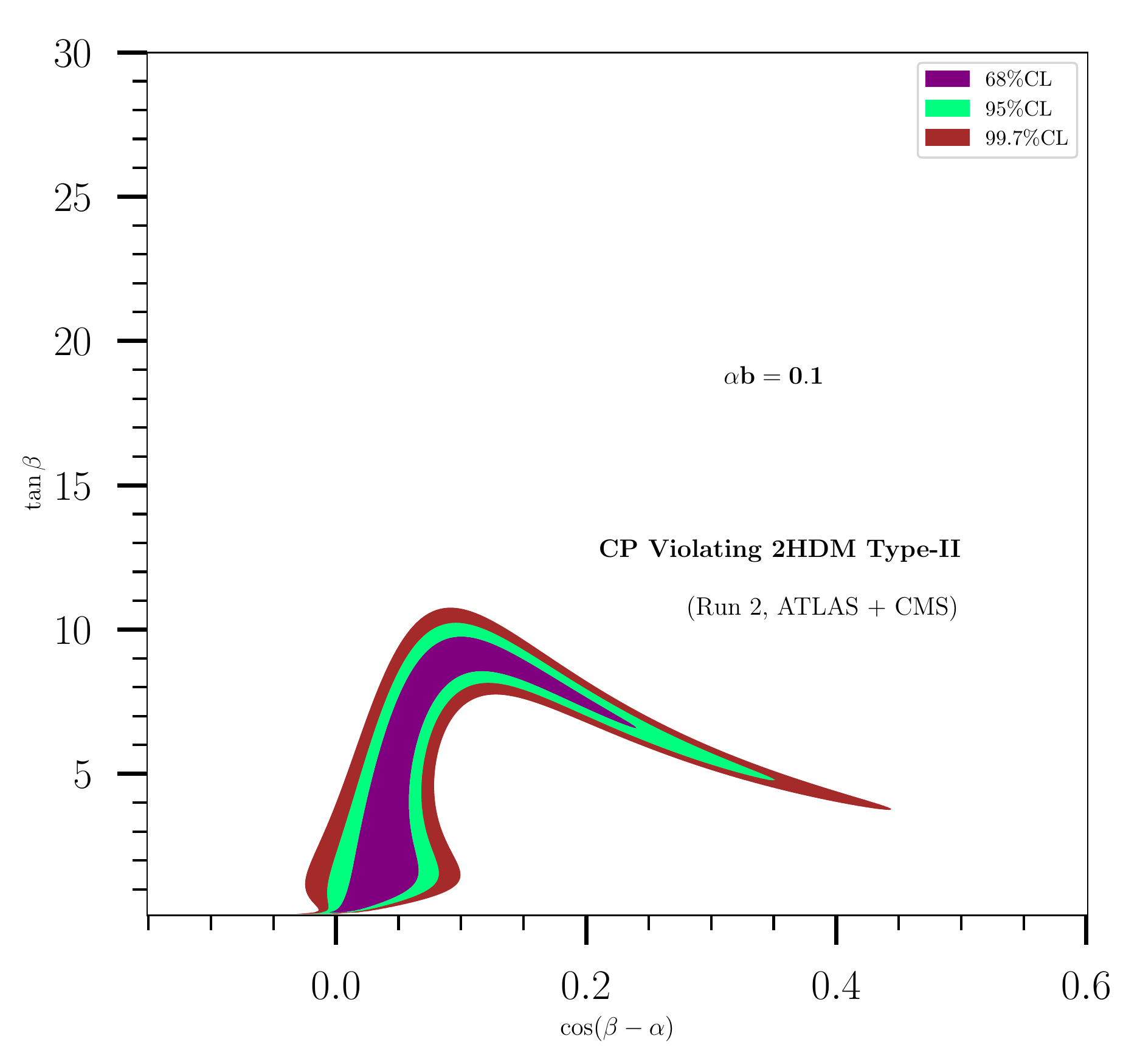}
\includegraphics[width=8.1cm]{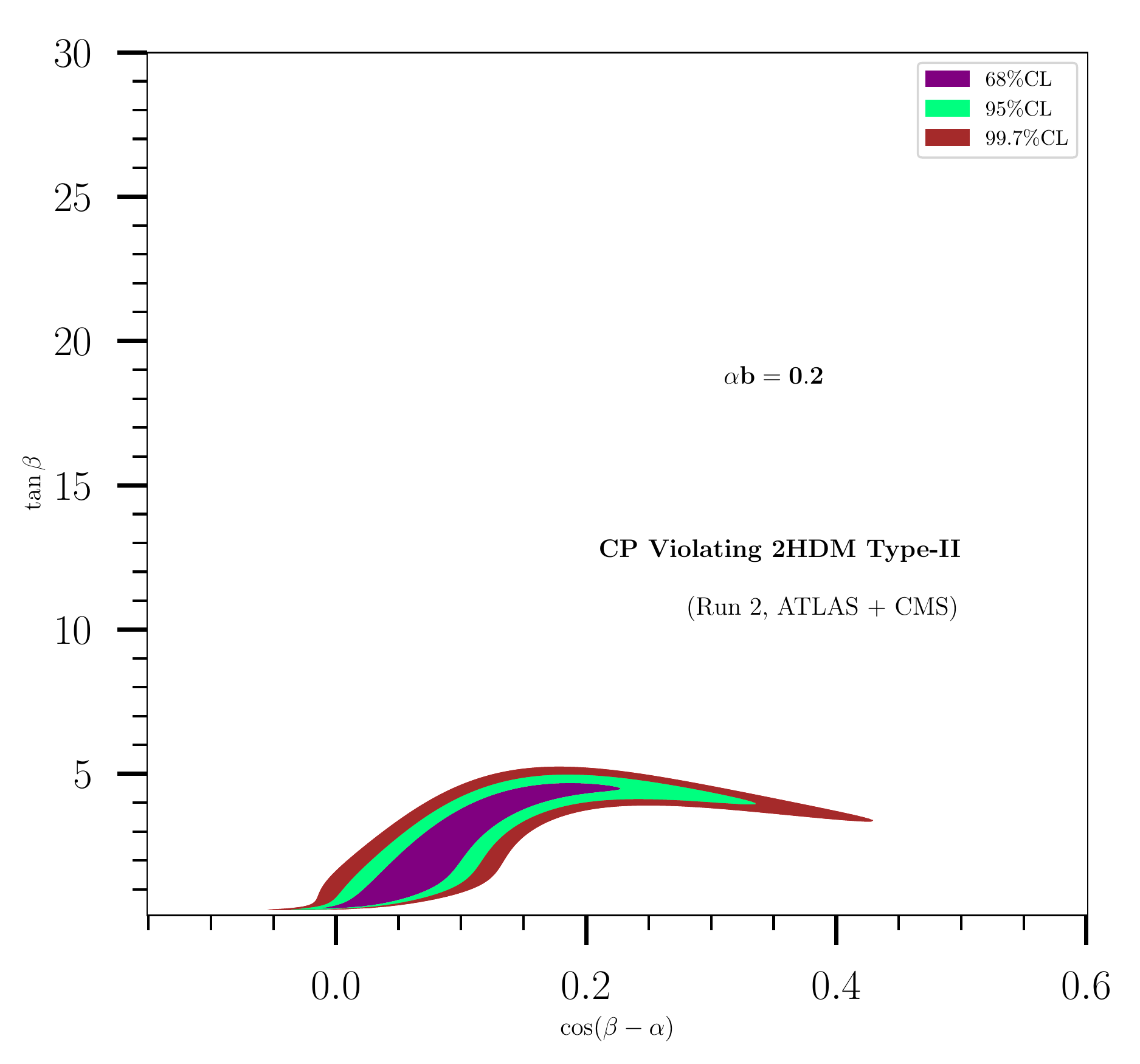}
\includegraphics[width=8.1cm]{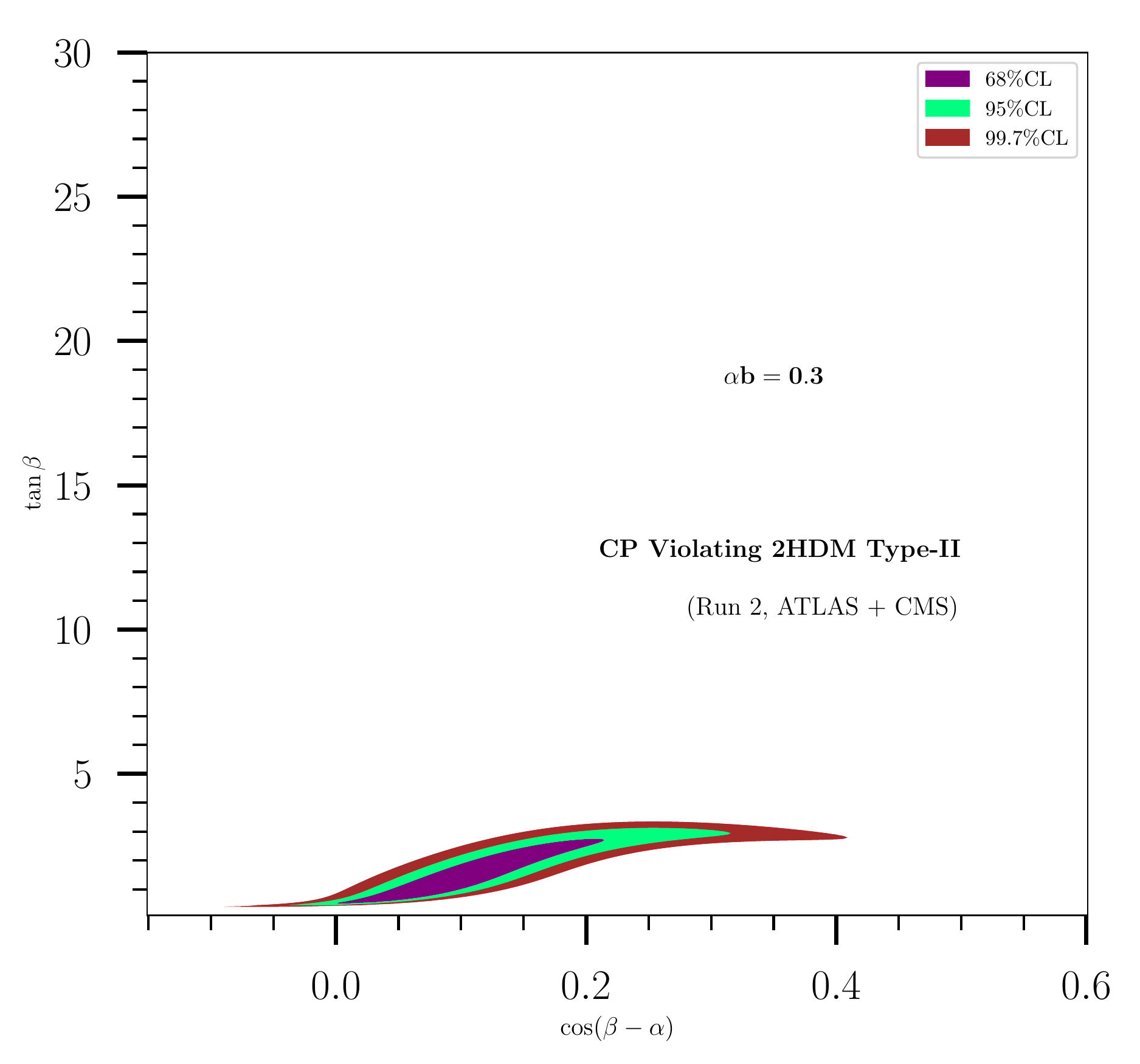}
\includegraphics[width=8.1cm]{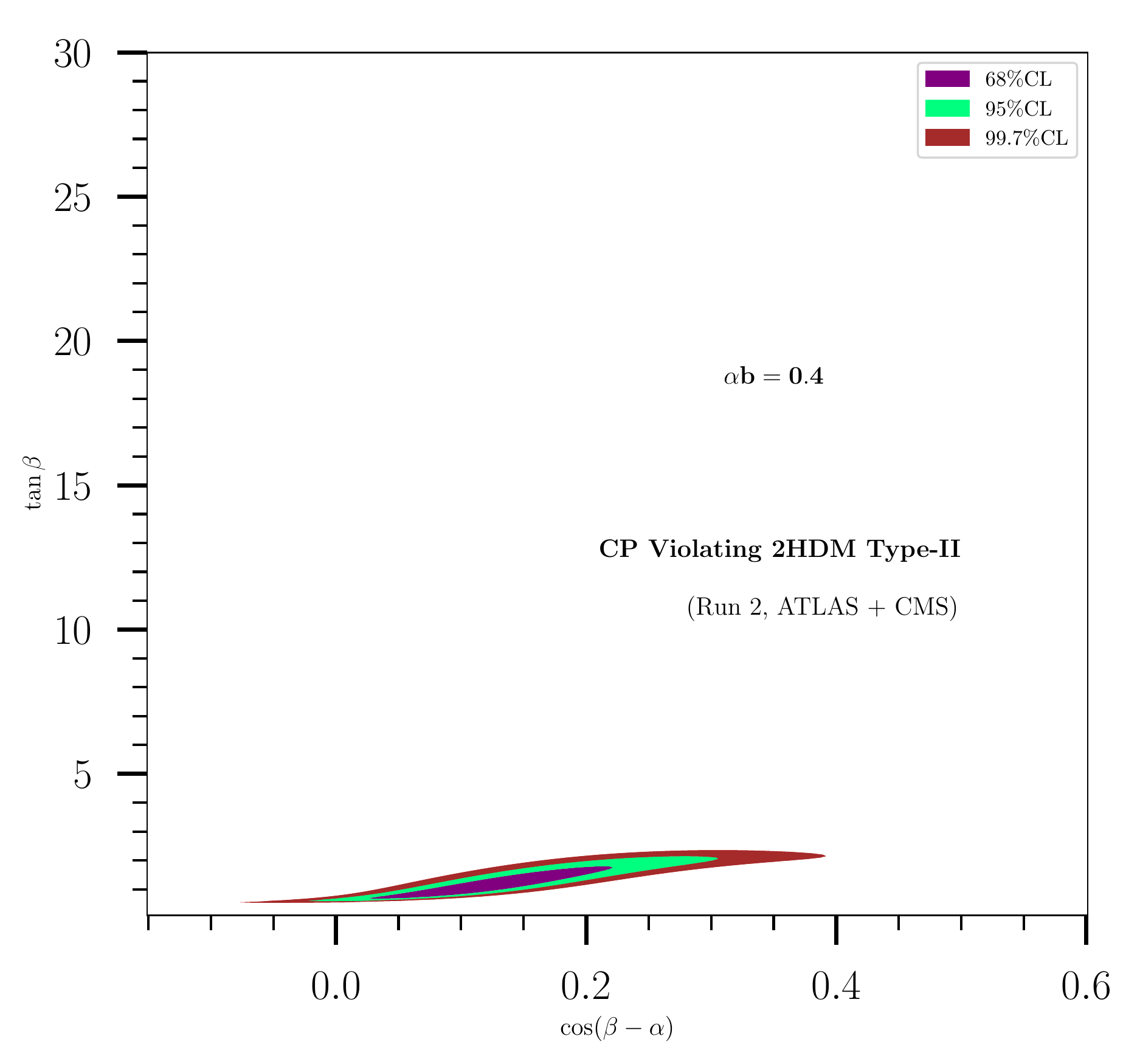}
\caption{Allowed parameter space for the 2HDM with CP violation from
  LHC-Higgs data. We show restrictions on the parameter space coming
  from varying $\tan \beta$ and $\cos(\beta-\alpha)$. Each plot has a
  fixed value of $ \alpha_b$ (measuring the amount of CP-violating
  admixture within the SM-like Higgs boson). Top left:
  $\alpha_b=0$ (CP-conserving SM-like Higgs), top right:
  $\alpha_b=0.05$, middle left: $\alpha_b=0.1$,  middle right:
  $\alpha_b=0.2$, bottom left: $\alpha_b=0.3$, bottom right:
  $\alpha_b=0.4$.  We included all CMS and ATLAS data at 36 fb$^{-1}$,
  as well as data on the Higgs production and diboson decays at 140
  fb$^{-1}$ as in the latest {\tt Lilith} software
  \cite{Bertrand:2020lyb}.
  %We vary $-{\pi\over 2} \le \alpha_c \le
   %{\pi\over   %2}$, but only a limited range survives, given in each plot.
} 
\label{fig:Higgs_space}
\end{centering}
\end{figure}
%\FloatBarrier
%%%%%%%%%%%%%%%%%%%%%%%%%%%%%%%%%%%%%%%%%%%%%%%%%%%%%%%%%%%%%%%%

In this section, we describe the methodology employed to explore the
parameter space.
%% as described previously in Sec. \ref{sec:CPV}, subject
%% to the constraints  in Sec. \ref{sec:constraints} .
We first investigated the parameter space region by performing detailed calculations of the
reduced couplings associated with CPV 2HDM and obtained the different
signal strengths associated to the SM Higgs. Using {\tt Lilith}  we
explored constraints on the model parameters in the $\cos(\beta-\alpha)$ and $\tan \beta$ plane.
%% We impose constraints from signal strength measurements of the observed
%% 125.09 GeV Higgs boson at the LHC.
%The library used by {\tt Lilith},
%available also in C and
%C++\ /{\tt ROOT}programs,
%includes the latest experimental  measurements from both the ATLAS and
%CMS collaborations at the LHC.
Once the SM Higgs bounds were considered in this way, we further employed {\tt ScannerS}
\cite{Muhlleitner:2020wwk}, a code that performs
  parameter scans and checks parameter points in BSM theories  with
  extended scalar sectors.
  {\tt ScannerS} incorporates theoretical (perturbative
    unitarity, vacuum stability, boundedness from below) and
  experimental constraints (electroweak precision, flavour
    constraints, Higgs searches and measurements, EDMs) from
  several different sources in order to  determine whether a parameter
  point is allowed or excluded at  approximately 95\% CL.

The constraints on the $\cos(\beta-\alpha)$ and $\tan \beta$
plane due to LHC-Higgs data are presented in Fig. \ref{fig:Higgs_space}  which shows the
allowed parameter space for different choices of  the $\alpha_b$ angle, a
measurement of the amount of CPV admixture within the SM-like Higgs. Note
that the composition of the SM-like Higgs, and thus the allowed
parameter region plotted, is independent of $\alpha_c$, which affects
only the amount of CPV admixture in the other neutral Higgs
bosons. The top left-handed panel shows the parameter restrictions for
the CP-conserving case $\alpha_b=0$, where we see that a significant
region of the parameter space survives away from the alignment region
(for which $\cos(\beta-\alpha)=0$). The interesting particularity of this away-from-alignment region is
that the bottom quark Yukawa coupling $c_{hbb}$ is negative while the top Yukawa
coupling does not flip sign \cite{Ginzburg:2001ss, Ferreira:2014sld}. Also note that this plot is consistent
with more recent phenomenological studies of 2HDM such as \cite{Accomando:2019jrb}.
The parameter $\cos(\beta-\alpha)$ determines the $c_{hVV}$ and $c_{HVV}$ couplings
and is a measure of deviations from alignment. As usual, $\tan \beta$,
the ratio of VEVs,  is important in determining the relative strength
of the couplings $c_{hbb}$ and $c_{htt}$ (and similarly for
$c_{H_{2,3}bb}$ and $c_{H_{2,3}tt}$).  Increasing $\alpha_b$ changes
significantly the parameter space, in particular, it reduces  the
allowed values for $\tan \beta$, which can be large at $\alpha_b=0$,
but has a maximum value of about 20 for $\alpha_b=0.05$ (top right
panel), decreasing to about 10 for $\alpha_b=0.1$ (middle left panel),  about 5 for
$\alpha_b=0.2$ (middle right panel),  about 3 for $\alpha_b=0.3$ (bottom
left panel) and  about 2 for $\alpha_b=0.4$ (bottom right
panel).

In all graphs the regions of the allowed parameter space shrink
considerably when increasing $\alpha_b$ (as expected) but noticeably,
the region away from alignment survives. This is significant because
the away-from-alignment region is a harbinger of non-standard
behaviour in the Higgs sector. In the figure, the maroon points
correspond to 3$\sigma$ agreement with the data, green points
correspond to 2$\sigma$ and purple points to 1$\sigma$
agreement. Increasing $\alpha_b$ leads to slowly shrinking of the
parameter regions consistent with experiment, and for $\alpha_b=0.4$
almost the whole parameter space is excluded, with only a few points
near the alignment region surviving. Still, it is interesting to note
that, for $\alpha_b=0.1$, parameter points consistent with the Higgs
data at 2$\sigma$ survive, in both the alignment and
away-from-alignment regions.    

%%%%%%%%%%%%%%%%%%%%%%%%%%%%%%%%%%%%%%%%%%%%%%%
 As mentioned before, the {\tt Lilith} code, which is self-contained,  combines different Higgs signal
constraints from published data at 36 fb$^{-1}$, with the addition
of the measurement at 139 fb$^{-1}$ for the process $gg\to h \to ZZ$. The
combination of all the most recently published signals has not been
fully implemented yet in {\tt Lilith}. Nevertheless it is
possible to show the evolution of each individual bound and gain an
insight  on the type of pressure that each Higgs signal puts on the
parameter space considered here. We find that the allowed parameter
space is most sensitive to three main Higgs signals, namely the
$b\bar{b}$, $VV$, and $\gamma\gamma$ channels.
Sensitivity to the rest of Higgs decay channels is less important as their bounds on the
parameter space under consideration are always less restrictive than the mentioned signals.
\begin{figure}[t]
  %\begin{centering}
 % \includegraphics[width=8.1cm]{bbZZalphabzero.pdf}\ \ \
  \includegraphics[width=8.1cm]{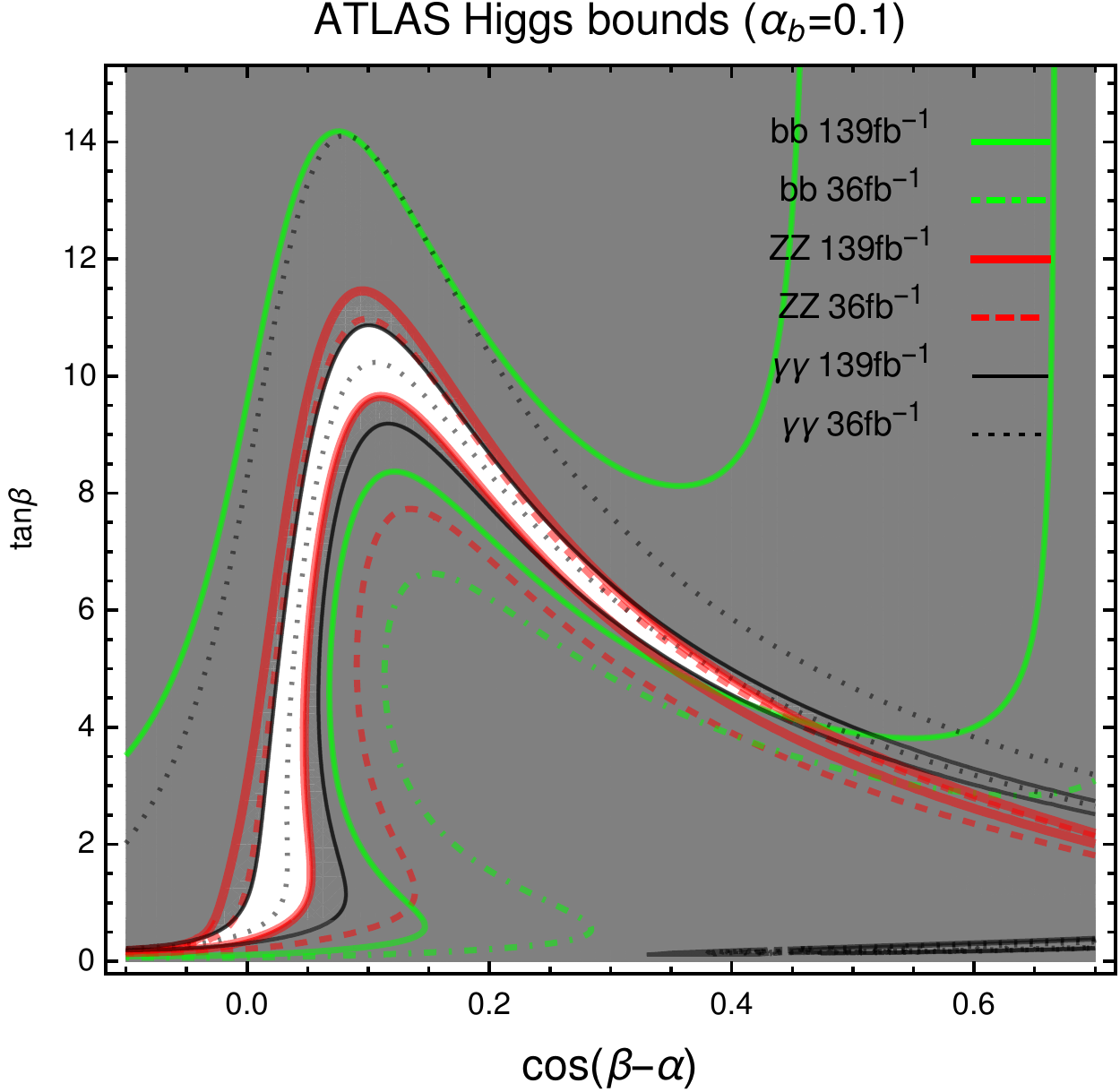}  
\includegraphics[width=8.1cm]{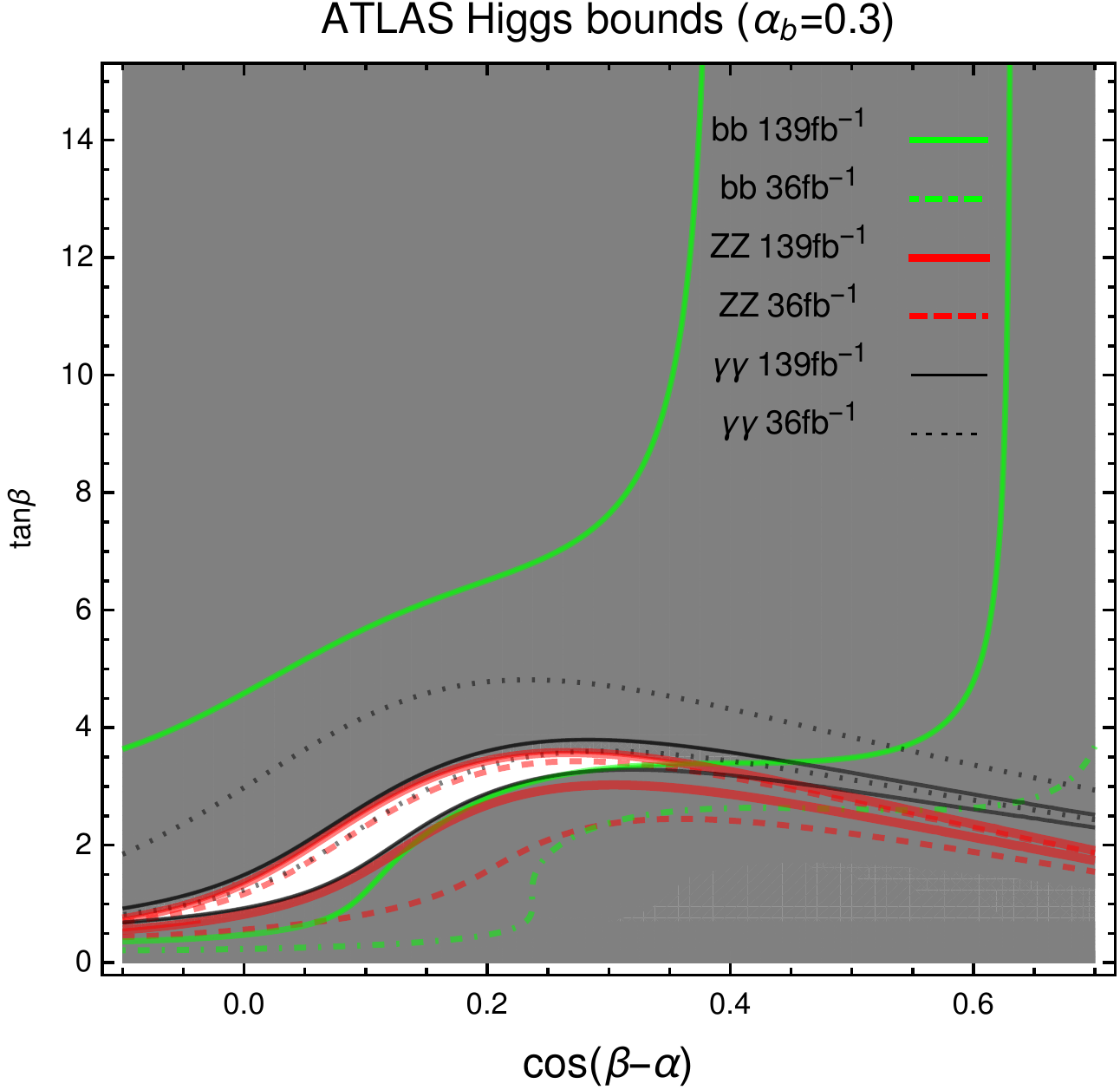}
\caption{Contours of the signal strengths for the three Higgs channels
  $gg\to h \to ZZ$ (red-thick), $Vh \to (h\to b\bar{b})$
  (green-light) and $gg\to h \to \gamma \gamma$ (black-thin), 
  using the upper and lower bounds from the ATLAS detector obtained with 36 fb$^{-1}$
  (dashed, dot-dashed and dotted respectively) and with 139 fb$^{-1}$
  (all solid) of integrated luminosity. The white regions are points
  in parameter space that individually lie within the $b\bar{b}$, the $ZZ$ and the
  $\gamma\gamma$ bounds using the 139 fb$^{-1}$ of luminosity from
  ATLAS. The gray regions represent points that are outside of either
  of the bounds, {\it  i.e.} these regions do {\bf not} represent  a combination of
  bounds. We take $ \alpha_b=0.1$ in the left panel and $ \alpha_b=0.3$ in the right
  panel. In both cases the individually allowed regions from each of the three
 Higgs signals is located in between their respective contour curves and clearly shrink when going from 
  36 fb$^{-1}$ to  139 fb$^{-1}$ of integrated luminosity (although
 note that the upper bound from $b\bar{b}$ at 36 fb$^{-1}$ does  does
 not appear in the plot). 
  }
\label{fig:individualchannels}
\end{figure}

To  display  the effect of reduced experimental
error bars thanks to improved statistics and analysis, we will focus
on three individual studies from the ATLAS collaboration, comparing their older results at 36 fb$^{-1}$ with their most recent bounds with 139 fb$^{-1}$. 
In Fig. \ref{fig:individualchannels} we chose two examples within the same parameter space 
as Fig.~\ref{fig:Higgs_space}, one for $\alpha_b=0.1$, the other for $ \alpha_b=0.3$. We show contours for the signal strengths of the
Higgs channels $gg\to h \to ZZ$ (red-thick),  $Vh \to (h\to
b\bar{b})$ (green-light) and $gg\to h \to \gamma\gamma$ (black-thin),
using the bounds set by ATLAS, with 36 fb$^{-1}$
(dashed/dot-dashed/dotted) and with 139 fb$^{-1}$ (solid) of
integrated luminosity \cite{ATLAS:2022fnp,ATLAS:2021tbi,ATLAS:2020wny}.
 In both cases, the white
regions are points in parameter space that  lie {\it individually} within
the bounds from $b\bar{b}$, $ZZ$ and $\gamma\gamma$, using the ATLAS
data at 139 fb$^{-1}$ of luminosity. Note that these regions do {\bf not}
represent a combination of bounds, but are merely representative of the
parameter space that will likely still be allowed once the bounds are
properly combined\footnote{Performing such a combination is beyond the scope of the
present paper.}. The improved statistics observed  are clearly apparent
in the three signal contours with  shrinking of the allowed regions as
data is improved from 36 fb$^{-1}$ to 139 fb$^{-1}$. This indicates that while our chosen parameter points are not ruled out, they are  under more  pressure, and may fit the new data with less C.L.
%%%%%%%%%%%%%%%%%%%%%%%%%%%%%%%%%%%%%%%%%%%%%%%%

After delineating the allowed and excluded regions of parameter space
using the LHC Higgs data, we proceed to investigate the constraints
of the CP-violating phases $\alpha_b$ and $\alpha_c$ on the masses of
the heavier neutral Higgs bosons $H_2$ and $H_3$ and on the mass of the charged Higgs
boson $H^\pm$.
%% from the constraints in Sec. \ref{sec:constraints}, in
%% particular from the electric dipole moments and oblique parameters
%% ($S$, $T$ and $U$).

\begin{figure} [t]
\begin{center}
\includegraphics[width=0.49\textwidth]{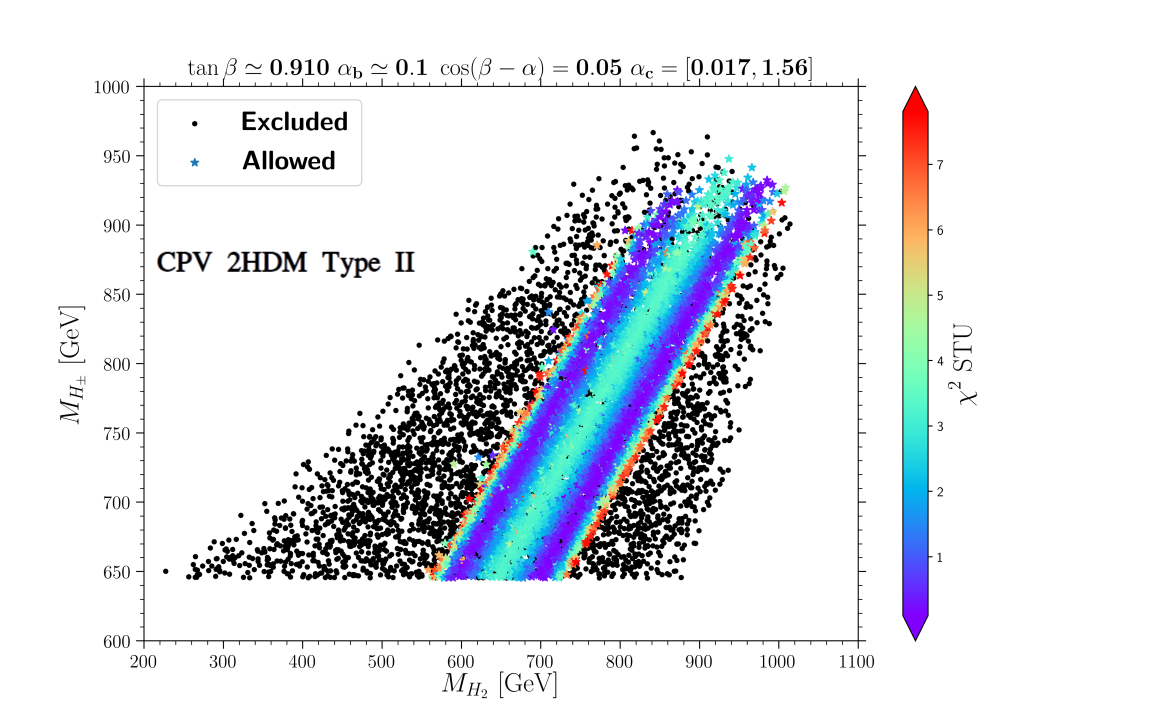}
\includegraphics[width=0.49\textwidth]{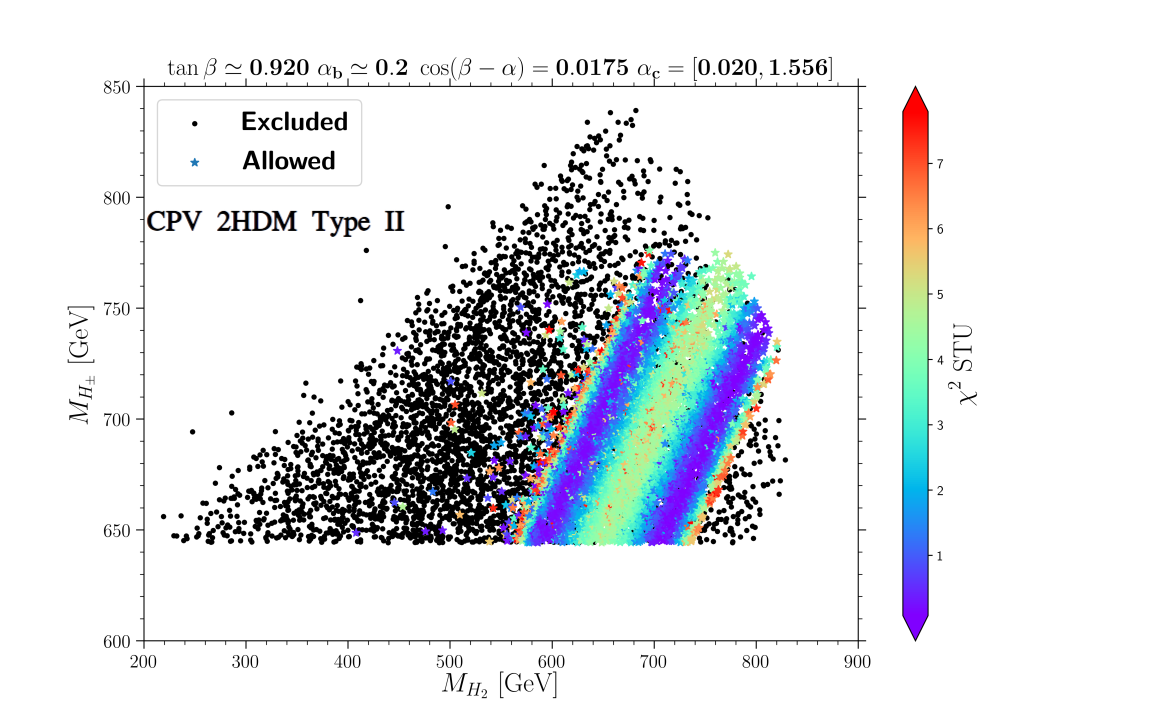}
\includegraphics[width=0.49\textwidth]{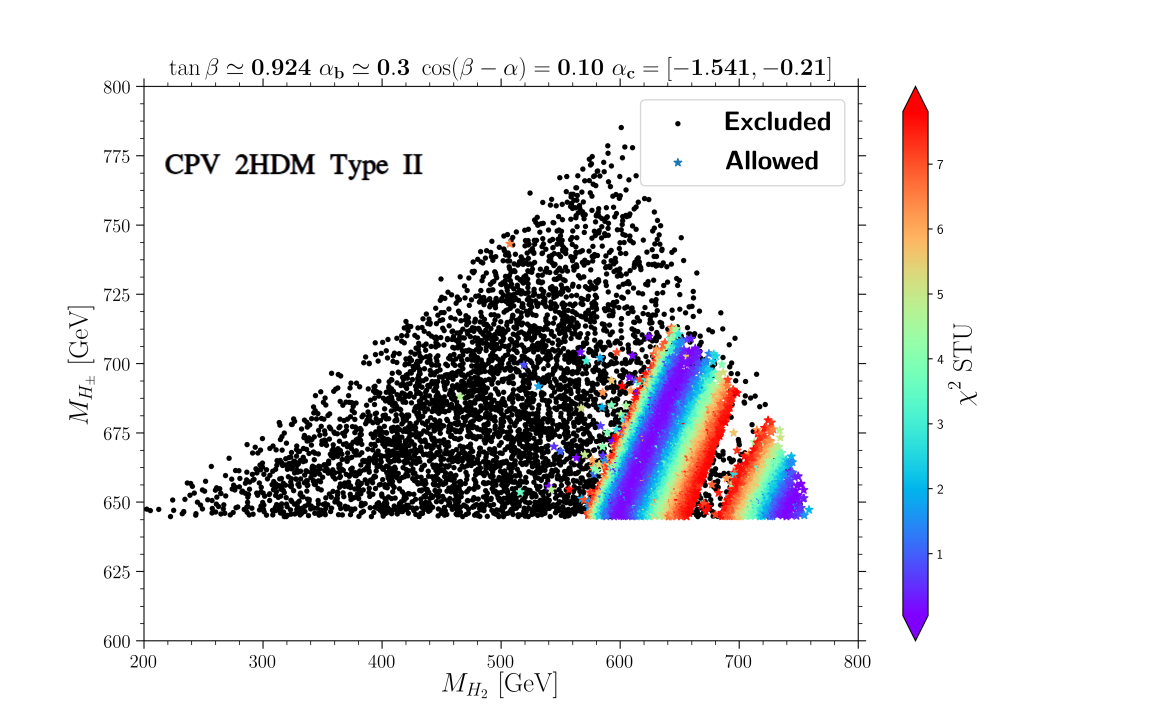}
\includegraphics[width=0.49\textwidth]{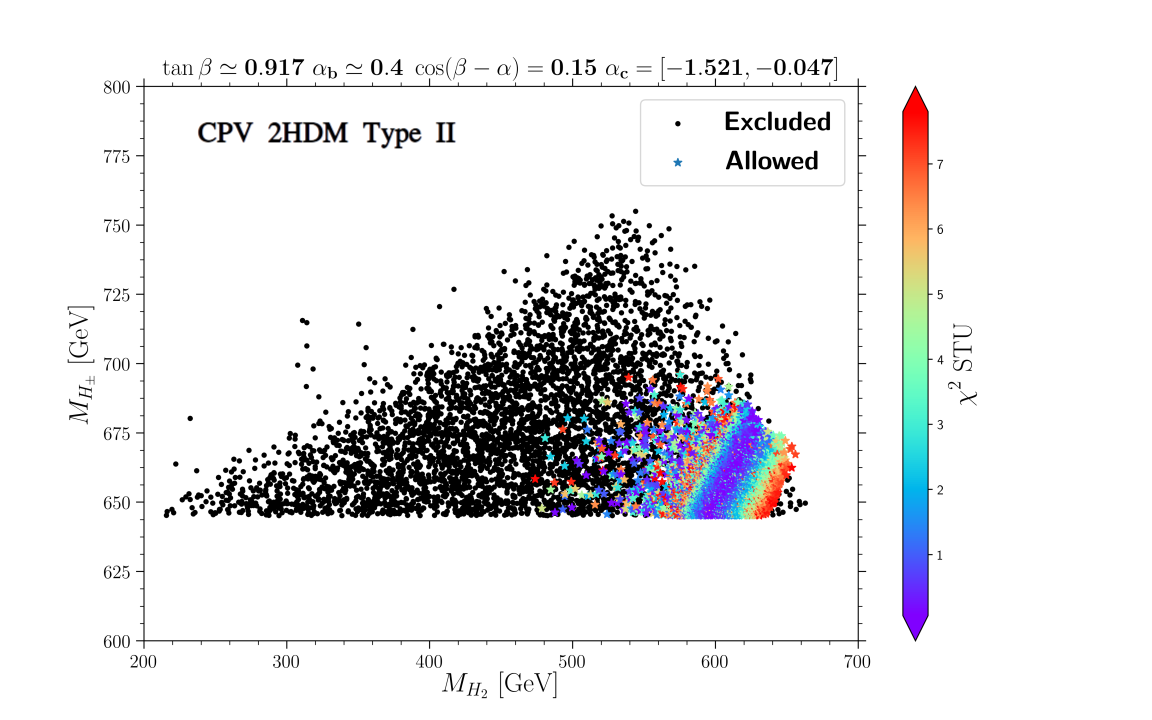}
\end{center}\vspace*{-2mm}
\caption{Restrictions for the CPV 2HDM  on the parameter space in
  $M_{H^{\pm}} - M_{H_2}$ plane from oblique parameters and EDMs. We
  show plots for different values of  $\sin \alpha_b$. Top left:
  $\alpha_b=0.1$, top right: $\alpha_b=0.2$, bottom left:
  $\alpha_b=0.3$, bottom right: $\alpha_b=0.4$. The black dotted
  region is populated by points that pass all experimental and
  theoretical bounds, except for precision electroweak tests and
  EDMs. The rainbow colored region contains points passing all tests
  described in Sec. \ref{sec:constraints}.} 
\label{fig:MHplusMH2forall}
\end{figure}
%\FloatBarrier

In Fig. \ref{fig:MHplusMH2forall} we show the
surviving parameter space in the $M_{H^\pm}-M_{H_2}$ plane, where $H_2$ is
the lightest of the two non-SM like neutral Higgs bosons. In these plots
$\alpha_b$, $\cos{(\beta-\alpha})$ and $\tan{\beta}$ are fixed such that
we are within an allowed parameter space point from
Fig. \ref{fig:Higgs_space}.
%We then vary $\tan \beta$ within the allowed region, as well as
%$-{\pi\over 2} \le \alpha_c \le {\pi\over 2}$.
In the four plots we fix $\alpha_b= 0.1$ (top left panel),
$\alpha_b= 0.2$ (top right panel), $\alpha_b= 0.3$ (bottom left) and
$\alpha_b= 0.4$ (bottom right panel). The other CP violating mixing
angle $\alpha_c$ is scanned over all possible values but the
experimental constraints restrict $\alpha_c$ to lie in the range given in each plot. 
%\FloatBarrier

The black dots represent the regions allowed by all constraints in
Sec. \ref{sec:constraints}, except for EDMs and oblique
parameters (we use {\tt  ScannerS}). We have chosen a parameter space region allowed by
Higgs data (see Fig. \ref{fig:Higgs_space}) and the main bounds on the
black dotted region come from B-physics constraints which are responsible for the requirement
$M_{H^\pm} \ge 650$ GeV. In addition, perturbativity and unitarity
requirements which restrict the parameter space to within the diagonal
border-lines at the left, right and top of each plot. In particular
one can see that direct unitarity and perturbativity constraints
involve the coupling $\lambda_4$, which depends explicitly on both the
charged Higgs mass and the neutral Higgs masses (see
Eqs.~(\ref{eq:lambda_3}) and (\ref{eq:lambda_4})).  Within
  the CP-violating parameters considered here, these
  unitarity and perturbativity requirements make $M_{H^\pm}$ bound from
  above, disallowing it to be much larger than $M_{H_2}$ and
  $M_{H_3}$, as  shown in the discussion in Sec. \ref{subsec:theory}.

 Inside the black dotted region, we show as the rainbow colored
 region,  the physical domain that passes all tests, including
 electroweak precision measurements constraints from the 
oblique parameters ($STU$), as in \ref{subsec:stu}, and CPV
constraints from electric dipole moments, as in  \ref{subsec:edm}.
The legend at the right shows the deviation from the prescribed
$STU$ and EDM parameters as measured by $\chi^2$, restricted to be
$<7$. We find that, within the allowed parameter region of
Fig. \ref{fig:Higgs_space}, points that pass $STU$ and EDM constraints
require an almost fixed value for $\tan \beta$, 
consistently around $\tan \beta \sim 0.9$ and {\it always} smaller
than 1{\footnote{Note that while in MSSM, $\tan \beta$ must be
  greater than 1,  $\tan \beta$ is not restricted in
  2HDMs.}}.  The reason is that a relative mass degeneracy
  among the heavy scalars is necessary to avoid precision tests bounds, and with
non-zero CP violation, the masses of the three neutral scalars are all
connected via $\tan \beta$ (see Eq.~\ref{eq:tanbeta}). One can check  that for
small values of $\cos(\beta -\alpha)$ (i.e.  for $\beta \sim \alpha
+\pi/2$) a small mass splitting between  between $M_{H_2}$ and 
$M_{H_3}$ can only happen when $\tan \beta \sim 1$.

The parameter regions chosen in all four graphs correspond to regions
slightly away from alignment in which $\cos(\beta - \alpha) > 0$.  In
the case of CP conservation, and in the decoupling limit, the mass
$M_h$ is independent of the other two neutral boson masses (one
CP-even, one CP-odd) which are not restricted. This is not the case
for the CPV case, and slightly away from the decoupling limit.  Here
the parameter $\alpha_c$ which determines the CP admixture
in the heavier neutral bosons $H_2$ and $H_3$ is varied but
only values within a limited range lead to allowed points, the exact interval depending on the value
$\alpha_b$ chosen. Increasing $\alpha_b$ enhances the amount of CPV
admixture in the SM-like Higgs boson  resulting in a decrease in the
allowed range by EDMs and $STU$ constraints in the
$M_{H^\pm}-M_{H_2}$ plane. The allowed regions (in rainbow colors) are separated by
straight diagonal lines in  $M_{H^\pm}-M_{H_2}$,  and while the
allowed window for $M_{H_2}$ is approximately constant,  $M_{H_2} \sim 550 \to
700-1000$ GeV. The  range for $M_{H^\pm}$, while being run into the
TeV region, is restricted by unitarity and perturbativity
requirements, as explained above. $M_{H^\pm}$  shrinks considerably
with increasing $\alpha_b$, from 650-950 GeV for $\alpha_b=0.1$ in the top
left hand panel, to 650-685 GeV for $\alpha_b=0.4$, in the bottom
right hand panel. We also note that regions slightly more separated from
alignment  survive for larger values of $\alpha_b$, such as for 
$\alpha_b=0.4$, where a small region of parameter space is allowed for
$\cos(\beta-\alpha) \simeq 0.15$. In this region, $\Gamma(H_{2,3} \to VV)
\neq 0$ and hence, the decay of the heavy neutral Higgs into the
electroweak gauge bosons is allowed.  
\begin{figure} [htbp]
\begin{center}
\includegraphics[width=0.50\textwidth]{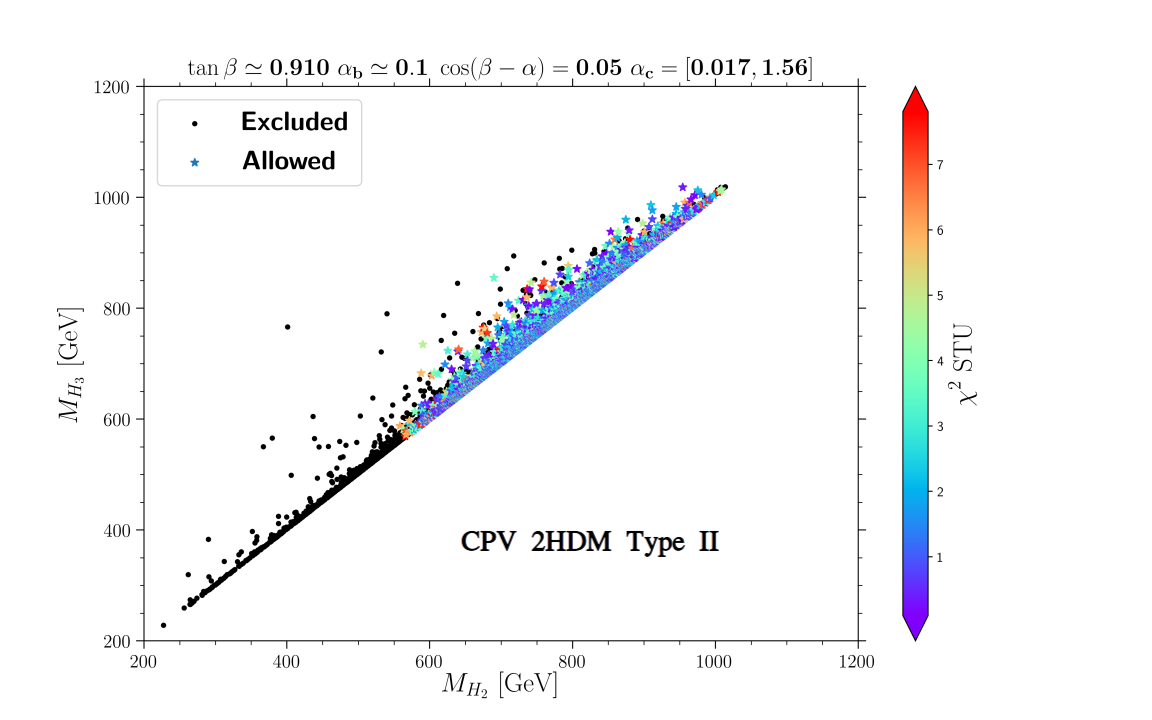}
\includegraphics[width=0.49\textwidth]{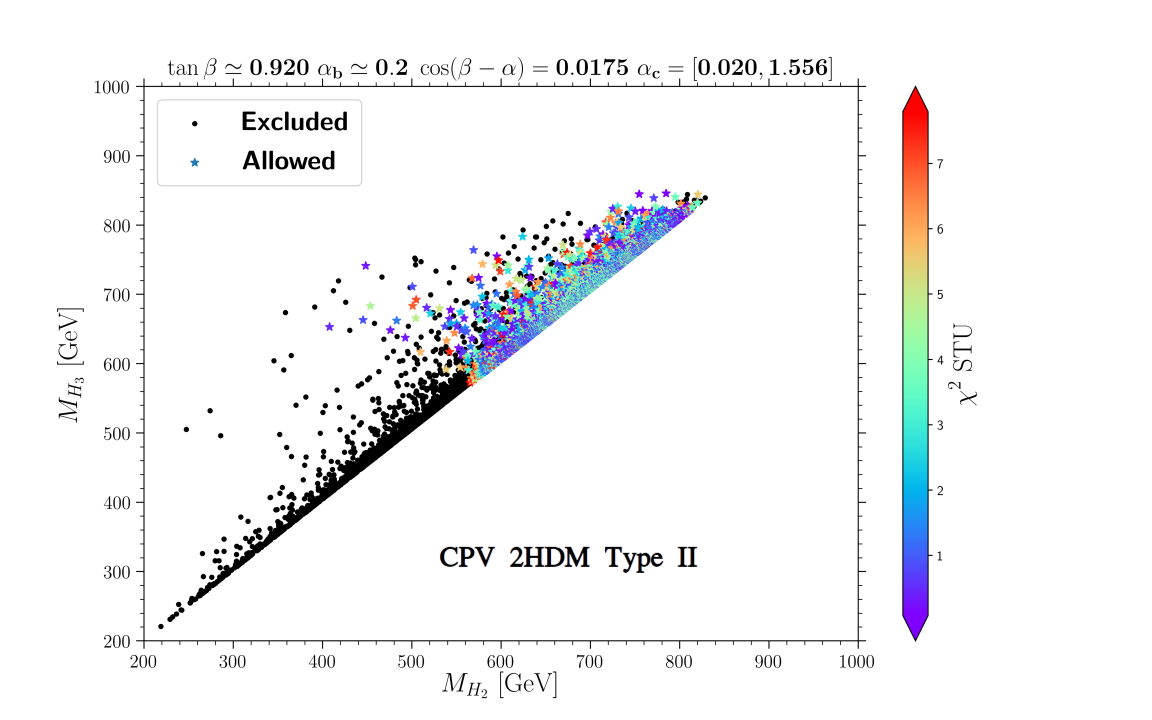}
\includegraphics[width=0.50\textwidth]{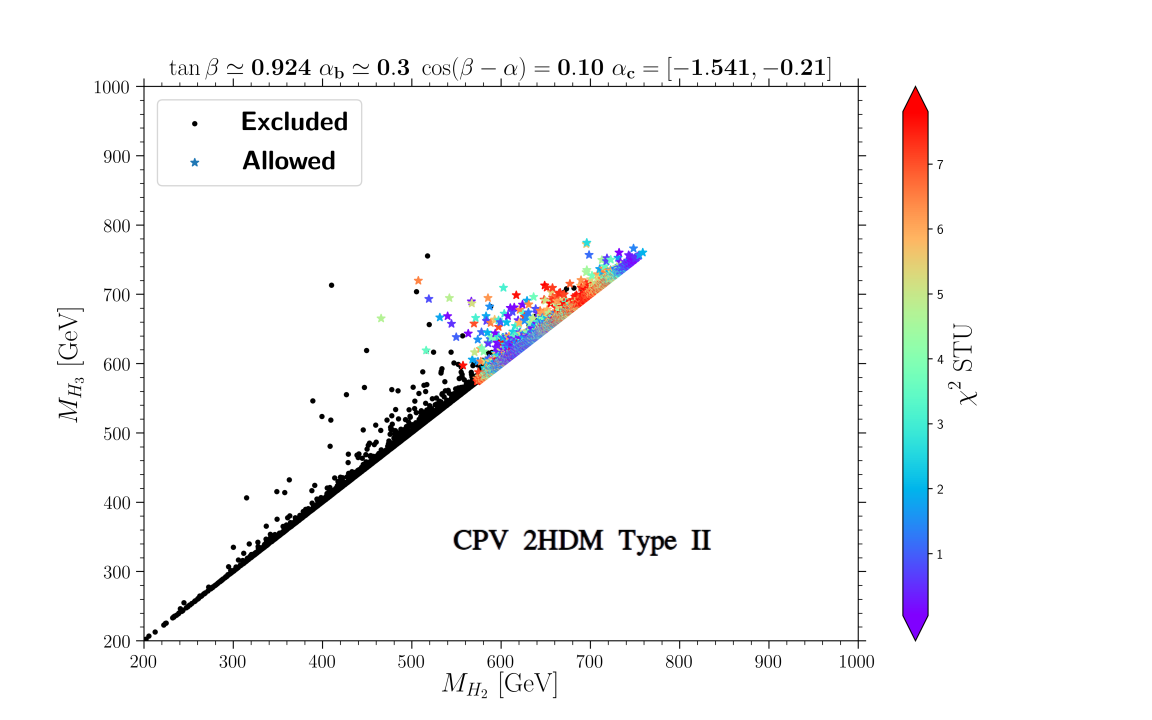}
\includegraphics[width=0.49\textwidth]{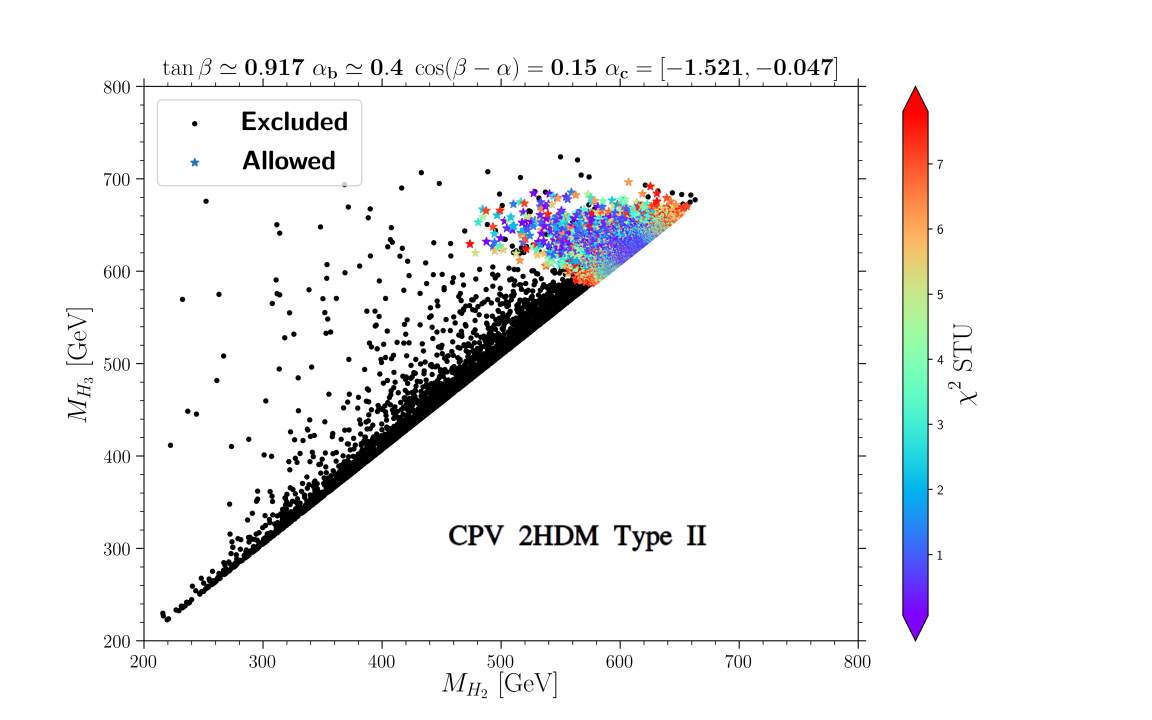}
\end{center}\vspace*{5mm}
\caption{Restrictions for the 2HDM with CP violation on the parameter
  space for $M_{H_{3}}$ and $M_{H_2}$ plane from oblique
  parameters. We show plots for different values of  $\sin
  \alpha_b$. Top left: $\alpha_b=0$.1, top right: $\alpha_b=0.2$,
  bottom left: $\alpha_b=0.3$, bottom right: $\alpha_b=0.4$.
The black dots are points that pass all experimental and
  theoretical bounds, except for precision electroweak tests and
  EDMs. The rainbow colored dots are points that pass all tests
  described in Sec. \ref{sec:constraints}.
} 
\label{fig:MH3MH2space}
\end{figure}

In Fig.~\ref{fig:MH3MH2space}, we show the allowed parameter space in
the $M_{H_3}$ versus $M_{H_2}$ plane from  imposing all constraints
except the EDMs and the $STU$  parameters (black points), and with
EDMS and oblique parameters (rainbow colored points). Here, the mass
of the charged Higgs was varied in the allowed ranges from
Fig. \ref{fig:MHplusMH2forall}. The relationship between the masses
(squared) of the two neutral heavy scalars is linear, and lies in a restricted range. In all cases we find
that these two masses must be relatively degenerate in mass in order
to satisfy experimental constraints, 
%lose within at
%most $\sim50$ GeV. %We also note that
%the masses of $H_2$ and $H_3$ are required to be almost degenerate to
%satisfy the constraints,
in particular restrictions from the $STU$ parameters. The allowed
ranges for both masses shrink with increasing $\alpha_b$, but note
also that for larger values of $\alpha_b$  the  degeneracy requirement
between the two masses is slightly relaxed.  

%\FloatBarrier
In the CPV 2HDM model, a main source of CP violation is encapsulated
in ${\Im}\lambda_5$,  given in Eq.~\ref{eq:Imlambda_5}.  From
\cite*{Chen:2018shg,Ginzburg:2005dt,Grinstein:2015rtl,Cacchio:2016qyh,Inoue:2014nva},
the unitarity bound on ${\Im}\lambda_5$ should be ${\Im}\lambda_5 <
4\pi$.  
The value depends on the CP violating mixing angles  $\alpha_b\, ,
\alpha_c$ and on the masses $M_{H_2}$ and $M_{H_3}$. We show in
Fig. \ref{fig:ImLambda5space} below,  ${\Im}\lambda_5$ as a function
of $M_{H_2}$. As before the mass of the SM-like Higgs boson $h$ is
kept at 125.09 GeV, and we also take the same input parameters as in
Fig. \ref{fig:MH3MH2space} above. We show here only the
rainbow-colored points, that is, points that survive all constraints,
including EDMs and $STU$ bounds. 
%\FloatBarrier
\begin{figure} [t]
\begin{center}
\includegraphics[width=0.49\textwidth]{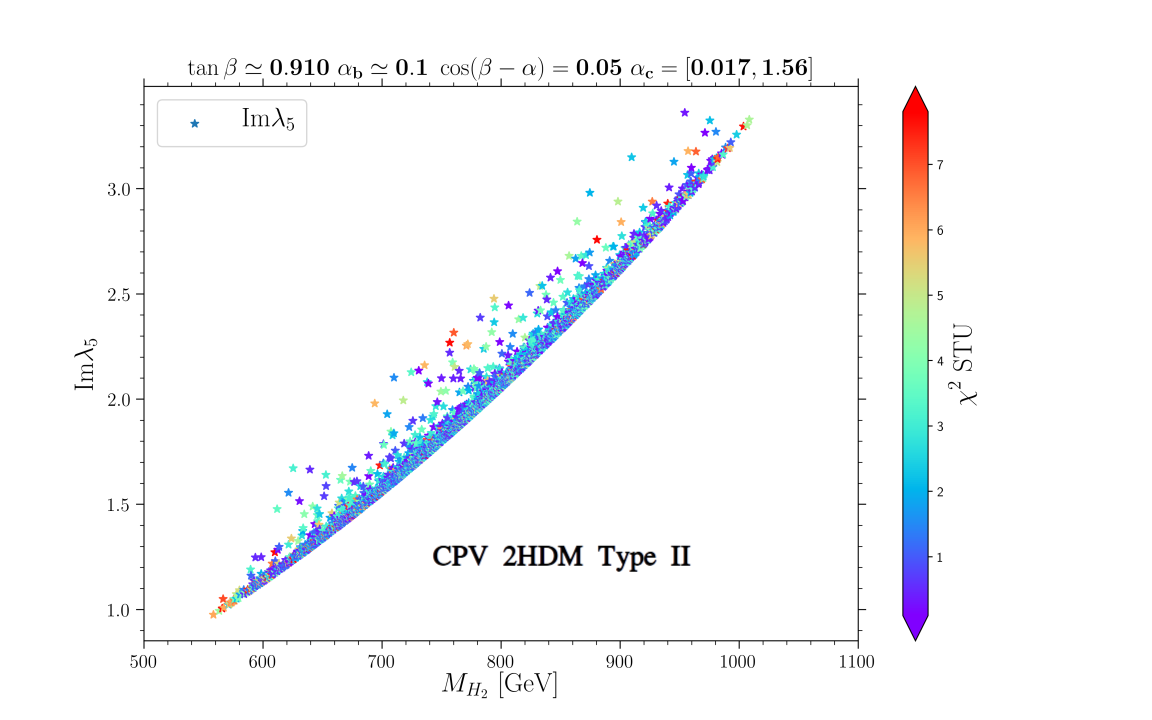}
\includegraphics[width=0.49\textwidth]{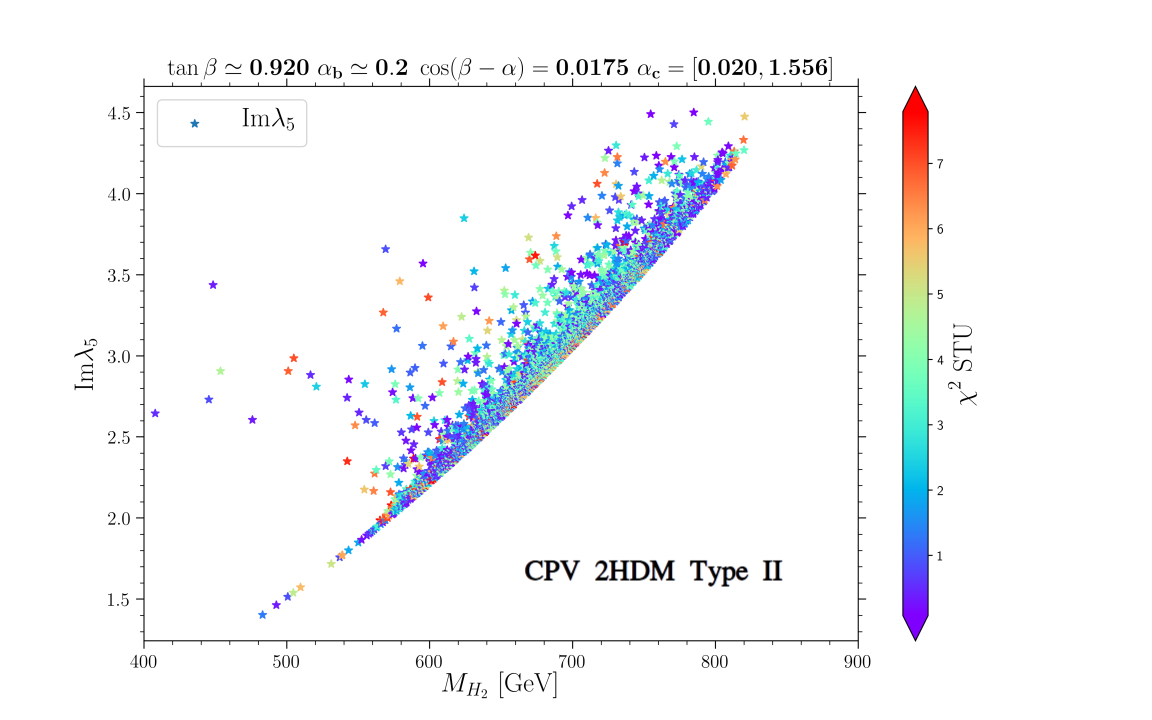}
\includegraphics[width=0.49\textwidth]{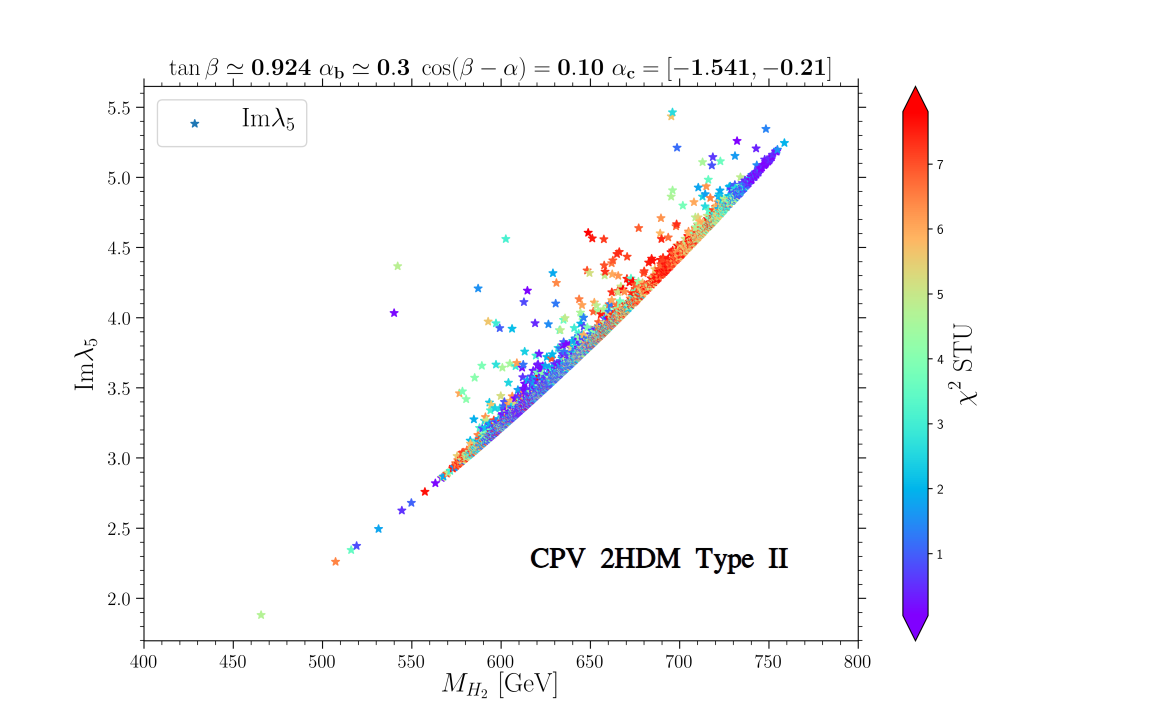}
\includegraphics[width=0.49\textwidth]{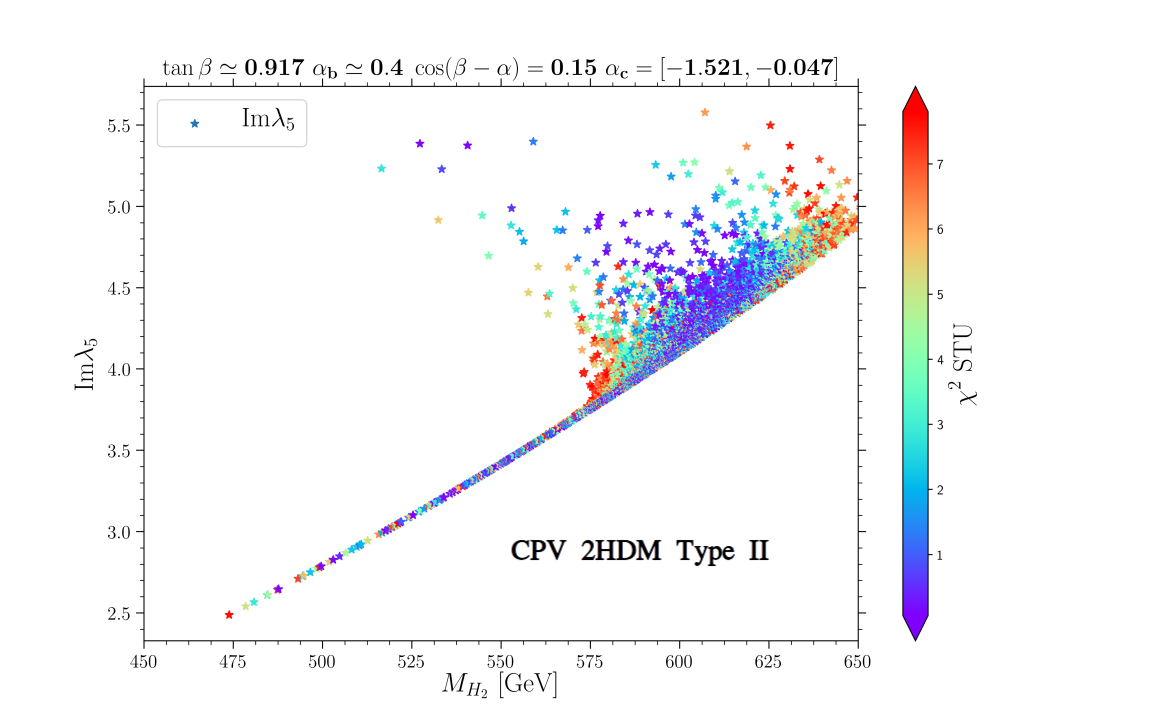}
\end{center}\vspace*{-5mm}
\caption{ Constraints on $ {\Im}\lambda_5 $, a source of CPV in the
  Lagrangian of the 2HDM, versus $M_{H_2}$ from oblique parameters. We
  show plots for different values of  $ \alpha_b$. Top left:
  $\alpha_b=0$.1, top right: $\alpha_b=0.2$, bottom left:
  $\alpha_b=0.3$, bottom right: $\alpha_b=0.4$.
The rainbow colored dots are points that pass all tests
described in Sec. \ref{sec:constraints}, including precision tests and
EDM bounds.
}
\label{fig:ImLambda5space}
\end{figure} 
\vspace*{-5mm}

%%%%%%%%%%%%%%%%%%%%%%%%%%%%%%%%%%
\section{Production and Decay Of Heavy Higgs Bosons at the LHC}
\label{sec:results}
%%%%%%%%%%%%%%%%%%%%%%%%%%%%%%%%%%%%%%%
We now proceed to analyze the implications of the restrictions on the
parameter space of $M_{H_2}\, , M_{H_3}\, , M_{H^\pm}$ on the
production and decay rates of the heavy gauge bosons at the LHC. We
analyse regions of parameter space that pass all restrictions, as
shown in Figs. \ref{fig:MHplusMH2forall} and \ref{fig:MH3MH2space}. 

%------------------------------------------------------------
\subsection{Neutral Higgs Bosons Production}
\label{subsec:production}
%------------------------------------------------------------
The heavier Higgs bosons are expected to be  produced at the LHC
dominantly via the gluon fusion process 
$\textsl{g}\textsl{g}\rightarrow H_{2,3}$.  Additional production
processes include vector boson fusion $pp\rightarrow H_{2,3}qq'$,
vector boson associated production $pp \rightarrow H_{2,3}V$, and  top
associated production  $\textsl{g}\textsl{g} \rightarrow
H_{2,3}t\bar{t}$.

%% A detailed discussion on the Higgs pair production
%% via gluon fusion processes in the 2HDM can be found in
%% \cite{Hespel:2014sla}, while \cite{Bernreuther:2015fts} studied the
%% production of the heavy Higgs boson and its decay into top quarks in
%% the strong coupling regime,  focussing on the total decay widths of
%% the heavy Higgs in Type II 2HDM  with three favorable parameter
%% scenarios, for  heavy Higgs bosons mass above the $t\bar{t}$ threshold
%% and unsuppressed Yukawa coupling to the top quarks.  
 
 In this work we employ {\tt SusHi} \cite{Harlander:2012pb} to
 calculate cross sections for gluon fusion to next to leading order
 (NLO),  At one loop, for the gluon fusion process, the ratio of the
 heavy Higgs boson production cross section in a CPV 2HDM to that of a
 SM-like Higgs is
 
 \begin{eqnarray}
R_{gg}^i=\frac{\sigma (gg\to H_i)}{\sigma (gg\to h)} = 
\frac{\left | c_{H_i tt} A_{1/2}^{H}(\tau^i_{t}) + c_{H_i bb} A_{1/2}^{H} (\tau^i_{b})\right|^2 + \left|\tilde c_{H_i tt} A_{1/2}^{A}(\tau^i_{t}) + \tilde c_{H_i bb} A_{1/2}^{A}(\tau^i_{b})\right|^2}{\left|A_{1/2}^h(\tau_{t}) + A_{1/2}^h(\tau_{b})\right|^2},
\end{eqnarray}
where $\tau^i_f=M_{H_i}^2/(4m_f^2), \tau_f=M_{h}^2/(4m_f^2)$ and $i=2,3$, $f=t,b$. The functions $A_{1/2}^H$, $A_{1/2}^A$ are given by
\begin{eqnarray}
A_{1/2}^H(\tau) &=& 2\left(\tau +(\tau-1) f(\tau) \right) \tau^{-2} \ ,  \\
A_{1/2}^A(\tau) &=& 2 f(\tau) \tau^{-1} \ ,  \\
f(\tau) &=& \left\{ \begin{array}{ll}
{\rm arcsin}^2\left( \sqrt{\tau} \right), & \hspace{0.5cm} \tau\leq 1 \\
\frac{1}{4} \left[ \log\left( \frac{1+\sqrt{1-\tau^{-1}}}{1-\sqrt{1-\tau^{-1}}} \right) -i\pi \right]^2, & \hspace{0.5cm} \tau> 1
\end{array}\right.\, .
\end{eqnarray}
The above expressions are given at LO. In our numerical evaluation, the production cross sections via gluon fusion are calculated at NLO using {\tt SusHiv1.6.0} \cite{Harlander:2012pb} and interfaced with {\tt ScannerS} \cite{Muhlleitner:2020wwk}. As neutral Higgs bosons in our model have no definite CP assignment, CP-even and CP-odd contributions are summed over \cite{Fontes:2017zfn,Cacchio:2016qyh}. In addition, all decay channels including the dominant higher-order effects such as radiative corrections and multi-body channels are included as in {\tt HDECAY}\cite{Djouadi:1997yw,Djouadi:2018xqq}.
%------------------------------------------------------------
\subsection{Neutral heavy Higgs Boson decay rates}
\label{subsec:decays}
%------------------------------------------------------------
The heavy neutral scalar Higgs bosons decay rates into electroweak gauge bosons are

\begin{eqnarray}
\Gamma(H_i \to VV) = \left( a_{H_i} \right)^2 \frac{G_F M_{H_i}^3 }{ 16\sqrt{2}\pi} \delta_V \left( 1 - \frac{ 4m_V^2 }{M_{H_i}^2} \right)^{1/2} 
\left[ 1 - \frac{4m_V^2}{M_{H_i}^2} + \frac{3}{4} \left( \frac{4m_V^2}{M_{H_i}^2} \right)^2 \right] \ ,
\end{eqnarray}
where $V=W,Z$,  $\delta_W=2$, $\delta_Z=1$, and $i=2,3$.
We note that in the alignment limit, $\Gamma(H_{2,3} \to VV)=0$ when
$\sin\alpha_b=\sin\alpha_c=0$, or away from alignment. These channels open up for non-zero  CP
violation. The decay rates of the neutral Higgs bosons  to SM fermions are
\begin{eqnarray}
\Gamma(H_i \to \bar f f) =\frac{N_cG_F m_f^2 M_{H_i}}{4\sqrt{2}\pi} \left\{  (c_{H_i ff})^2   \left( 1 - \frac{4m_f^2}{M_{H_i}^2} \right)^{3/2} +  (\tilde c_{H_i ff})^2  \left( 1 - \frac{4m_f^2}{M_{H_i}^2} \right)^{1/2} \right\} \ ,
\end{eqnarray}
where $N_c=3$ for quarks and 1 for charged leptons. 

\begin{figure} [htbp]
\begin{center}
\includegraphics[width=0.497\textwidth]{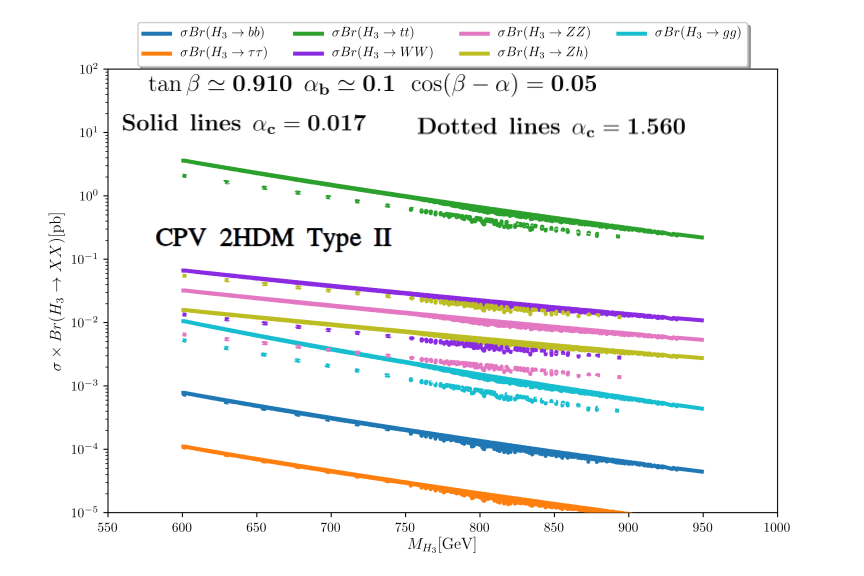}
\includegraphics[width=0.497\textwidth]{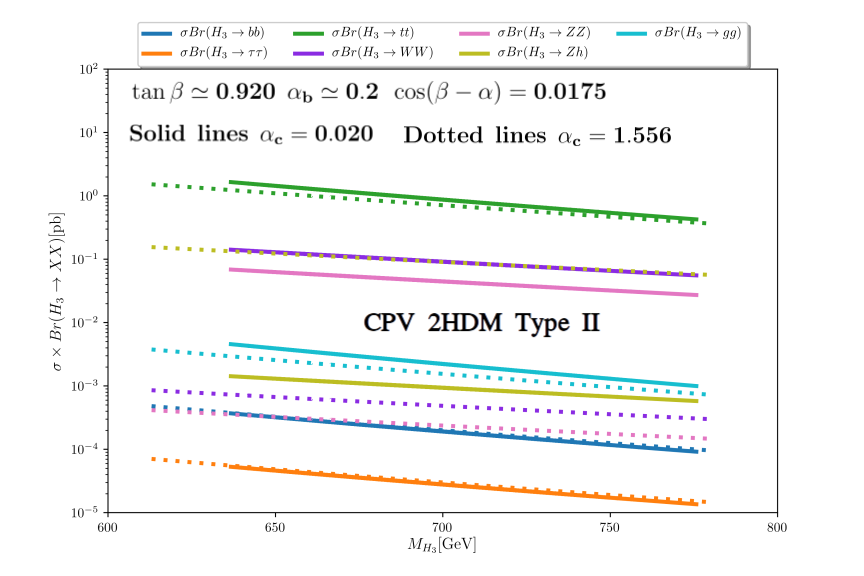}
\includegraphics[width=0.497\textwidth]{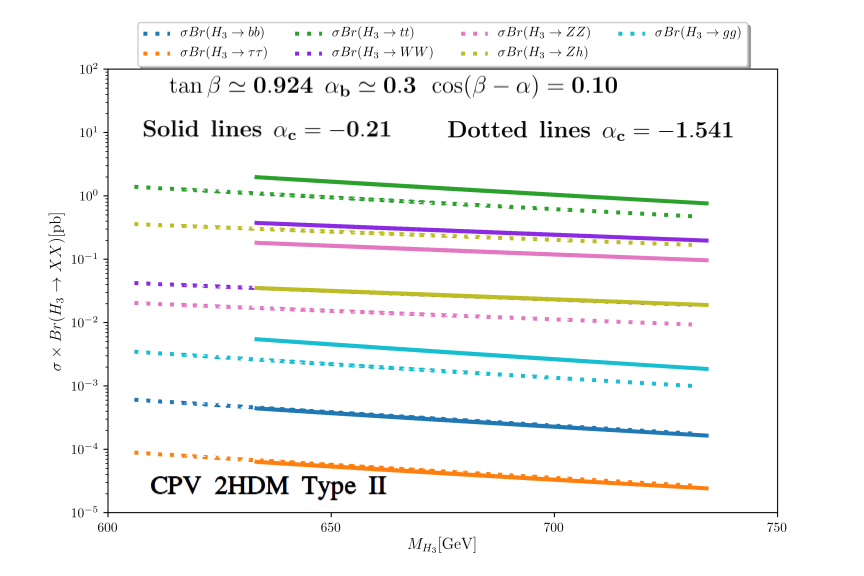}
\includegraphics[width=0.497\textwidth]{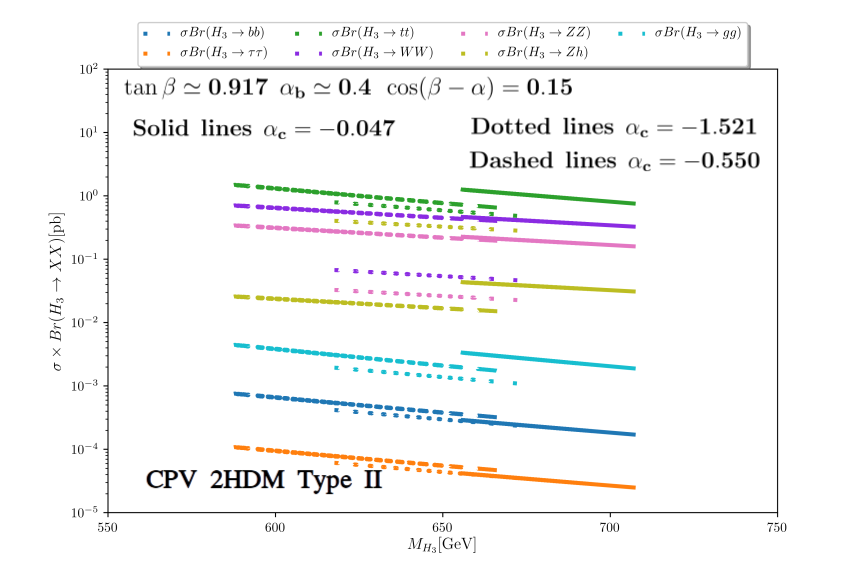}
\end{center}\vspace*{-5mm}
\caption{$\sigma_{gg_{H_3}}$ times branching ratios versus $M_{H_3}$,
  in pb,  allowed by both STU and EDM constraints, for different
  values of  $\sin \alpha_b$. Top left: $\alpha_b=0$.1, top right:
  $\alpha_b=0.2$, bottom left: $\alpha_b=0.3$, bottom right:
  $\alpha_b=0.4$. We plot, in each panel, the  $M_{H_3}$ region and
  $\alpha_c$ range surviving the constraints for the chosen values of
  $\alpha_b$ and $\cos(\beta-\alpha)$, consistent with the allowed
  ones in  Fig. \ref{fig:MHplusMH2forall}.} 
\label{fig:XSECBRH3CPV}
\end{figure}

The heavy scalars can also decay to a pair of gluons via a loop of top or bottom quarks, and the rates are
\begin{eqnarray}
\Gamma(H_i \to gg) = \frac{\alpha_s^2 G_F M_{H_i}^3}{64\sqrt{2}\pi^3} \left[ \left|c_{H_i tt} A_{1/2}^H(\tau^i_{t}) +c_{H_i bb} A_{1/2}^H(\tau^i_{b})\right|^2 + \left|\tilde c_{H_i tt} A_{1/2}^A(\tau^i_{t}) + \tilde c_{H_i bb} A_{1/2}^A(\tau^i_{b})\right|^2 \right] . \ \ \ 
\end{eqnarray}
As the decay rate is a CP even quantity, in all the above decay rates, the CP even coefficient $c_{H_i ff}$ and the CP odd one $\tilde c_{H_i ff}$ always contribute incoherently.
In addition the heavy neutral scalars, $H_2,H_3$ can decay
 into the $Z$ boson and the SM-like Higgs boson $h$,
\begin{eqnarray}\label{gAZH}
\Gamma(H_i \to Zh) &=& \frac{|g_{H_iZh}|^2}{16\pi M_{H_i}^3} \sqrt{\left( M_{H_i}^2 - (M_{h}+m_Z)^2 \right)\left( M_{H_i}^2 - (M_{h}-m_Z)^2 \right)}  \nonumber \\
&&\times \left[ - (2M_{H_i}^2+2M_{h}^2-m_Z^2) +\frac{1}{M_Z^2} (M_{H_i}^2-M_{h}^2)^2 \right]\, ,
\end{eqnarray}
where $g_{H_iZh}=(e/\sin2\theta_W)\left[(-\sin\beta {\cal R}_{11}+\cos\beta {\cal R}_{12}) {\cal R}_{i3} + (\sin\beta {\cal R}_{i1}-\cos\beta {\cal R}_{i2}) {\cal R}_{13} \right]$.
We have also calculated the decay rate of $H_i \to hh$ from the Higgs self-interactions. The decay rate is \cite{Yagyu:2012qp}
\begin{eqnarray}
\Gamma(H_i \to h h ) &=& \frac{g^2_{H_i hh} }{8\pi M_{H_i}} \sqrt{1-\frac{4 M_{h}^2}{M_{H_i}^2}} \ ,
\label{trihiggs}
\end{eqnarray}
where $g_{H_i hh}\, (i=2,3)$ are \cite{Kanemura:2004mg} %defined in \cite{Chen:2015gaa}.
\begin{eqnarray}
g_{H_i hh}&=&  \frac{1}{4v \sin 2 \beta} \left\{ 4\cos^2 (\beta-\alpha) \cos(\beta+\alpha) |m_{12}|^2- \left[ \cos(3 \alpha-\beta) +3 \cos(\beta+\alpha)\right] M_h^2\right \} {\cal R}_{i1}  \nonumber \\
&+&  \frac{\cos(\beta-\alpha)}{2v \sin 2 \beta} \left\{  (3 \sin 2\alpha -\sin2 \beta) |m_{12}|^2-\sin 2\alpha (2M_h^2+M_{H_i}^2)\right \} {\cal R}_{i2}
\end{eqnarray}
To obtain the branching ratios, we include the total width of the heavy Higgs,
\begin{eqnarray}
\Gamma_{\rm tot}(H_i) &=& \Gamma(H_i \to W^+W^-) + \Gamma(H_i \to ZZ) + \Gamma(H_i \to t\bar t) + \Gamma(H_i \to b\bar b) \nonumber \\
&+& \Gamma(H_i \to \tau^+ \tau^-) + \Gamma(H_i \to gg) + \Gamma(H_i \to Z h) + \Gamma(H_i \to h h) \ .
\end{eqnarray}
\FloatBarrier
%At this point, it will be very important to study the effect of CP violation on the heavy Higgs decays to gauge bosons channels, 
%rch for heavy Higgs decays into both gauge bosons and two SM-like Higgs states.\\
The branching ratios of the various physical scalars in CPV 2HDM were
calculated using {\tt C2HDM HDECAY} \cite{Fontes:2017zfn} through the
{\tt HDECAY} interface \cite{Djouadi:1997yw, Djouadi:2018xqq,
  Butterworth:2010ym}. Predictions for gluon-fusion Higgs production
at hadron colliders for the scalars were obtained  using tabulated
results from {\tt SusHi} \cite{Harlander:2012pb} at NLO. We insert the
CPV 2HDM model by turning on the CP violating angles, $\alpha_b$ and
$\alpha_c$. From our parameter space investigations, we find that the maximum CP
violating angle $\alpha_b$ allowed by the Higgs signal strength
measurements from the LHC is around $\alpha_b \simeq 0.4$ as shown on the bottom
right panel in Fig. \ref{fig:Higgs_space}. 
\FloatBarrier
\begin{figure} [htbp]
\begin{center}
\includegraphics[width=0.497\textwidth]{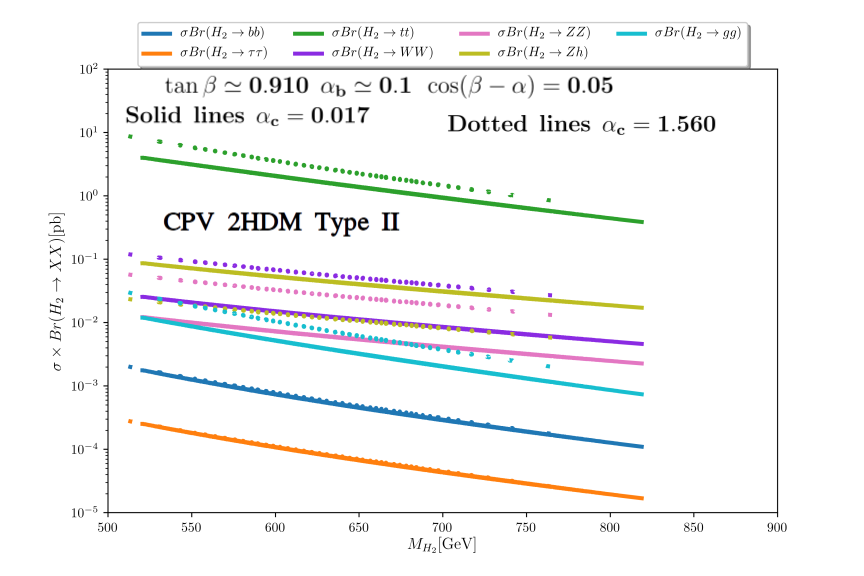}
\includegraphics[width=0.497\textwidth]{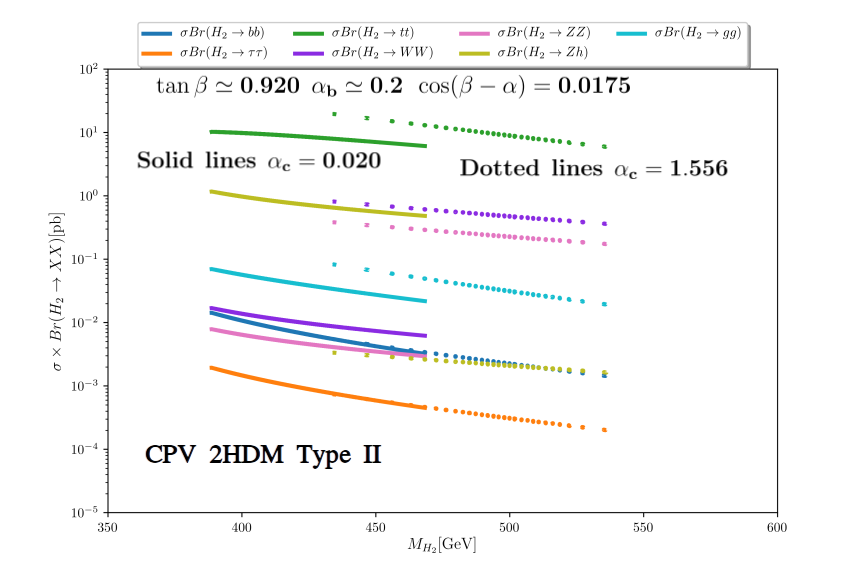}
\includegraphics[width=0.497\textwidth]{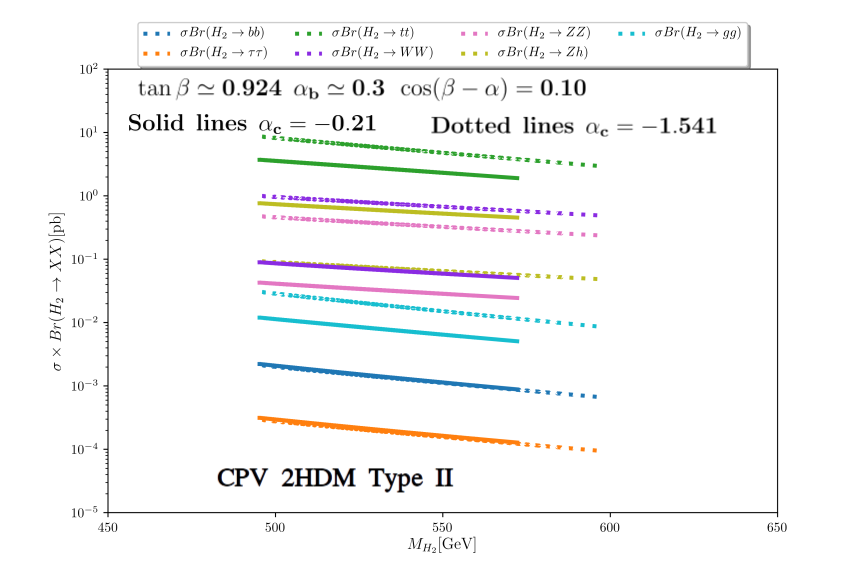}
\includegraphics[width=0.497\textwidth]{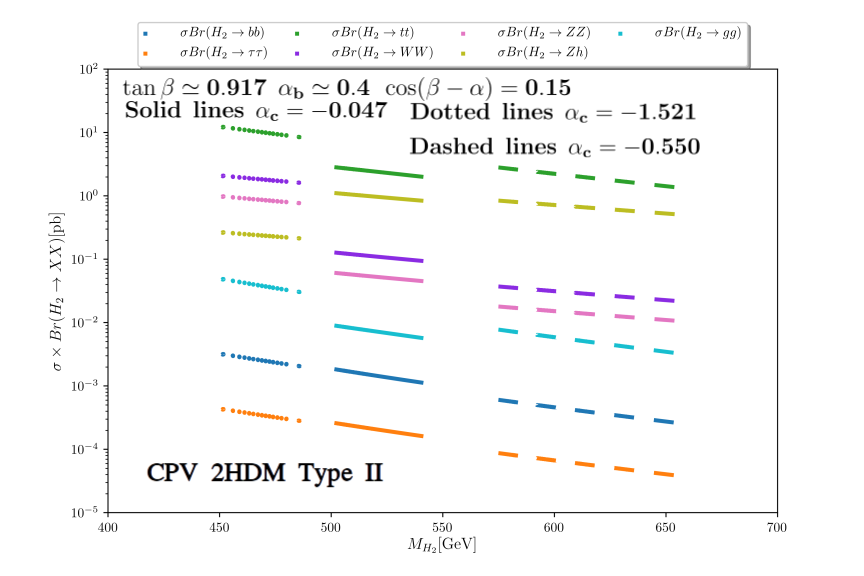}
\end{center}
\caption{$\sigma_{gg_{H_2}}$ times branching ratios versus $M_{H_2}$, in pb,  allowed by both STU and EDM constraints for different values of  $\sin \alpha_b$. Top left: $\alpha_b=0$.1, top right: $\alpha_b=0.2$, bottom left: $\alpha_b=0.3$, bottom right: $\alpha_b=0.4$. We plot, in each panel, the  $M_{H_2}$ region and $\alpha_c$ range surviving the constraints for the chosen values of $\alpha_b$ and $\cos(\beta-\alpha)$, consistent with the allowed ones in Fig. \ref{fig:MHplusMH2forall}.}
\label{fig:XSECBRH2CPV}
\end{figure}
In Fig. \ref{fig:XSECBRH2CPV} and Fig. \ref{fig:XSECBRH3CPV}, we
present the results for the the production cross sections times the
branching ratios of $H_2$ and $H_3$, respectively, for various mixing
angles $\cos (\beta-\alpha)$ and CPV angles. For consistency, we maintain the same parameter space analysed in Fig. \ref{fig:MHplusMH2forall}.

In our analysis,  the
second and third neutral Higgs are close in mass, as a
result  of the $STU$ constraints, and because of this, the heavy
neutral Higgs bosons decay only into SM particles. The mass of the charged
Higgs boson is initially randomly scanned although its 
value is restricted to lie within the allowed range (note that the
third neutral Higgs mass is not an independent parameter, but 
computed from the mass of the lighter neutral Higgs bosons and the
various  angles as shown in Eq. \ref{eq:mh3}).

We  show results for the production and decay of the heavy Higgs slightly
away from the alignment limit (in which  $\cos\left(\beta
-\alpha\right) = 0$). %and in which the heavy Higgs' have vanishing
%couplings with gauge bosons).
The results shown in Fig. \ref{fig:XSECBRH2CPV} and
Fig. \ref{fig:XSECBRH3CPV} are such that  $\cos(\beta - \alpha) =
0.05$ on the top left,  $\cos(\beta - \alpha) = 0.0175$ on the top right,
$\cos(\beta - \alpha) = 0.1$ on the bottom left and $\cos(\beta - \alpha)
= 0.15$ on the bottom right, and are all in a parameter space region
allowed by LHC Higgs data from Fig. \ref{fig:Higgs_space}.

 Note that the mass regions in each plot are different. This is
 because we show,  in each panel, the  $M_{H_2}$ and $M_{H_3}$ regions
 and $\alpha_c$ range surviving the constraints for the chosen values
 of $\alpha_b$ and $\cos(\beta-\alpha)$, consistent with the allowed
 ones in  Fig. \ref{fig:MHplusMH2forall}. 

In this regime, the decays into gauge bosons are open, in both the CP-even
final states ($ZZ$ and $WW$) and in the CP-odd final state 
($Zh$). We plot as solid (dotted) lines, the production times branching ratios
for $H_2$ for the minimum and maximum 
allowed values for $\alpha_c$ from Fig. \ref{fig:MHplusMH2forall},
while keeping the same colour convention for each channel.

In all plots, $H_2\rightarrow t {\bar t}$ is the dominant decay mode, owing to small
values of $\tan \beta$.
Production and decay rates into gauge bosons can 
reach  ${\cal O}(1)$ pb depending on the values of $\alpha_b$ and
$\alpha_c$; similar cross-section values can be reached for the CP-odd
final state $Zh$,  for different values of the CPV mixing angles.
In fact, in all panels we observe how the cross sections
between CP-even gauge boson final states and CP-odd final states can
flip by changing the value of $\alpha_c$ from its minimum value (solid
lines) to its maximum value (dotted lines).

In the two top panels, with  $\alpha_b=0.1, \alpha_b=0.2$, when the value of
$\alpha_c$ is close to $0$ (solid lines), the decays into $Zh$ represent the main
decay channel involving gauge bosons, with the CP-even decay channels
into $WW$ and $ZZ$ being between one to two orders of magnitude smaller.
This shows that in this limit of small CPV mixing angles, the state
$H_2$ is ``mostly'' CP-odd.
However, as we increase the value of $\alpha_c$ (dotted lines), the decays into CP-even gauge
boson channels become larger than into the CP-odd channel $Zh$, and so the
state $H_2$ can be considered as ``mostly'' CP-even.

In the lower panels the situation is similar, even though the value of $\alpha_b$ is larger, with
$\alpha_b=0.3, \alpha_c=0.4$. The allowed values of $\alpha_c$ are constrained
to be negative, but again when the absolute value of $\alpha_c$ is
small (solid lines) the decays in to $Zh$ represent the main 
decay channel with a gauge boson, so that again, for smaller CPV
angles,  the state $H_2$ is ``mostly'' CP-odd.
For a larger absolute value of $\alpha_c$ (dotted lines), the CP-even gauge
boson decays become larger than the CP-odd channel $Zh$, and so the
state $H_2$ is now ``mostly''  CP-even.

In the top panels and the bottom left panel, intermediate values of
$\alpha_c$ do allow for relatively similar windows for the mass of
$H_2$. However for the case $\alpha_b=0.4$ only a limited range of 
mass values for $H_2$ satisfies the constraints, and this is distinct
for every value of $\alpha_c$. To showcase this effect in the bottom
right panel, we also plot the production and decay cross section of
$H_2$ for  an intermediate  value of $\alpha_c$, shown 
by dashed lines. The effect of changing  $\alpha_c$ is entirely
non-linear: while the minimum to maximum are  slightly shifted from
the left (lower masses) to the  right (higher  masses), the plots for
intermediate values of $\alpha_c$ do not fall  in-between but are
shifted further to the right. However both the dashed and solid lines
correspond to a ``mostly'' CP-odd scalar, whereas the
``mostly'' CP-even scalar corresponds to the dotted line, for the lowest
allowed scalar masses.

We expect that  plotting  all allowed values of $\alpha_c$ will result in
 uninterrupted lines, each segment corresponding to different
 $\alpha_c$'s. The slight shifting in masses is noticeable in the
 other panels, but not as pronounced as for $\alpha_b=0.4$. 
 
Fig. \ref{fig:XSECBRH3CPV} analyses the same production and decay
cross sections for $H_3$, the heaviest neutral boson. We vary the
parameters over the same allowed region as in
Fig. \ref{fig:XSECBRH2CPV}. The plots show the expected behavior of
the heaviest neutral scalar relative to the scalar $H_2$ with respect
to its couplings to gauge bosons.
For the parameter values where the scalar $H_2$ was found to be ``mostly''
CP-odd (small values of $\alpha_c$), the scalar $H_3$ is ``mostly''
CP-even, meaning its decay cross section into $WW$ or $ZZ$ is dominant
over the decay into $Zh$. And conversely when $H_2$ is ``mostly'' CP-even,
$H_3$ then behaves as ``mostly'' CP-odd.

In all cases the dominant decay channels are $t{\bar t}$, $WW$, $ZZ$ and $Zh$.

\FloatBarrier
%\vspace{-0.5cm}
%%%%%%%%%%%%%%%%%%%%%%%%%%%%%%%%%%%%%%%%%%%%%%%%%%%%%%%%

The approximate degeneracy of the masses $M_{H_3}$ and $M_{H_2}$ has
attracted some interest from phenomenologists. At the LHC, experiments
looked for the four lepton final states, resulting from decays into
$ZZ$, often referred to as the golden channel when searching for
additional heavy scalar resonances. This is  motivated by the fact
that the corresponding SM backgrounds are small and
controllable. Encouragingly, the four-lepton analyses from ATLAS
\cite{ATLAS:2017tlw} and CMS \cite{CMS:2018mmw,CMS:2017jkd} indicate
some enhancement in the event rates in final states with high
invariant masses. These analyses were interpreted to be consistent
with a broad resonance structure around 700 GeV \cite{Cosmai:2021hvc},
and most recently, also interpreted   as a\ double peak of degenerate
states around 680 GeV in the 2HDM  \cite{Antusch:2021oit}. In their
case, the degeneracy is imposed to fit the data, while in our analysis, this scenario
is completely consistent and imposed by the experimental and
theoretical constraints. 
\section{SUMMARY and CONCLUSION}
\label{sec:summary}
%%%%%%%%%%%%%%%%%%%%%%%%%%%%%%%%%%%%%%%%%%%%%%%%%%%%%%%%
In the present study, we have explored the allowed parameter space of the
Type-II Two Higgs Doublet Model with CP violation, and the
implications for the production and decays of the additional neutral
Higgs bosons at the LHC.
%% , allowing one doublet to couple to up-type
%% quarks and neutrinos, the other to down-type and changed leptons. This
%% model extends the SM particle content 
%% minimally by one additional Higgs doublet. With CP violation there are
%% three neutral scalars Higgs boson mixing with each other, and we
%% require one of them to be consistent with having the mass and
%% properties of the SM Higgs, as restricted from the LHC data.  Due to
%% the complex nature of the Yukawa structure in the model, the Higgs
%% couplings with fermions  and bosons are modified.
After implementing the model with approximate $Z_2$ symmetry, we set some parameters to
zero to forbid flavor-changing neutral currents, but still allowing
for complex parameters in the potential. We then proceed by restricting the
CP-violating parameter space using both theoretical considerations and
experimental data constraints.  

First, we showed that after imposing constraints from the LHC-Higgs data
%(that is, we require the lightest neutral scalar to have the
%properties consistent with the SM-like Higgs boson to at least
%3$\sigma$ precision)
some regions of parameter space  very close to
alignment as well as relatively away from it survive (the
alignment region is such that the heavy Higgs fields decouple from the
rest of the spectrum leaving only a single SM-like Higgs within the scalar
sector).

Within the allowed regions of parameter space, we imposed theoretical constraints involving  vacuum stability,
unitarity, and perturbativity, electroweak precision constraints, as
well as experimental constraints from $B$-physics and electric dipole
moments. We compute the relevant couplings and perform a random scan
over two of the 2HDM physical masses,  $M_{H_2}$ (the second lightest
neutral Higgs) and $M_{H^{\pm}}$, within parameter regions which
satisfy the Higgs data  and all other constraints.

\vskip0.2in
The main findings presented in this work, representing new
  contributions to the study of CVP 2HDM can be listed as follows. 

\begin{itemize}
\item We clearly presented restrictions on the parameter space
  of the CPV 2HDM coming from LHC Higgs data in  the plane $\tan
  \beta$ - $\cos (\beta-\alpha)$ (the mixing angle in the neutral
  Higgs boson sector), for fixed values of $\alpha_b$, the amount of CPV admixture in the lightest
  (SM-like) Higgs. Our plots, which for $\alpha_b=0$ coincide with
  available analyses for the CP-conserving 2HDM, also show how the parameter space allowed by Higgs data shrinks
  with increasing $\alpha_b$. 
\item Throughout the CPV parameter space explored, $\tan \beta \sim
  0.9$ emerges as a solid constraint caused by the necessity of having
  close to degenerate neutral heavy scalar masses in order to pass
  precision electroweak tests and EDM's. 
  \item  We proved analytically (and confirmed numerically) that in the CPV 2HDM the masses of the heavy neutral Higgs, $M_{H_2}$ and $M_{H_3}$, as well as that of the charged Higgs, $M_{H^\pm}$, are bounded from above. We demonstrated that this is a feature specific to CPV only, and is absent in the CPC 2HDMs.
  %, independent of masses, mixing
 % angles and amount of CPV.   
\item We also show that, for most of the parameter region allowed, and except for small corners of the parameter
  space (for the largest $\alpha_b$ allowed, and away from alignment),
  the two heavier neutral Higgs bosons $H_2$ and $H_3$ are constrained
  to be approximately  degenerate in mass. This is a condition imposed
  by $S, T$ parameters. The latter also affect $\Im \lambda_5$, a
  measure of CP violation in the potential, which is quite large in
  these scenarios. 
\item While the masses of the charged and neutral Higgs bosons are varied in a
  large parameter range, they are constrained to lie in a small mass
  window. %For a wide choice of fixed $\alpha_b$, the mass
  %splitting between the charged Higgs $H^\pm$ and the second lightest
  %neutral Higgs $ H_2$, is severely constrained, a consequence of requirements of unitarity and perturbativity of the scalar potential. 
  We also separate clearly restrictions coming from
  oblique $S, T$ parameters and electric dipole moments, from the
  region which obeys all other experimental and theoretical bounds
  (such as Higgs data, and $B$-physics
  constraints).  %Throughout there are strong correlations between the masses in the
  %neutral and charged sectors due to perturbativity bounds,
 % disallowing $M_{H^\pm}$ to be arbitrarily heavy.  
  \item While the exact
  value of the mass window $M_{H^\pm}-M_{H_2}$ depends on $\cos (\beta-\alpha)$,
  $\alpha_b$ and $\alpha_c$ (the amount of CPV admixture in $H_2$ and
  $H_3$), the window is usually only 80-250 GeV wide. There is lower bound on $M_{H^\pm}> 650$ GeV  as required by
  $B$ physics. Thus our analysis restricts $M_{H^\pm}$, for a given choice of parameters, {\it both} from above and below.
\item Our plots also show how, increasing the value for $\alpha_c$ the
  two heavier Higgs boson flip from ``mostly CP-even'' to ``mostly
  CP-odd''. And while the dominant decay is into $t{\bar t}$, the
  decays into bosonic states are next largest and can be used to
  distinguish this scenario. 
 \end{itemize}

In conclusion, our work shows that the parameter space of the CPV 2HDM is very
constrained, and the neutral Higgs bosons in the CPV Type-II 2HDM show
some interesting features, which could distinguish them from Higgs
bosons in other models at the colliders. 

\bigskip
%\clearpage
\noindent
%%%%%%%%%%%%%%%%%%%%%%%%%%%%%%%%%%%%%%%%%%%%%%%%%%%%%%%%
{\Large \bf Acknowledgements}\\[5mm]
\label{acknowledgement}
%%%%%%%%%%%%%%%%%%%%%%%%%%%%%%%%%%%%%%%%%%%%%%%%%%%%%%%%
The work of MF  has been partly supported by NSERC
through the grant number SAP105354.
M.T. would like to thank FRQNT for financial support under              
grant number PRC-290000. 
%%%%%%%%%%%%%%%%%%%%%%%%%%%%%%%%%%%%%%%%%%%%%%%
%%%%%%%%%%%%%%%%%
\bibliographystyle{JHEP}
\bibliography{ref}

\providecommand{\href}[2]{#2}\begingroup\raggedright\begin{thebibliography}{100}

\bibitem{Pich:2012sx}
A.~Pich, \emph{{The Standard Model of Electroweak Interactions}},  in
  \emph{{2010 European School of High Energy Physics}}, pp.~1--50, 1, 2012
  [\href{https://arxiv.org/abs/1201.0537}{{\ttfamily 1201.0537}}].

\bibitem{Furey:2018drh}
C.~Furey, \emph{{$SU(3)_C\times SU(2)_L\times U(1)_Y\left( \times U(1)_X
  \right) $ as a symmetry of division algebraic ladder operators}},
  \href{https://doi.org/10.1140/epjc/s10052-018-5844-7}{\emph{Eur. Phys. J. C}
  {\bfseries 78} (2018) 375}
  [\href{https://arxiv.org/abs/1806.00612}{{\ttfamily 1806.00612}}].

\bibitem{Diaz:2003dk}
R.A.~Diaz, R.~Martinez and F.~Ochoa, \emph{{The Scalar sector of the SU(3)(c) x
  SU(3)(L) x U(1)(X) model}},
  \href{https://doi.org/10.1103/PhysRevD.69.095009}{\emph{Phys. Rev. D}
  {\bfseries 69} (2004) 095009}
  [\href{https://arxiv.org/abs/hep-ph/0309280}{{\ttfamily hep-ph/0309280}}].

\bibitem{Novaes:1999yn}
S.F.~Novaes, \emph{{Standard model: An Introduction}},  in \emph{{10th Jorge
  Andre Swieca Summer School: Particle and Fields}}, pp.~5--102, 1, 1999
  [\href{https://arxiv.org/abs/hep-ph/0001283}{{\ttfamily hep-ph/0001283}}].

\bibitem{Logan:2014jla}
H.E.~Logan, \emph{{TASI 2013 lectures on Higgs physics within and beyond the
  Standard Model}},  \href{https://arxiv.org/abs/1406.1786}{{\ttfamily
  1406.1786}}.

\bibitem{Aad:2012tfa}
{\scshape ATLAS} collaboration, \emph{{Observation of a new particle in the
  search for the Standard Model Higgs boson with the ATLAS detector at the
  LHC}}, \href{https://doi.org/10.1016/j.physletb.2012.08.020}{\emph{Phys.
  Lett.} {\bfseries B716} (2012) 1}
  [\href{https://arxiv.org/abs/1207.7214}{{\ttfamily 1207.7214}}].

\bibitem{Chatrchyan:2012xdj}
{\scshape CMS} collaboration, \emph{{Observation of a New Boson at a Mass of
  125 GeV with the CMS Experiment at the LHC}},
  \href{https://doi.org/10.1016/j.physletb.2012.08.021}{\emph{Phys. Lett.}
  {\bfseries B716} (2012) 30}
  [\href{https://arxiv.org/abs/1207.7235}{{\ttfamily 1207.7235}}].

\bibitem{Lykken:2010mc}
J.D.~Lykken, \emph{{Beyond the Standard Model}},  in \emph{{CERN Yellow Report
  CERN-2010-002, 101-109}}, 2010,
  \href{http://lss.fnal.gov/archive/2010/conf/fermilab-conf-10-103-t.pdf}{http://lss.fnal.gov/archive/2010/conf/fermilab-conf-10-103-t.pdf}
  [\href{https://arxiv.org/abs/1005.1676}{{\ttfamily 1005.1676}}].

\bibitem{Gunion_2018}
J.F.~Gunion, H.E.~Haber, G.~Kane and D.~Sally, \emph{The Higgs
  Hunter{\textquotesingle}s Guide}, {CRC} Press (mar, 2018),
  \href{https://doi.org/10.1201/9780429496448}{10.1201/9780429496448}.

\bibitem{HABER198575}
H.~Haber and G.~Kane, \emph{The search for supersymmetry: Probing physics
  beyond the standard model},
  \href{https://doi.org/https://doi.org/10.1016/0370-1573(85)90051-1}{\emph{Physics
  Reports} {\bfseries 117} (1985) 75}.

\bibitem{Gunion:1984yn}
J.F.~Gunion and H.E.~Haber, \emph{{Higgs Bosons in Supersymmetric Models. 1.}},
  \href{https://doi.org/10.1016/0550-3213(86)90340-8}{\emph{Nucl. Phys. B}
  {\bfseries 272} (1986) 1}.

\bibitem{Branco:2011iw}
G.C.~Branco, P.M.~Ferreira, L.~Lavoura, M.N.~Rebelo, M.~Sher and J.P.~Silva,
  \emph{{Theory and phenomenology of two-Higgs-doublet models}},
  \href{https://doi.org/10.1016/j.physrep.2012.02.002}{\emph{Phys. Rept.}
  {\bfseries 516} (2012) 1} [\href{https://arxiv.org/abs/1106.0034}{{\ttfamily
  1106.0034}}].

\bibitem{Nilles:1983ge}
H.P.~Nilles, \emph{{Supersymmetry, Supergravity and Particle Physics}},
  \href{https://doi.org/10.1016/0370-1573(84)90008-5}{\emph{Phys. Rept.}
  {\bfseries 110} (1984) 1}.

\bibitem{Accomando:2006ga}
E.~Accomando et~al., \emph{{Workshop on CP Studies and Non-Standard Higgs
  Physics}},  \href{https://arxiv.org/abs/hep-ph/0608079}{{\ttfamily
  hep-ph/0608079}}.

\bibitem{Pilaftsis:1999qt}
A.~Pilaftsis and C.E.M.~Wagner, \emph{{Higgs bosons in the minimal
  supersymmetric standard model with explicit CP violation}},
  \href{https://doi.org/10.1016/S0550-3213(99)00261-8}{\emph{Nucl. Phys. B}
  {\bfseries 553} (1999) 3}
  [\href{https://arxiv.org/abs/hep-ph/9902371}{{\ttfamily hep-ph/9902371}}].

\bibitem{Haber:2006ue}
H.E.~Haber and D.~O'Neil, \emph{{Basis-independent methods for the
  two-Higgs-doublet model. II. The Significance of tan$\beta$}},
  \href{https://doi.org/10.1103/PhysRevD.74.015018}{\emph{Phys. Rev. D}
  {\bfseries 74} (2006) 015018}
  [\href{https://arxiv.org/abs/hep-ph/0602242}{{\ttfamily hep-ph/0602242}}].

\bibitem{Sakharov:1967dj}
A.~Sakharov, \emph{{Violation of CP Invariance, C asymmetry, and baryon
  asymmetry of the universe}},
  \href{https://doi.org/10.1070/PU1991v034n05ABEH002497}{\emph{Sov. Phys. Usp.}
  {\bfseries 34} (1991) 392}.

\bibitem{Kuzmin:1985mm}
V.~Kuzmin, V.~Rubakov and M.~Shaposhnikov, \emph{{On the Anomalous Electroweak
  Baryon Number Nonconservation in the Early Universe}},
  \href{https://doi.org/10.1016/0370-2693(85)91028-7}{\emph{Phys. Lett. B}
  {\bfseries 155} (1985) 36}.

\bibitem{Gunion:2002zf}
J.F.~Gunion and H.E.~Haber, \emph{{The CP conserving two Higgs doublet model:
  The Approach to the decoupling limit}},
  \href{https://doi.org/10.1103/PhysRevD.67.075019}{\emph{Phys. Rev. D}
  {\bfseries 67} (2003) 075019}
  [\href{https://arxiv.org/abs/hep-ph/0207010}{{\ttfamily hep-ph/0207010}}].

\bibitem{Haber:2015pua}
H.E.~Haber and O.~St\r{a}l, \emph{{New LHC benchmarks for the $\mathcal{CP}$
  -conserving two-Higgs-doublet model}},
  \href{https://doi.org/10.1140/epjc/s10052-015-3697-x}{\emph{Eur. Phys. J. C}
  {\bfseries 75} (2015) 491}
  [\href{https://arxiv.org/abs/1507.04281}{{\ttfamily 1507.04281}}].

\bibitem{Grzadkowski:2015zma}
B.~Grzadkowski, O.~Ogreid and P.~Osland, \emph{{Testing the presence of CP
  violation in the 2HDM}},
  \href{https://doi.org/10.22323/1.231.0086}{\emph{PoS} {\bfseries CORFU2014}
  (2015) 086} [\href{https://arxiv.org/abs/1504.06076}{{\ttfamily
  1504.06076}}].

\bibitem{Keus:2015hva}
V.~Keus, S.F.~King, S.~Moretti and K.~Yagyu, \emph{{CP Violating
  Two-Higgs-Doublet Model: Constraints and LHC Predictions}},
  \href{https://doi.org/10.1007/JHEP04(2016)048}{\emph{JHEP} {\bfseries 04}
  (2016) 048} [\href{https://arxiv.org/abs/1510.04028}{{\ttfamily
  1510.04028}}].

\bibitem{Darvishi:2016gvm}
N.~Darvishi and M.~Krawczyk, \emph{{CP violation in the extension of SM with a
  complex singlet scalar and vector quarks}},
  \href{https://doi.org/10.1016/j.nuclphysb.2020.115242}{\emph{Nucl. Phys. B}
  {\bfseries 962} (2021) 115}
  [\href{https://arxiv.org/abs/1603.00598}{{\ttfamily 1603.00598}}].

\bibitem{Emmanuel-Costa:2016opd}
D.~Emmanuel-Costa, O.~Ogreid, P.~Osland and M.~Rebelo, \emph{{CP Violation in
  the scalar sector}}, \href{https://doi.org/10.22323/1.263.0044}{\emph{PoS}
  {\bfseries CORFU2015} (2016) 044}
  [\href{https://arxiv.org/abs/1604.00637}{{\ttfamily 1604.00637}}].

\bibitem{Pich:2009sp}
A.~Pich and P.~Tuzon, \emph{{Yukawa Alignment in the Two-Higgs-Doublet Model}},
  \href{https://doi.org/10.1103/PhysRevD.80.091702}{\emph{Phys. Rev. D}
  {\bfseries 80} (2009) 091702}
  [\href{https://arxiv.org/abs/0908.1554}{{\ttfamily 0908.1554}}].

\bibitem{Arhrib:2010ju}
A.~Arhrib, E.~Christova, H.~Eberl and E.~Ginina, \emph{{CP violation in charged
  Higgs production and decays in the Complex Two Higgs Doublet Model}},
  \href{https://doi.org/10.1007/JHEP04(2011)089}{\emph{JHEP} {\bfseries 04}
  (2011) 089} [\href{https://arxiv.org/abs/1011.6560}{{\ttfamily 1011.6560}}].

\bibitem{Jung:2010ik}
M.~Jung, A.~Pich and P.~Tuzon, \emph{{Charged-Higgs phenomenology in the
  Aligned two-Higgs-doublet model}},
  \href{https://doi.org/10.1007/JHEP11(2010)003}{\emph{JHEP} {\bfseries 11}
  (2010) 003} [\href{https://arxiv.org/abs/1006.0470}{{\ttfamily 1006.0470}}].

\bibitem{Bao:2010sz}
S.-S.~Bao, Y.~Tang and Y.-L.~Wu, \emph{{$W^{\pm}H^{\mp}$ associated production
  at LHC in the general 2HDM with Spontaneous CP Violation}},
  \href{https://doi.org/10.1103/PhysRevD.83.075006}{\emph{Phys. Rev. D}
  {\bfseries 83} (2011) 075006}
  [\href{https://arxiv.org/abs/1011.1409}{{\ttfamily 1011.1409}}].

\bibitem{Basso:2012st}
L.~Basso, A.~Lipniacka, F.~Mahmoudi, S.~Moretti, P.~Osland, G.~Pruna et~al.,
  \emph{{Probing the charged Higgs boson at the LHC in the CP-violating type-II
  2HDM}}, \href{https://doi.org/10.1007/JHEP11(2012)011}{\emph{JHEP} {\bfseries
  11} (2012) 011} [\href{https://arxiv.org/abs/1205.6569}{{\ttfamily
  1205.6569}}].

\bibitem{Jung:2013hka}
M.~Jung and A.~Pich, \emph{{Electric Dipole Moments in Two-Higgs-Doublet
  Models}}, \href{https://doi.org/10.1007/JHEP04(2014)076}{\emph{JHEP}
  {\bfseries 04} (2014) 076} [\href{https://arxiv.org/abs/1308.6283}{{\ttfamily
  1308.6283}}].

\bibitem{Brod:2013cka}
J.~Brod, U.~Haisch and J.~Zupan, \emph{{Constraints on CP-violating Higgs
  couplings to the third generation}},
  \href{https://doi.org/10.1007/JHEP11(2013)180}{\emph{JHEP} {\bfseries 11}
  (2013) 180} [\href{https://arxiv.org/abs/1310.1385}{{\ttfamily 1310.1385}}].

\bibitem{Inoue:2014nva}
S.~Inoue, M.J.~Ramsey-Musolf and Y.~Zhang, \emph{{CP-violating phenomenology of
  flavor conserving two Higgs doublet models}},
  \href{https://doi.org/10.1103/PhysRevD.89.115023}{\emph{Phys. Rev. D}
  {\bfseries 89} (2014) 115023}
  [\href{https://arxiv.org/abs/1403.4257}{{\ttfamily 1403.4257}}].

\bibitem{Gaitan:2015hga}
R.~Gait\'an, J.H.~Montes~de Oca, E.A.~Garc\'es and R.~Martinez, \emph{{Rare top
  decay $t \rightarrow c \gamma$ with flavor changing neutral scalar
  interactions in two Higgs doublet model}},
  \href{https://doi.org/10.1103/PhysRevD.94.094038}{\emph{Phys. Rev. D}
  {\bfseries 94} (2016) 094038}
  [\href{https://arxiv.org/abs/1503.04391}{{\ttfamily 1503.04391}}].

\bibitem{Chen:2015gaa}
C.-Y.~Chen, S.~Dawson and Y.~Zhang, \emph{{Complementarity of LHC and EDMs for
  Exploring Higgs CP Violation}},
  \href{https://doi.org/10.1007/JHEP06(2015)056}{\emph{JHEP} {\bfseries 06}
  (2015) 056} [\href{https://arxiv.org/abs/1503.01114}{{\ttfamily
  1503.01114}}].

\bibitem{Glashow:1976nt}
S.L.~Glashow and S.~Weinberg, \emph{{Natural Conservation Laws for Neutral
  Currents}}, \href{https://doi.org/10.1103/PhysRevD.15.1958}{\emph{Phys. Rev.}
  {\bfseries D15} (1977) 1958}.

\bibitem{Georgi:1978ri}
H.~Georgi and D.V.~Nanopoulos, \emph{{Suppression of Flavor Changing Effects
  From Neutral Spinless Meson Exchange in Gauge Theories}},
  \href{https://doi.org/10.1016/0370-2693(79)90433-7}{\emph{Phys. Lett.}
  {\bfseries 82B} (1979) 95}.

\bibitem{Barger:1989fj}
V.D.~Barger, J.L.~Hewett and R.J.N.~Phillips, \emph{{New Constraints on the
  Charged Higgs Sector in Two Higgs Doublet Models}},
  \href{https://doi.org/10.1103/PhysRevD.41.3421}{\emph{Phys. Rev. D}
  {\bfseries 41} (1990) 3421}.

\bibitem{Grossman:1994jb}
Y.~Grossman, \emph{{Phenomenology of models with more than two Higgs
  doublets}}, \href{https://doi.org/10.1016/0550-3213(94)90316-6}{\emph{Nucl.
  Phys. B} {\bfseries 426} (1994) 355}
  [\href{https://arxiv.org/abs/hep-ph/9401311}{{\ttfamily hep-ph/9401311}}].

\bibitem{Akeroyd:1996he}
A.G.~Akeroyd, \emph{{Nonminimal neutral Higgs bosons at LEP-2}},
  \href{https://doi.org/10.1016/0370-2693(96)00330-9}{\emph{Phys. Lett. B}
  {\bfseries 377} (1996) 95}
  [\href{https://arxiv.org/abs/hep-ph/9603445}{{\ttfamily hep-ph/9603445}}].

\bibitem{Djouadi:2005gj}
A.~Djouadi, \emph{{The Anatomy of electro-weak symmetry breaking. II. The Higgs
  bosons in the minimal supersymmetric model}},
  \href{https://doi.org/10.1016/j.physrep.2007.10.005}{\emph{Phys. Rept.}
  {\bfseries 459} (2008) 1}
  [\href{https://arxiv.org/abs/hep-ph/0503173}{{\ttfamily hep-ph/0503173}}].

\bibitem{Demir:1999hj}
D.A.~Demir, \emph{{Effects of the supersymmetric phases on the neutral Higgs
  sector}}, \href{https://doi.org/10.1103/PhysRevD.60.055006}{\emph{Phys. Rev.
  D} {\bfseries 60} (1999) 055006}
  [\href{https://arxiv.org/abs/hep-ph/9901389}{{\ttfamily hep-ph/9901389}}].

\bibitem{Choi:2000wz}
S.Y.~Choi, M.~Drees and J.S.~Lee, \emph{{Loop corrections to the neutral Higgs
  boson sector of the MSSM with explicit CP violation}},
  \href{https://doi.org/10.1016/S0370-2693(00)00421-4}{\emph{Phys. Lett. B}
  {\bfseries 481} (2000) 57}
  [\href{https://arxiv.org/abs/hep-ph/0002287}{{\ttfamily hep-ph/0002287}}].

\bibitem{Carena:2000yi}
M.~Carena, J.R.~Ellis, A.~Pilaftsis and C.E.M.~Wagner, \emph{{Renormalization
  group improved effective potential for the MSSM Higgs sector with explicit CP
  violation}}, \href{https://doi.org/10.1016/S0550-3213(00)00358-8}{\emph{Nucl.
  Phys. B} {\bfseries 586} (2000) 92}
  [\href{https://arxiv.org/abs/hep-ph/0003180}{{\ttfamily hep-ph/0003180}}].

\bibitem{Carena:2001fw}
M.~Carena, J.R.~Ellis, A.~Pilaftsis and C.E.M.~Wagner, \emph{{Higgs Boson Pole
  Masses in the MSSM with Explicit CP Violation}},
  \href{https://doi.org/10.1016/S0550-3213(02)00014-7}{\emph{Nucl. Phys. B}
  {\bfseries 625} (2002) 345}
  [\href{https://arxiv.org/abs/hep-ph/0111245}{{\ttfamily hep-ph/0111245}}].

\bibitem{King:2015oxa}
S.~King, M.~Muhlleitner, R.~Nevzorov and K.~Walz, \emph{{Exploring the
  CP-violating NMSSM: EDM Constraints and Phenomenology}},
  \href{https://doi.org/10.1016/j.nuclphysb.2015.11.003}{\emph{Nucl. Phys. B}
  {\bfseries 901} (2015) 526}
  [\href{https://arxiv.org/abs/1508.03255}{{\ttfamily 1508.03255}}].

\bibitem{Choi:1999aj}
S.Y.~Choi and J.S.~Lee, \emph{{MSSM Higgs boson production at hadron colliders
  with explicit CP violation}},
  \href{https://doi.org/10.1103/PhysRevD.61.115002}{\emph{Phys. Rev. D}
  {\bfseries 61} (2000) 115002}
  [\href{https://arxiv.org/abs/hep-ph/9910557}{{\ttfamily hep-ph/9910557}}].

\bibitem{Choi:2001pg}
S.Y.~Choi, K.~Hagiwara and J.S.~Lee, \emph{{Higgs boson decays in the minimal
  supersymmetric standard model with radiative Higgs sector CP violation}},
  \href{https://doi.org/10.1103/PhysRevD.64.032004}{\emph{Phys. Rev. D}
  {\bfseries 64} (2001) 032004}
  [\href{https://arxiv.org/abs/hep-ph/0103294}{{\ttfamily hep-ph/0103294}}].

\bibitem{Choi:2002zp}
S.Y.~Choi, M.~Drees, J.S.~Lee and J.~Song, \emph{{Supersymmetric Higgs boson
  decays in the MSSM with explicit CP violation}},
  \href{https://doi.org/10.1007/s10052-002-0997-8}{\emph{Eur. Phys. J. C}
  {\bfseries 25} (2002) 307}
  [\href{https://arxiv.org/abs/hep-ph/0204200}{{\ttfamily hep-ph/0204200}}].

\bibitem{Lee:2008eqa}
J.S.~Lee, \emph{{Manifestations of CP Violation in the MSSM Higgs Sector}},
  \href{https://doi.org/10.1063/1.3051964}{\emph{AIP Conf. Proc.} {\bfseries
  1078} (2009) 36} [\href{https://arxiv.org/abs/0808.2014}{{\ttfamily
  0808.2014}}].

\bibitem{Bechtle:2013wla}
P.~Bechtle, O.~Brein, S.~Heinemeyer, O.~St\r{a}l, T.~Stefaniak, G.~Weiglein
  et~al., \emph{{$\mathsf{HiggsBounds}-4$: Improved Tests of Extended Higgs
  Sectors against Exclusion Bounds from LEP, the Tevatron and the LHC}},
  \href{https://doi.org/10.1140/epjc/s10052-013-2693-2}{\emph{Eur. Phys. J. C}
  {\bfseries 74} (2014) 2693}
  [\href{https://arxiv.org/abs/1311.0055}{{\ttfamily 1311.0055}}].

\bibitem{Kraml:2019sis}
S.~Kraml, T.Q.~Loc, D.T.~Nhung and L.D.~Ninh, \emph{{Constraining new physics
  from Higgs measurements with Lilith: update to LHC Run 2 results}},
  \href{https://doi.org/10.21468/SciPostPhys.7.4.052}{\emph{SciPost Phys.}
  {\bfseries 7} (2019) 052} [\href{https://arxiv.org/abs/1908.03952}{{\ttfamily
  1908.03952}}].

\bibitem{Fontes:2017zfn}
D.~Fontes, M.~M\"uhlleitner, J.C.~Rom\~ao, R.~Santos, J.a.P.~Silva and
  J.~Wittbrodt, \emph{{The C2HDM revisited}},
  \href{https://doi.org/10.1007/JHEP02(2018)073}{\emph{JHEP} {\bfseries 02}
  (2018) 073} [\href{https://arxiv.org/abs/1711.09419}{{\ttfamily
  1711.09419}}].

\bibitem{Sirunyan:2020xwk}
{\scshape CMS} collaboration, \emph{{A measurement of the Higgs boson mass in
  the diphoton decay channel}},
  \href{https://doi.org/10.1016/j.physletb.2020.135425}{\emph{Phys. Lett. B}
  {\bfseries 805} (2020) 135425}
  [\href{https://arxiv.org/abs/2002.06398}{{\ttfamily 2002.06398}}].

\bibitem{Muhlleitner:2020wwk}
M.~M\"uhlleitner, M.O.~Sampaio, R.~Santos and J.~Wittbrodt, \emph{{ScannerS:
  Parameter Scans in Extended Scalar Sectors}},
  \href{https://arxiv.org/abs/2007.02985}{{\ttfamily 2007.02985}}.

\bibitem{Bernon:2015hsa}
J.~Bernon and B.~Dumont, \emph{{Lilith: a tool for constraining new physics
  from Higgs measurements}},
  \href{https://doi.org/10.1140/epjc/s10052-015-3645-9}{\emph{Eur. Phys. J. C}
  {\bfseries 75} (2015) 440}
  [\href{https://arxiv.org/abs/1502.04138}{{\ttfamily 1502.04138}}].

\bibitem{Djouadi:2013qya}
A.~Djouadi and G.~Moreau, \emph{{The couplings of the Higgs boson and its CP
  properties from fits of the signal strengths and their ratios at the 7+8 TeV
  LHC}}, \href{https://doi.org/10.1140/epjc/s10052-013-2512-9}{\emph{Eur. Phys.
  J. C} {\bfseries 73} (2013) 2512}
  [\href{https://arxiv.org/abs/1303.6591}{{\ttfamily 1303.6591}}].

\bibitem{Ginzburg:2004vp}
I.F.~Ginzburg and M.~Krawczyk, \emph{{Symmetries of two Higgs doublet model and
  CP violation}}, \href{https://doi.org/10.1103/PhysRevD.72.115013}{\emph{Phys.
  Rev. D} {\bfseries 72} (2005) 115013}
  [\href{https://arxiv.org/abs/hep-ph/0408011}{{\ttfamily hep-ph/0408011}}].

\bibitem{Bertrand:2020lyb}
M.~Bertrand, S.~Kraml, T.Q.~Loc, D.T.~Nhung and L.D.~Ninh, \emph{{Constraining
  new physics from Higgs measurements with Lilith-2}},
  \href{https://doi.org/10.22323/1.392.0040}{\emph{PoS} {\bfseries TOOLS2020}
  (2021) 040} [\href{https://arxiv.org/abs/2012.11408}{{\ttfamily
  2012.11408}}].

\bibitem{kernighan2006c}
B.W.~Kernighan and D.M.~Ritchie, \emph{The C programming language} (2006).

\bibitem{stroustrup2000c++}
B.~Stroustrup, \emph{The C++ programming language}, Pearson Education India
  (2000).

\bibitem{brun1997root}
R.~Brun and F.~Rademakers, \emph{Root—an object oriented data analysis
  framework}, {\emph{Nuclear Instruments and Methods in Physics Research
  Section A: Accelerators, Spectrometers, Detectors and Associated Equipment}
  {\bfseries 389} (1997) 81}.

\bibitem{Bechtle:2015pma}
P.~Bechtle, S.~Heinemeyer, O.~Stal, T.~Stefaniak and G.~Weiglein,
  \emph{{Applying Exclusion Likelihoods from LHC Searches to Extended Higgs
  Sectors}}, \href{https://doi.org/10.1140/epjc/s10052-015-3650-z}{\emph{Eur.
  Phys. J. C} {\bfseries 75} (2015) 421}
  [\href{https://arxiv.org/abs/1507.06706}{{\ttfamily 1507.06706}}].

\bibitem{Amhis:2019ckw}
{\scshape HFLAV} collaboration, \emph{{Averages of b-hadron, c-hadron, and
  $\tau $-lepton properties as of 2018}},
  \href{https://doi.org/10.1140/epjc/s10052-020-8156-7}{\emph{Eur. Phys. J. C}
  {\bfseries 81} (2021) 226}
  [\href{https://arxiv.org/abs/1909.12524}{{\ttfamily 1909.12524}}].

\bibitem{Antusch:2020ngh}
S.~Antusch, O.~Fischer, A.~Hammad and C.~Scherb, \emph{{Testing CP Properties
  of Extra Higgs States at the HL-LHC}},
  \href{https://doi.org/10.1007/JHEP03(2021)200}{\emph{JHEP} {\bfseries 03}
  (2021) 200} [\href{https://arxiv.org/abs/2011.10388}{{\ttfamily
  2011.10388}}].

\bibitem{Baak:2012kk}
M.~Baak, M.~Goebel, J.~Haller, A.~Hoecker, D.~Kennedy, R.~Kogler et~al.,
  \emph{{The Electroweak Fit of the Standard Model after the Discovery of a New
  Boson at the LHC}},
  \href{https://doi.org/10.1140/epjc/s10052-012-2205-9}{\emph{Eur. Phys. J. C}
  {\bfseries 72} (2012) 2205}
  [\href{https://arxiv.org/abs/1209.2716}{{\ttfamily 1209.2716}}].

\bibitem{Enomoto:2015wbn}
T.~Enomoto and R.~Watanabe, \emph{{Flavor constraints on the Two Higgs Doublet
  Models of Z$_{2}$ symmetric and aligned types}},
  \href{https://doi.org/10.1007/JHEP05(2016)002}{\emph{JHEP} {\bfseries 05}
  (2016) 002} [\href{https://arxiv.org/abs/1511.05066}{{\ttfamily
  1511.05066}}].

\bibitem{ParticleDataGroup:2020ssz}
{\scshape Particle Data Group} collaboration, \emph{{Review of Particle
  Physics}}, \href{https://doi.org/10.1093/ptep/ptaa104}{\emph{PTEP} {\bfseries
  2020} (2020) 083C01}.

\bibitem{Funk:2011ad}
G.~Funk, D.~O'Neil and R.M.~Winters, \emph{{What the Oblique Parameters S, T,
  and U and Their Extensions Reveal About the 2HDM: A Numerical Analysis}},
  \href{https://doi.org/10.1142/S0217751X12500212}{\emph{Int. J. Mod. Phys. A}
  {\bfseries 27} (2012) 1250021}
  [\href{https://arxiv.org/abs/1110.3812}{{\ttfamily 1110.3812}}].

\bibitem{Grimus:2007if}
W.~Grimus, L.~Lavoura, O.M.~Ogreid and P.~Osland, \emph{{A Precision constraint
  on multi-Higgs-doublet models}},
  \href{https://doi.org/10.1088/0954-3899/35/7/075001}{\emph{J. Phys. G}
  {\bfseries 35} (2008) 075001}
  [\href{https://arxiv.org/abs/0711.4022}{{\ttfamily 0711.4022}}].

\bibitem{Abe:2013qla}
T.~Abe, J.~Hisano, T.~Kitahara and K.~Tobioka, \emph{{Gauge invariant Barr-Zee
  type contributions to fermionic EDMs in the two-Higgs doublet models}},
  \href{https://doi.org/10.1007/JHEP01(2014)106}{\emph{JHEP} {\bfseries 01}
  (2014) 106} [\href{https://arxiv.org/abs/1311.4704}{{\ttfamily 1311.4704}}].

\bibitem{BowserChao:1997bb}
D.~Bowser-Chao, D.~Chang and W.-Y.~Keung, \emph{{Electron electric dipole
  moment from CP violation in the charged Higgs sector}},
  \href{https://doi.org/10.1103/PhysRevLett.79.1988}{\emph{Phys. Rev. Lett.}
  {\bfseries 79} (1997) 1988}
  [\href{https://arxiv.org/abs/hep-ph/9703435}{{\ttfamily hep-ph/9703435}}].

\bibitem{Barr:1990vd}
S.M.~Barr and A.~Zee, \emph{{Electric Dipole Moment of the Electron and of the
  Neutron}}, \href{https://doi.org/10.1103/PhysRevLett.65.21}{\emph{Phys. Rev.
  Lett.} {\bfseries 65} (1990) 21}.

\bibitem{Chun:2019oix}
E.J.~Chun, J.~Kim and T.~Mondal, \emph{{Electron EDM and Muon anomalous
  magnetic moment in Two-Higgs-Doublet Models}},
  \href{https://doi.org/10.1007/JHEP12(2019)068}{\emph{JHEP} {\bfseries 12}
  (2019) 068} [\href{https://arxiv.org/abs/1906.00612}{{\ttfamily
  1906.00612}}].

\bibitem{Altmannshofer:2020shb}
W.~Altmannshofer, S.~Gori, N.~Hamer and H.H.~Patel, \emph{{Electron EDM in the
  complex two-Higgs doublet model}},
  \href{https://doi.org/10.1103/PhysRevD.102.115042}{\emph{Phys. Rev. D}
  {\bfseries 102} (2020) 115042}
  [\href{https://arxiv.org/abs/2009.01258}{{\ttfamily 2009.01258}}].

\bibitem{Wang:2018hnw}
L.~Wang, J.M.~Yang, M.~Zhang and Y.~Zhang, \emph{{Revisiting lepton-specific
  2HDM in light of muon $g?2$ anomaly}},
  \href{https://doi.org/10.1016/j.physletb.2018.11.045}{\emph{Phys. Lett. B}
  {\bfseries 788} (2019) 519}
  [\href{https://arxiv.org/abs/1809.05857}{{\ttfamily 1809.05857}}].

\bibitem{Chun:2015xfx}
E.J.~Chun, \emph{{The muon g\ensuremath{-}2 in two-Higgs-doublet models}},
  \href{https://doi.org/10.1051/epjconf/201611801006}{\emph{EPJ Web Conf.}
  {\bfseries 118} (2016) 01006}
  [\href{https://arxiv.org/abs/1511.05225}{{\ttfamily 1511.05225}}].

\bibitem{Wang:2014sda}
L.~Wang and X.-F.~Han, \emph{{A light pseudoscalar of 2HDM confronted with muon
  g-2 and experimental constraints}},
  \href{https://doi.org/10.1007/JHEP05(2015)039}{\emph{JHEP} {\bfseries 05}
  (2015) 039} [\href{https://arxiv.org/abs/1412.4874}{{\ttfamily 1412.4874}}].

\bibitem{Broggio:2014mna}
A.~Broggio, E.J.~Chun, M.~Passera, K.M.~Patel and S.K.~Vempati, \emph{{Limiting
  two-Higgs-doublet models}},
  \href{https://doi.org/10.1007/JHEP11(2014)058}{\emph{JHEP} {\bfseries 11}
  (2014) 058} [\href{https://arxiv.org/abs/1409.3199}{{\ttfamily 1409.3199}}].

\bibitem{Ilisie:2015tra}
V.~Ilisie, \emph{{New Barr-Zee contributions to $\mathbf{(g-2)_\mu}$ in
  two-Higgs-doublet models}},
  \href{https://doi.org/10.1007/JHEP04(2015)077}{\emph{JHEP} {\bfseries 04}
  (2015) 077} [\href{https://arxiv.org/abs/1502.04199}{{\ttfamily
  1502.04199}}].

\bibitem{Abe:2015oca}
T.~Abe, R.~Sato and K.~Yagyu, \emph{{Lepton-specific two Higgs doublet model as
  a solution of muon g \ensuremath{-} 2 anomaly}},
  \href{https://doi.org/10.1007/JHEP07(2015)064}{\emph{JHEP} {\bfseries 07}
  (2015) 064} [\href{https://arxiv.org/abs/1504.07059}{{\ttfamily
  1504.07059}}].

\bibitem{Chen:2017com}
C.-Y.~Chen, H.-L.~Li and M.~Ramsey-Musolf, \emph{{CP-Violation in the Two Higgs
  Doublet Model: from the LHC to EDMs}},
  \href{https://doi.org/10.1103/PhysRevD.97.015020}{\emph{Phys. Rev. D}
  {\bfseries 97} (2018) 015020}
  [\href{https://arxiv.org/abs/1708.00435}{{\ttfamily 1708.00435}}].

\bibitem{Andreev:2018ayy}
{\scshape ACME} collaboration, \emph{{Improved limit on the electric dipole
  moment of the electron}},
  \href{https://doi.org/10.1038/s41586-018-0599-8}{\emph{Nature} {\bfseries
  562} (2018) 355}.

\bibitem{Mendez:1991gp}
A.~Mendez and A.~Pomarol, \emph{{Signals of CP violation in the Higgs sector}},
  \href{https://doi.org/10.1016/0370-2693(91)91836-K}{\emph{Phys. Lett. B}
  {\bfseries 272} (1991) 313}.

\bibitem{Bernreuther:1993hq}
W.~Bernreuther and A.~Brandenburg, \emph{{Tracing CP violation in the
  production of top quark pairs by multiple TeV proton proton collisions}},
  \href{https://doi.org/10.1103/PhysRevD.49.4481}{\emph{Phys. Rev. D}
  {\bfseries 49} (1994) 4481}
  [\href{https://arxiv.org/abs/hep-ph/9312210}{{\ttfamily hep-ph/9312210}}].

\bibitem{Ogreid:2018xdx}
O.M.~Ogreid, P.~Osland and M.N.~Rebelo, \emph{{CP violation in extended Higgs
  sectors}}, \href{https://doi.org/10.22323/1.318.0044}{\emph{PoS} {\bfseries
  CORFU2017} (2018) 044} [\href{https://arxiv.org/abs/1804.02529}{{\ttfamily
  1804.02529}}].

\bibitem{Ginzburg:2001ss}
I.F.~Ginzburg, M.~Krawczyk and P.~Osland, \emph{{Resolving SM like scenarios
  via Higgs boson production at a photon collider. 1. 2HDM versus SM}},
  \href{https://arxiv.org/abs/hep-ph/0101208}{{\ttfamily hep-ph/0101208}}.

\bibitem{Ferreira:2014sld}
P.M.~Ferreira, R.~Guedes, J.F.~Gunion, H.E.~Haber, M.O.P.~Sampaio and
  R.~Santos, \emph{{The CP-conserving 2HDM after the 8 TeV run}},  in
  \emph{{22nd International Workshop on Deep-Inelastic Scattering and Related
  Subjects}}, 7, 2014 [\href{https://arxiv.org/abs/1407.4396}{{\ttfamily
  1407.4396}}].

\bibitem{Accomando:2019jrb}
E.~Accomando, D.~Englert, J.~Hays and S.~Moretti, \emph{{Voyage Across the 2HDM
  Type-II with Magellan}},  \href{https://arxiv.org/abs/1905.07313}{{\ttfamily
  1905.07313}}.

\bibitem{ATLAS:2022fnp}
{\scshape ATLAS} collaboration, \emph{{Measurements of the Higgs boson
  inclusive and differential fiducial cross-sections in the diphoton decay
  channel with $pp$ collisions at $\sqrt{s} = 13$ TeV with the ATLAS
  detector}},  \href{https://arxiv.org/abs/2202.00487}{{\ttfamily 2202.00487}}.

\bibitem{ATLAS:2021tbi}
{\scshape ATLAS} collaboration, \emph{{Constraints on Higgs boson production
  with large transverse momentum using $H\rightarrow b\bar{b}$ decays in the
  ATLAS detector}},
  \href{https://doi.org/10.1103/PhysRevD.105.092003}{\emph{Phys. Rev. D}
  {\bfseries 105} (2022) 092003}
  [\href{https://arxiv.org/abs/2111.08340}{{\ttfamily 2111.08340}}].

\bibitem{ATLAS:2020wny}
{\scshape ATLAS} collaboration, \emph{{Measurements of the Higgs boson
  inclusive and differential fiducial cross sections in the 4$\ell$ decay
  channel at $\sqrt{s}$ = 13 TeV}},
  \href{https://doi.org/10.1140/epjc/s10052-020-8223-0}{\emph{Eur. Phys. J. C}
  {\bfseries 80} (2020) 942}
  [\href{https://arxiv.org/abs/2004.03969}{{\ttfamily 2004.03969}}].

\bibitem{Chen:2018shg}
N.~Chen, T.~Han, S.~Su, W.~Su and Y.~Wu, \emph{{Type-II 2HDM under the
  Precision Measurements at the $Z$-pole and a Higgs Factory}},
  \href{https://doi.org/10.1007/JHEP03(2019)023}{\emph{JHEP} {\bfseries 03}
  (2019) 023} [\href{https://arxiv.org/abs/1808.02037}{{\ttfamily
  1808.02037}}].

\bibitem{Ginzburg:2005dt}
I.F.~Ginzburg and I.P.~Ivanov, \emph{{Tree-level unitarity constraints in the
  most general 2HDM}},
  \href{https://doi.org/10.1103/PhysRevD.72.115010}{\emph{Phys. Rev. D}
  {\bfseries 72} (2005) 115010}
  [\href{https://arxiv.org/abs/hep-ph/0508020}{{\ttfamily hep-ph/0508020}}].

\bibitem{Grinstein:2015rtl}
B.~Grinstein, C.W.~Murphy and P.~Uttayarat, \emph{{One-loop corrections to the
  perturbative unitarity bounds in the CP-conserving two-Higgs doublet model
  with a softly broken $ {\mathrm{\mathbb{Z}}}_2 $ symmetry}},
  \href{https://doi.org/10.1007/JHEP06(2016)070}{\emph{JHEP} {\bfseries 06}
  (2016) 070} [\href{https://arxiv.org/abs/1512.04567}{{\ttfamily
  1512.04567}}].

\bibitem{Cacchio:2016qyh}
V.~Cacchio, D.~Chowdhury, O.~Eberhardt and C.W.~Murphy, \emph{{Next-to-leading
  order unitarity fits in Two-Higgs-Doublet models with soft $\mathbb{Z}_2$
  breaking}}, \href{https://doi.org/10.1007/JHEP11(2016)026}{\emph{JHEP}
  {\bfseries 11} (2016) 026}
  [\href{https://arxiv.org/abs/1609.01290}{{\ttfamily 1609.01290}}].

\bibitem{Harlander:2012pb}
R.V.~Harlander, S.~Liebler and H.~Mantler, \emph{{SusHi: A program for the
  calculation of Higgs production in gluon fusion and bottom-quark annihilation
  in the Standard Model and the MSSM}},
  \href{https://doi.org/10.1016/j.cpc.2013.02.006}{\emph{Comput. Phys. Commun.}
  {\bfseries 184} (2013) 1605}
  [\href{https://arxiv.org/abs/1212.3249}{{\ttfamily 1212.3249}}].

\bibitem{Djouadi:1997yw}
A.~Djouadi, J.~Kalinowski and M.~Spira, \emph{{HDECAY: A Program for Higgs
  boson decays in the standard model and its supersymmetric extension}},
  \href{https://doi.org/10.1016/S0010-4655(97)00123-9}{\emph{Comput. Phys.
  Commun.} {\bfseries 108} (1998) 56}
  [\href{https://arxiv.org/abs/hep-ph/9704448}{{\ttfamily hep-ph/9704448}}].

\bibitem{Djouadi:2018xqq}
A.~Djouadi, J.~Kalinowski, M.~Muehlleitner and M.~Spira, \emph{{HDECAY:
  Twenty$_{++}$ years after}},
  \href{https://doi.org/10.1016/j.cpc.2018.12.010}{\emph{Comput. Phys. Commun.}
  {\bfseries 238} (2019) 214}
  [\href{https://arxiv.org/abs/1801.09506}{{\ttfamily 1801.09506}}].

\bibitem{Yagyu:2012qp}
K.~Yagyu, \emph{{Studies on Extended Higgs Sectors as a Probe of New Physics
  Beyond the Standard Model}}, Ph.D. thesis, Toyama U., 2012.
\newblock \href{https://arxiv.org/abs/1204.0424}{{\ttfamily 1204.0424}}.

\bibitem{Kanemura:2004mg}
S.~Kanemura, Y.~Okada, E.~Senaha and C.P.~Yuan, \emph{{Higgs coupling constants
  as a probe of new physics}},
  \href{https://doi.org/10.1103/PhysRevD.70.115002}{\emph{Phys. Rev. D}
  {\bfseries 70} (2004) 115002}
  [\href{https://arxiv.org/abs/hep-ph/0408364}{{\ttfamily hep-ph/0408364}}].

\bibitem{Butterworth:2010ym}
J.M.~Butterworth et~al., \emph{{THE TOOLS AND MONTE CARLO WORKING GROUP Summary
  Report from the Les Houches 2009 Workshop on TeV Colliders}},  in \emph{{6th
  Les Houches Workshop on Physics at TeV Colliders}}, 3, 2010
  [\href{https://arxiv.org/abs/1003.1643}{{\ttfamily 1003.1643}}].

\bibitem{ATLAS:2017tlw}
{\scshape ATLAS} collaboration, \emph{{Search for heavy ZZ resonances in the
  $\ell ^+\ell ^-\ell ^+\ell ^-$ and $\ell ^+\ell ^-\nu \bar{\nu }$ final
  states using proton\textendash{}proton collisions at $\sqrt{s}= 13$ $\text
  {TeV}$ with the ATLAS detector}},
  \href{https://doi.org/10.1140/epjc/s10052-018-5686-3}{\emph{Eur. Phys. J. C}
  {\bfseries 78} (2018) 293}
  [\href{https://arxiv.org/abs/1712.06386}{{\ttfamily 1712.06386}}].

\bibitem{CMS:2018mmw}
{\scshape CMS} collaboration, \emph{{Measurements of properties of the Higgs
  boson in the four-lepton final state at $\sqrt{s}=13~\mathrm{TeV}$}}, .

\bibitem{CMS:2017jkd}
{\scshape CMS} collaboration, \emph{{Measurements of properties of the Higgs
  boson decaying into four leptons in pp collisions at sqrt{s} = 13 TeV}}, .

\bibitem{Cosmai:2021hvc}
L.~Cosmai and M.~Consoli, \emph{{Experimental signals for a second resonance of
  the Higgs field}},  \href{https://arxiv.org/abs/2111.08962}{{\ttfamily
  2111.08962}}.

\bibitem{Antusch:2021oit}
S.~Antusch, O.~Fischer, A.~Hammad and C.~Scherb, \emph{{Explaining excesses in
  four-leptons at the LHC with a double peak from a CP violating Two Higgs
  Doublet Model}},  \href{https://arxiv.org/abs/2112.00921}{{\ttfamily
  2112.00921}}.

\end{thebibliography}\endgroup

%----------------------------------------------------------------------------------------

\end{document}